%% file: ms.tex
\documentclass[aps,10pt,twocolumn,floats,floatfix,prd,nofootinbib,superscriptaddress,showpacs,showkeys]{revtex4-1}

\usepackage{graphicx}
\usepackage{amsmath}
\usepackage{amssymb}
\usepackage{siunitx}
\usepackage{verbatim}
\usepackage{rotate}
\usepackage{color}
\usepackage{enumitem}
\usepackage{threeparttable}

\usepackage{amsmath}
\usepackage{amssymb}
\usepackage{siunitx}

\usepackage{tikz}
\usetikzlibrary{graphs,positioning}

\DeclareFontFamily{OT1}{pzc}{}
\DeclareFontShape{OT1}{pzc}{m}{it}%
            {<-> s * [1.10] pzcmi7t}{}
\DeclareMathAlphabet{\mathscr}{OT1}{pzc}%
                                {m}{it}

\definecolor{RedWine}{rgb}{0.743,0,0}
\definecolor{GrassGreen}{rgb}{0.125,0.75,0.125}
\definecolor{RoyalBlue}{rgb}{0.25,0.41,0.88}

\providecommand{\jcap}{JCAP}
\providecommand{\mnras}{MNRAS}
\providecommand{\prd}{Phys. Rev. D}
\providecommand{\physrep}{Phys.~Rep.}

\newcommand{\be}{\begin{equation}}
\newcommand{\ee}{\end{equation}}
\newcommand{\bea}{\begin{eqnarray}}
\newcommand{\eea}{\end{eqnarray}}
\def\ba#1\ea{\begin{align}#1\end{align}}

\newcommand{\refeq}[1]{Eq.~(\ref{eq:#1})}          
\newcommand{\refeqs}[2]{Eqs.~(\ref{eq:#1})--(\ref{eq:#2})}          
\newcommand{\reffig}[1]{Fig.~\ref{fig:#1}}          
\newcommand{\refsec}[1]{Sec.~\ref{sec:#1}}
\newcommand{\refapp}[1]{App.~\ref{app:#1}}
\newcommand{\reftab}[1]{Tab.~\ref{tab:#1}}
\newcommand{\vs}{\nonumber\\}

\newcommand{\bfx}{\mathbf{x}}

\newcommand{\bfk}{\mathbf{k}}

\newcommand{\nhat}{\hat{\mathbf{n}}}
\newcommand{\khat}{\hat{\mathbf{k}}}
\newcommand{\xhat}{\hat{\mathbf{x}}}

\newcommand{\orderof}{\mathcal{O}}

\DeclareSIUnit \parsec {pc}
\DeclareSIUnit \h {\text{$h$}}

\renewcommand{\min}{\mathrm{min}}
\renewcommand{\max}{\mathrm{max}}

\newcommand{\Q}{\mathcal{Q}}
\newcommand{\dd}{\mathrm{d}}
\newcommand{\cW}{\mathcal{W}}

\usepackage{standalone}
\usetikzlibrary{calc,graphs,positioning}
\usepackage{ifthen}

\newcommand{\changed}[1]{#1}

\usepackage[pdfborder={0 0 0}]{hyperref}
\usepackage{cleveref}

\begin{document}

\title{Fast and accurate computation of projected two-point functions}

\author{Henry S. \surname{Grasshorn Gebhardt}}
\email{hsg113@psu.edu}
\author{Donghui Jeong}
\affiliation{Department of Astronomy and Astrophysics, and Institute
	for Gravitation and the Cosmos,
	The Pennsylvania State University, University Park, Pennsylvania 16802, USA}

\begin{abstract}
 We present the two-point function from fast and accurate spherical Bessel
 transformation (2-FAST) algorithm\changed{\footnote{\changed{Our code is
 available at \url{https://github.com/hsgg/twoFAST}}.}}
 for a fast and accurate computation of
 integrals involving one or two spherical Bessel functions. These types of
 integrals occur when projecting the galaxy power spectrum $P(k)$ onto the
 configuration space, $\xi_\ell^\nu(r)$, or spherical harmonic space,
 $C_\ell(\chi,\chi')$. First, we employ the FFTLog transformation of the power
 spectrum to divide the calculation into $P(k)$-dependent coefficients and
 $P(k)$-independent integrations of basis functions multiplied by
 spherical Bessel functions.
 We find analytical expressions for the latter integrals in terms of special
 functions, for which recursion provides a fast and accurate evaluation.
 The algorithm, therefore, circumvents direct integration of highly
 oscillating spherical Bessel functions.
\end{abstract}
\date{\today}

\maketitle

\section{Introduction}\label{sec:intro}
In standard cosmology, the large-scale structure of the Universe is
statistically homogeneous and isotropic and evolved from nearly
Gaussian \citep{maldacena:2003} primordial curvature perturbations
\citep{mukhanov/chibisov:1981,hawking:1982,guth/pi:1982,bardeen/etal:1983}
generated during inflation \citep{starobinsky:1979,starobinsky:1982,guth:1981,sato:1981,linde:1982,albrecht/steinhardt:1982}.
The statistics of large-scale structure, therefore, are often predicted
in terms of the power spectrum $P(k)$ (the two-point correlation function in
Fourier space) that reflects the underlying spatial symmetry of the Universe,
and that connects directly with the primordial curvature power spectrum.
The predicted power spectrum at late times responds sensitively to key
cosmological parameters such as the dark energy equation of state, primordial
non-Gaussianity parameters, and total mass of neutrinos. This
makes the power spectrum a powerful cosmological probe \citep{percival/etal:2001,tegmark/etal:2004,tegmark/etal:2006,percival/etal:2007,blake/etal:2011,beutler/etal:2014,rota/etal:2017}.

The observed large-scale structure, however, does not enjoy full spatial
symmetry because all observations must be done within our past light cone;
at each cosmological distance, we observe the large-scale structure at a
different time. As a result, the time evolution of large-scale structure
genuinely breaks the homogeneity along the radial direction, and we are
left only with the spherical symmetry on the two-dimensional sky.

On the sky, the equivalent of the power spectrum $P(k)$ is the angular power
spectrum $C_\ell$, which is the two-point function in spherical harmonic space.
The harmonic-space basis $Y_{\ell m}(\xhat)$ is related to the Fourier basis
$e^{i\bfk\cdot\bfx}$ by Rayleigh's formula:
\ba
e^{i\bfk\cdot\bfx} &= 4\pi\sum_{\ell,m} i^\ell j_\ell(kx)\,
Y_{\ell m}(\khat)\,Y^*_{\ell m}(\xhat),
\ea
so that the angular power spectrum is related to the power spectrum $P(k)$
by integrals of the form
\begin{align}
\label{eq:well}
w_{\ell\ell'}(\chi,\chi')
&= \frac{2}{\pi} \int_0^\infty \dd{}k\,k^2\,P(k)\,j_\ell(k\chi)\,j_{\ell'}(k\chi'),
\end{align}
where $\chi$ and $\chi'$ are the comoving angular diameter distances at two
different epochs, and $j_\ell(z)$ are spherical Bessel functions. Note that we
consider the general case of $\ell'\neq\ell$, because the contribution from
vector or tensor quantities can couple adjacent $\ell$-modes. For example, in
order to account for the peculiar velocity effect on redshift-space
distortion to linear order,
one needs up to $\ell-\ell'=\pm4$.
We show an explicit expression of the angular power spectrum
$C_\ell(\chi,\chi')$ of galaxies in redshift space in terms of
$w_{\ell\ell'}(\chi,\chi')$ in \refapp{cl}.
The brute-force numerical integration of \refeq{well} is quite cumbersome and
time-consuming because it involves the evaluation of the spherical Bessel
functions with high degree $\ell$ and large arguments $k\chi$ at which
the $j_\ell(k\chi)$ functions are highly oscillatory. It is the oscillatory nature
of the spherical Bessel functions that delays the convergence of the numerical
integration. \changed{Additionally, these integrals are often needed to sample a
large area in $\chi$-$\chi'$-space.}

Although the spherical harmonic basis reflects the underlying spherical
symmetry and facilitates data analysis, intuition often works better in
configuration space. The prediction for the configuration-space galaxy
two-point correlation function that is valid on the spherical sky is often
called the \emph{wide-angle} formula
\citep{szalay/etal:1998,szapudi:2004,papai/szapudi:2008}
in contrast to the plane-parallel approximation \citep{kaiser:1987} that works
for small sky coverage. The building blocks of the wide-angle formula are
the configuration-space functions $\xi_\ell^\nu(r)$ defined
as
\ba
\xi^{\nu}_\ell(r) &\equiv \int_0^\infty \frac{k^2\dd{}k}{2\pi^2}\,P(k)\,
\frac{j_\ell(kr)}{(kr)^\nu}
\label{eq:xi}\,.
\ea
Using this notation, the linear two-point correlation function becomes
$\xi(r) = \xi_0^0(r)$, and calculation of the linear redshift-space galaxy
correlation function requires $\xi_2^0(r)$ and $\xi_4^0(r)$
\cite{hamilton:1992}. These functions also appear in calculating
the higher-order correlation functions \citep{slepian/eisenstein:2017},
the correlation functions of peaks \citep{BBKS:1986}, and nonlinear
correlation functions
\citep{McCullagh/szalay:2014,McEwen/etal:2016,schmittfull/vlah/McDonald:2016,fang/etal:2017}.
Note that, although not as cumbersome and time consuming as \refeq{well}, the
evaluation of \refeq{xi} also involves integrating over spherical Bessel
functions that are highly oscillatory in the $k\to \infty$ limit.

In this paper, we shall present a fast and accurate method of calculating the
integrations in \refeqs{well}{xi}. Specifically, we use the
fast Hankel transformation first proposed in \cite{talman:1978} and
\cite{siegman:1977}, and introduced
to the cosmology community in \cite{hamilton:2000}. Following
Ref.~\cite{hamilton:2000}, hereafter, we call it an FFTLog transformation.
The idea of \citet{talman:1978} is as follows. When changing the integration
variables to a logarithmic scale, the spherical Bessel integrations in
\refeqs{well}{xi} become convolutions. We then use the convolution theorem
to perform the integration: by first Fourier transforming the convolving
functions, then multiplying, and inverse Fourier transforming back.
The method requires no explicit computation and integration of
spherical Bessel functions. Instead, it requires the computation of
Gamma functions and the Gauss hypergeometric function ${}_2F_1$, which are the
FFTLog transformation of, respectively, one and two spherical
Bessel functions. Therefore, a fast and accurate calculation of
\refeqs{well}{xi} boils down to a fast and accurate computation of the
Gamma function and Gauss hypergeometric function for any $\ell$ and any ratio
$R=\chi'/\chi$. We shall achieve this goal by using a recursion.

A recent paper by \citet{assassi/etal:2017} has also proposed a similar
algorithm to efficiently calculate the angular two-point function
$w_{\ell\ell}(\chi,\chi')$.
Here, we have further extended the algorithm by studying a fast and
accurate method to calculate the Gauss hypergeometric functions, and by
including the cases for $\ell\neq\ell'$. We also study the choice of
parameters such as the biasing parameter $q$ and the size of the FFTLog
transformation $N$ in a systematic way.

This paper is organized as follows. In \refsec{phi} we introduce the FFTLog
transformation. We present the two-point function from fast and accurate spherical Bessel transformation (2-FAST)
algorithm for computing the real-space correlation functions $\xi^\nu_\ell(r)$
in \refsec{xi} and the harmonic-space two-point correlation functions
$w_{\ell\ell'}(\chi,\chi')$ in \refsec{well}. We then apply the 2-FAST
algorithm to the galaxy two-point correlation function,
to the angular power spectrum of the lensing potential, and
to the lensing-convergence-galaxy
cross-correlation functions in \refsec{applications}.
We conclude in \refsec{conclusion}. In \refapp{discrete}
we present discrete versions of the equations that we use for the
implementation. We study the effect of choosing a different biasing parameter
$q$, sampling $N$, and integration interval $G$ in \refapp{xiq}.
In \refapp{transmat} we compare our algorithm to a traditional integration method.
We lay out the details of our method of calculating the hypergeometric function
in \refapp{hyp2f1-contiguity}, \refapp{hyp2f1}, and \refapp{Mellell-recurrence}.
We show explicitly the relation between $w_{\ell\ell'}$ and the observed
galaxy correlation function $C_\ell$ in redshift space in \refapp{cl}.
We summarize a high-accuracy numerical algorithm (the Lucas algorithm
\cite{lucas1995}) that we use to benchmark our result in \refapp{lucas1995}.
Finally, we derive the extended Limber approximation for $\ell\neq\ell'$ cases
in \refapp{general_limber}.

Throughout, we use a flat $\Lambda$CDM universe with
$w=-1$,
$h=0.6778$,
$\Omega_bh^2 = 0.022307$,
$\Omega_ch^2 = 0.11865$,
$\Omega_\nu h^2 = 0.000638$,
$T_\text{CMB} = \SI{2.7255}{\kelvin}$,
$n_s = 0.9672$, and
$A_s = \num{2.147e-9}$
as the reference cosmology, \changed{where $h \equiv H_0 /
\SI{100}{\km\per\s\per\mega\parsec}$}.

All numerical implementations in this paper are done in the high-level
programming language \texttt{Julia}\footnote{\url{https://julialang.org}}, which
is a just-in-time compiled language developed specifically for scientific
numerical computations. We use \texttt{Julia} version 0.6.
We run the tests on a laptop with an Intel(R) Core(TM)
i7-4750HQ CPU, at \SI{3.1}{\giga\hertz} with \SI{11}{GiB\per\second} memory
access, and a \SI{360}{\mega B\per\second} SSD.
We have not yet parallelized the code, and all tests were run on a single core.
\changed{We make the code available publicly at
\url{https://github.com/hsgg/twoFAST}.}

\section{FFTLog transform of the power spectrum}
\label{sec:phi}
\begin{figure}
    \includegraphics[width=0.48\textwidth]{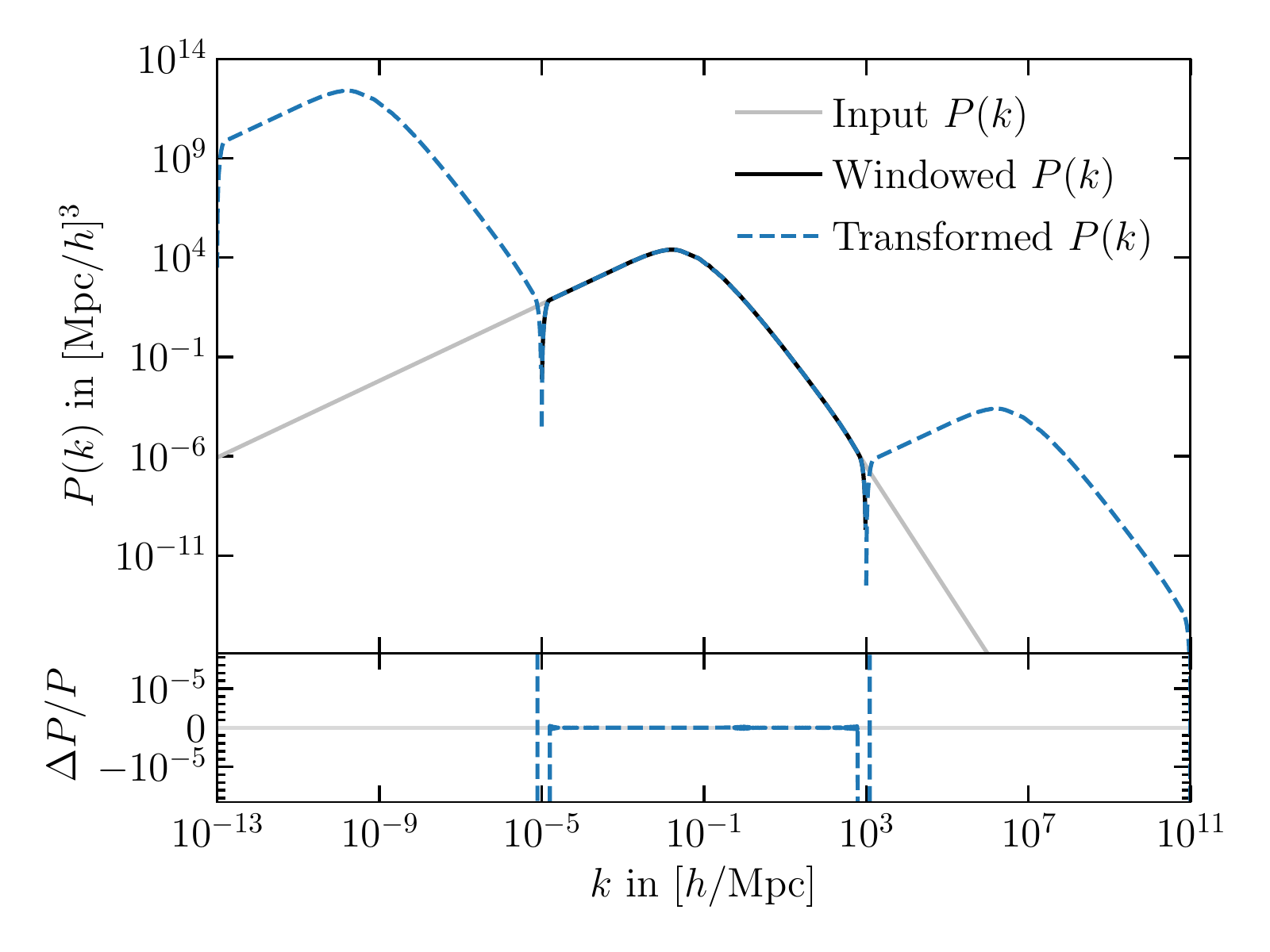}
    \caption{The linear matter power spectrum, with $q=2$, and
    $k_0=\SI{e-5}{\h\per\mega\parsec}$. In gray is the power spectrum, in black
    the windowed power spectrum [see \refeq{windowfn}], and in dashed blue the
    Fourier-transformed power spectrum as calculated by the discrete form of
    \refeq{pkdebiased}. The lower panel shows the relative difference to the
    input $P(k)$. Here, the number of sampling points between
    $k_\min=\SI{e-5}{\h\per\mega\parsec}$ and
    $k_\max=\SI{e3}{\h\per\mega\parsec}$ was taken to be $N=1024$.}
    \label{fig:pk}
\end{figure}
\begin{figure}
    \includegraphics[width=0.48\textwidth]{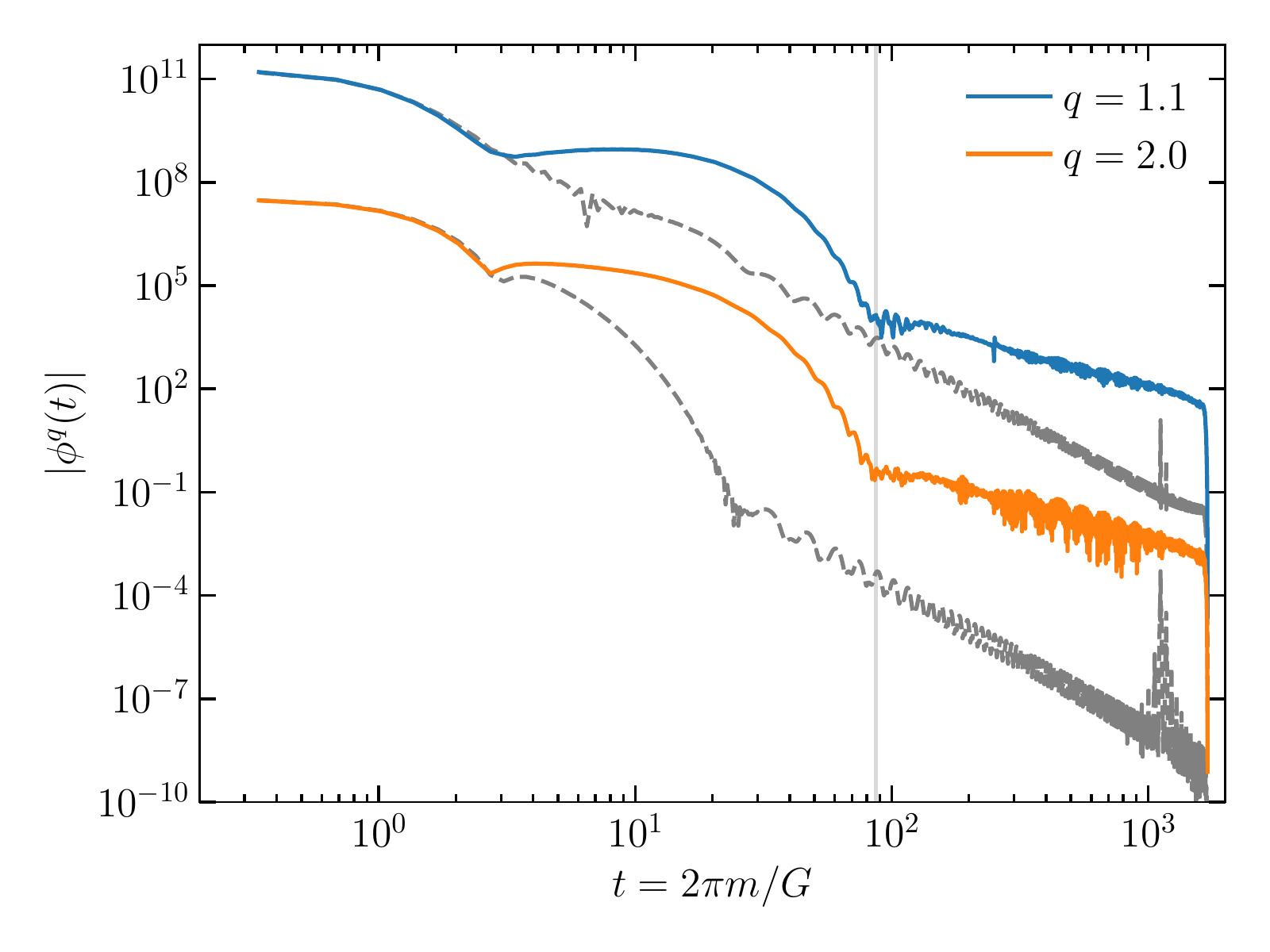}
    \caption{
        The absolute value $|\phi^{q}(t)|$ of the FFTLog
        transform of the biased power spectrum $k^{3-q}P(k)$ for two values
	of the biasing parameter ${q}$ (colored lines). For
        comparison, we also show the FFTLog transform of a power spectrum
        without the baryon acoustic oscillations (BAO) feature for each ${q}$
        (gray dashed lines with the same
        symbols as their BAO counterpart). The gray vertical line shows the
        maximum $t$ when using $N=512$ sampling points in the interval
        $k_\min=\SI{e-5}{\h\per\mega\parsec}$ to
	$k_\max=\SI{e3}{\h\per\mega\parsec}$. \changed{Here, $m$ is the mode of
	the Fourier transform, and $G=\ln(k_\max/k_\min)$.} }
    \label{fig:phiqt}
\end{figure}
The 2-FAST algorithm is based on the FFTLog transformation \cite{hamilton:2000}
of the power spectrum $P(k)$ which can be implemented by a fast Fourier
Transform (FFT) of the $P(k)$ sampled at wave numbers $k_i$ regularly sampled
in logarithmic space. In practice, we perform an FFTLog transformation of the
biased power spectrum
\ba
\label{eq:pkq}
\left(\frac{k}{k_0}\right)^{3-q} P(k) &= e^{(3-q)\kappa} P(k_0e^\kappa)\,,
\ea
in order to reduce numerical artifacts such as aliasing. Here, $q$ is the
biasing parameter and $\kappa$ is the logarithmic variable defined as
\ba
k &= k_0 e^\kappa
\ea
with some pivot wave number $k_0$. By defining the inverse Fourier transform of
the biased power spectrum as $\phi^{q}(x)$, we have the following Fourier
pair:
\ba
\label{eq:phiq}
\phi^{q}(x) &= \int \frac{\dd\kappa}{2\pi}\,e^{i\kappa x}\,
e^{(3-q)\kappa}\,P(k_0e^\kappa)
\\
P(k)
&=
e^{-(3-q)\kappa}
\int \dd{}x\,e^{-i\kappa x}\,\phi^{q}(x)
\label{eq:pkdebiased}
\ea
We present a discrete version of these equations suitable for numerical
implementation in \refeq{phi-discrete}.

Furthermore, in order to reduce ringing, we apply a window function to the
biased power spectrum before and after the Fourier transformation. We use the
same window function as \citet{McEwen/etal:2016} [their Eq.~(C.1), and
\refeq{windowfn} here]. \changed{This choice of the window function
ensures that the power spectrum vanishes smoothly at each end of the
integration interval, thus reducing ringing}.

We calculate the linear matter power spectrum by using \texttt{CAMB}
\cite{camb}.\footnote{http://camb.info/} However, we have modified \texttt{CAMB} so
that the output power spectrum prints more significant digits required for
a more accurate FFTLog transformation. Also, when the power spectrum is needed
outside of the range of the \texttt{CAMB} output, we extrapolate the linear power
spectrum by a power law for both high- and low-$k$ regions
\ba
\label{eq:pk-smallk}
\lim_{k\rightarrow0}P(k)      &= N_1\,k^{n_1} \\
\lim_{k\rightarrow\infty}P(k) &= N_2\,k^{n_2-4}\,,
\label{eq:pk-largek}
\ea
\changed{where the limits have been chosen so that both indices are similar to
the spectral index, i.e. $n_1 \simeq n_2 \simeq n_s$.
However,} we measure $n_1$ and $n_2$ to ensure that the extrapolated linear power
spectrum is smooth. Note that the asymptotic behavior in
\refeqs{pk-smallk}{pk-largek} implies that the FFTLog transform
\refeq{phiq} only converges when
\ba
\label{eq:phi:q}
n_2 - 1 < q < 3 + n_1\,.
\ea
For our reference cosmology, $n_1\simeq n_s=0.967$ and $n_2=0.85$, we find
$-0.15<q<3.967$.

\reffig{pk} shows the linear matter power spectrum for our fiducial
$\Lambda$CDM cosmology with the biasing parameter $q=2$. The blue dashed line
[``Transformed $P(k)$''] shows the result of \refeq{pkdebiased}, the solid gray
line [``Input $P(k)$''] shows the input $P(k)$, and the solid black line
[``Windowed $P(k)$''] shows the input $P(k)$ amputated by the window function
\changed{\refeq{windowfn}}. In this plot, the number of sample points is
$N=\num{1024}$ in the interval $k_\min=\SI{e-5}{\h\per\mega\parsec}$ to
$k_\max=\SI{e3}{\h\per\mega\parsec}$. The periodicity shown in the figure is
due to the use of the FFT. The global slope is due to the use of the biasing
parameter $q=2$, since in \refeq{pkdebiased} the integral is periodic, and it
is multiplied by $k^{-1}$.

In \reffig{phiqt} we show the FFTLog transformation $\phi^q(t)$ of the
linear matter power spectrum for two values of the biasing parameter:
$q=1.1$ (blue line) and $q=2$ (orange line).
In order to highlight the effect from the baryon acoustic oscillations (BAO),
we also show $\phi^{q}(t)$ for a linear power spectrum without BAO
(gray, dashed lines) that we have calculated from the fitting formula
given in Ref.~\cite{eisenstein/hu:1998}.
The BAO appears in $\phi^{q}(t)$ as the ``bump'' to the left of the gray
vertical line (indicating the Nyquist \emph{frequency} for the case $N=512$).

In principle, the choice of $q$ within the limits of \refeq{phi:q} should not
affect the result of the calculation. When implementing
\refeqs{phiq}{pkdebiased} as a finite sum, however, we can reduce the aliasing
effect by choosing a proper $q$ value. The rule of thumb is that the
Fourier-transformed function will decay quickly (thus, yielding
smaller aliasing) when the original function has a broader width (say,
measured by the full-width at half maximum).
With our parametrization in \refeqs{pk-smallk}{pk-largek}, the slopes of the
Fourier-transformed function $e^{(3-q)\kappa}P(k_0e^\kappa)$ are,
$e^{3+n_1-q}$ and $e^{n_2-1-q}$, respectively, at low- and high-$\kappa$
regions. A bigger $q$, therefore, would make the lower-$\kappa$ side shallower
and higher-$\kappa$ side steeper.
In \refapp{xiq}, we study the aliasing effect for different biasing
parameter $q$ and the resolution of FFTLog, $N$.
It turns out that the aliasing effect is smaller when the slopes on
both sides of the power spectrum are almost equal:
$q\simeq 1+ (n_1+n_2)/2 \simeq 1.9$
(see \reffig{phiqt_NN}). This is the choice of the $q$ value that we shall use
in \refsec{xi} when we calculate the overlapping of the power spectrum and one
spherical Bessel function. It turns out that, however, a smaller $q$-value is
desired when calculating $w_{\ell\ell'}(\chi,\chi')$.
We shall justify our choice of the biasing parameter $q$ in
\refapp{xiq}.

Note that in the implementation of 2-FAST, we shall use the ``coefficients''
$\phi^{q}(x)$ of the FFTLog transformation instead of the power spectrum;
thus, $\phi^{q}(x)$ is the only $P(k)$-dependent quantity of the
integration.

\section{Projection onto real space: power spectrum overlapping with one
spherical Bessel function}
\label{sec:xi}
We start from the integration of the power spectrum overlapping with one spherical
Bessel function:
\be
\xi_\ell^\nu(r) \equiv \int_0^\infty \frac{k^2\dd k}{2\pi^2} P(k)
\frac{j_\ell(kr)}{(kr)^\nu}\,.
\label{eq:xihere}
\ee
Here, we briefly outline the method and present some
examples, including the calculation of the real-space correlation function
$\xi(r)\equiv\xi^0_0(r)$ and its first and second derivatives.

The key observation is that, by introducing logarithmic variables $\kappa$
and $\rho$ such that
\ba
\label{eq:kappa-r}
k &= k_0 e^\kappa  &  r &= r_0 e^\rho\,,
\ea
with some pivot $k_0$ and $r_0$, the integration in \refeq{xihere} can be
expressed as a convolution:
\ba
\xi^\nu_\ell(r)
&= \frac{k_0^3e^{-(q_\nu+\nu)\rho}}{2\pi^2\alpha^\nu} \int_{-\infty}^\infty
\dd\kappa\,e^{(3-q_\nu-\nu)\kappa}P(k_0e^\kappa)
\vs&\qquad\qquad\qquad\quad\times
e^{q_\nu(\kappa+\rho)}\,j_\ell(\alpha e^{\kappa+\rho})\,.
\label{eq:xilog}
\ea
Here, we define $\alpha=k_0r_0$, and $q_\nu$ is the biasing parameter that may
depend on $\nu$. That the convolution in real space is a multiplication in
Fourier space motivates us to introduce the Fourier transform of the spherical
Bessel function $M_\ell^{q_\nu}(t)$:
\ba
e^{q_\nu\sigma}\,j_\ell(\alpha e^\sigma)
&= \int_{-\infty}^\infty \frac{\dd{}t}{2\pi}\,e^{i\sigma t}\,M_\ell^{q_\nu}(t)\,.
\label{eq:Mell}
\ea
Together with $\phi^{q_\nu+\nu}(t)$ that we defined earlier in
\refeq{phiq}, \refeq{xihere} becomes
\ba
\xi^\nu_\ell(r)
&= \frac{k_0^3e^{-(q_\nu+\nu)\rho}}{\pi\alpha^\nu}
\int_{-\infty}^\infty \frac{\dd{}t}{2\pi}\,
e^{i\rho t}\,
\phi^{q_\nu+\nu}(t)\,
M_\ell^{q_\nu}(t)\,.
\label{eq:xifftlog}
\ea
\refeq{xifftlog} is the key equation for the 2-FAST algorithm. The
cosmology-dependent part $\phi^{q_\nu+\nu}(t)$ is calculated as the FFTLog
transformation of the power spectrum as described in \refsec{phi}.
The cosmology-independent part $M^{q_\nu}_\ell(t)$ is calculated analytically by
inverting its definition \refeq{Mell}.
Defining a variable $s=\alpha e^\sigma$, the inverse Fourier
transformation may be written as
\ba
M_\ell^{q_\nu}(t)
&= \int_{-\infty}^\infty \dd{}\sigma\,e^{-it\sigma}\,e^{q_\nu\sigma}\,
j_\ell(\alpha e^\sigma)
\vs
&= \alpha^{it-q_\nu} \int_0^\infty \dd{}s\,s^{q_\nu-1-it}\,j_\ell(s)
\vs
&\equiv \alpha^{it-q_\nu} \, u_\ell(q_\nu-1-it)\,.
\label{eq:Ml}
\ea
The integral $u_\ell(n)$ is given by
\ba
u_\ell(n)
&\equiv \int_0^\infty \dd{}s\,s^n\,j_\ell(s)
= 2^{n-1} \sqrt{\pi}\,
\frac{\Gamma\big[\frac12(1+\ell+n)\big]}
{\Gamma\big[\frac12(2+\ell-n)\big]}
\label{eq:xiuell}
\ea
when $\Re(n-1)<0$ and $\Re(n+\ell)>-1$. Hereafter, $\Re(z)$ denotes the
real part of a complex number $z$. For our case,
\ba
n &= q_\nu - 1 - it\,.
\ea
\changed{For $r_0$ we recommend the choice $r_0\sim1/k_\max$.}

\subsection{The biasing parameter \texorpdfstring{$q$}{q}}\label{sec:xiq}
How do we need to choose the biasing parameter $q$? First, the integration
$M_{\ell}^{q_\nu}(t)$ restricts the biasing parameter $q_\nu$ to the range
\ba
\label{eq:Ml:q}
- \ell < q_\nu < 2\,.
\ea
In addition, the FFTLog transformation exists when
$n_2-1<q_\nu+\nu<3+n_1$ [\refeq{phi:q}]. Combining the two conditions, we find
\ba
\label{eq:xi:q}
\max(n_2 - 1 - \nu, -\ell) < q_\nu < \min(3 + n_1 - \nu, 2)\,,
\ea
or $ \max(-0.15-\nu,-\ell)<q_\nu<\min(3.967-\nu,2)$ for our reference cosmology.
Note that \refeq{xi:q} implies that a valid value of $q_\nu$ exists only if
\ba
\label{eq:nulimits}
n_2 - 3 < \nu < 3 + n_1 + \ell\,,
\ea
or $-2.15 < \nu < \ell + 3.967$ for our reference cosmology, and this is
the condition of convergence for the integral \refeq{xihere} when using the
asymptotic behavior of the power spectrum in \refeqs{pk-smallk}{pk-largek}.

As we show in \refapp{xiq}, there is an aliasing effect from the discrete
implementation of the integration in
\refeq{phiq}. We shall first choose a finer Fourier resolution $N$ in order to
ensure that all the relevant Fourier modes are summed over in \refeq{xifftlog}.
Then, our first choice for $q_\nu$ is $q_\nu = 1.9-\nu$,
because the aliasing effect in $\phi^q(t)$ is small for $q=q_\nu+\nu=1.9$.
If $1.9-\nu$ falls outside the range in \refeq{xi:q}, then we choose
\ba
q_\nu &= \frac13(q_{\nu,\min} + 2 q_{\nu,\max})\,,
\ea
where $q_{\nu,\min}$ and $q_{\nu,\max}$ are the boundaries given in
\refeq{xi:q}. Note that we weight slightly toward the higher-$q_\nu$ values.
We show that this choice of $q_\nu$ gives accurate results for a
wide range of $(\ell,\nu)$ combinations in \refapp{xiq}.

\subsection{Results: Accuracy}
\label{sec:xifftlog-accuracy}
\begin{figure*}
    \includegraphics[width=0.49\textwidth]{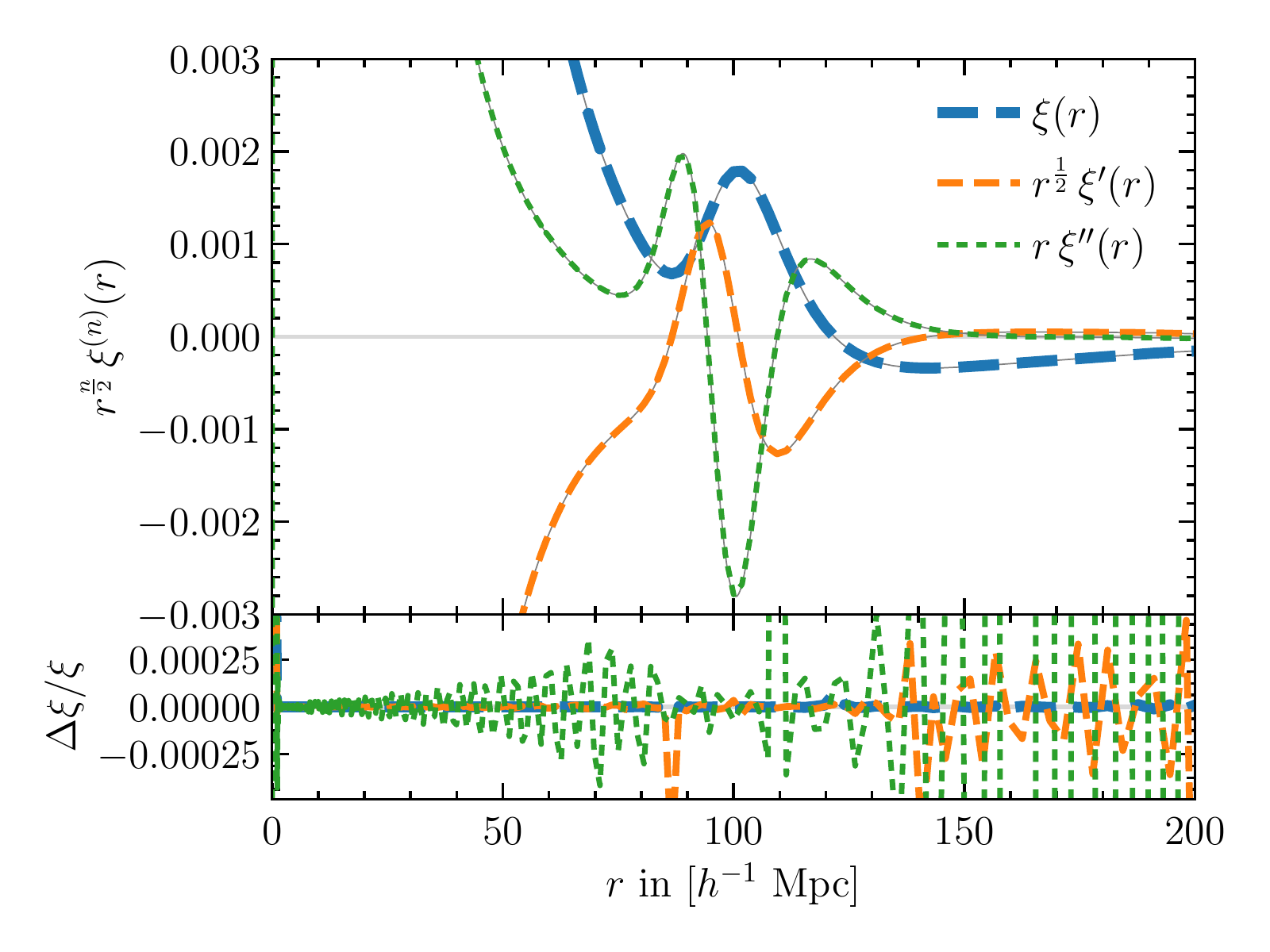}
    \includegraphics[width=0.49\textwidth]{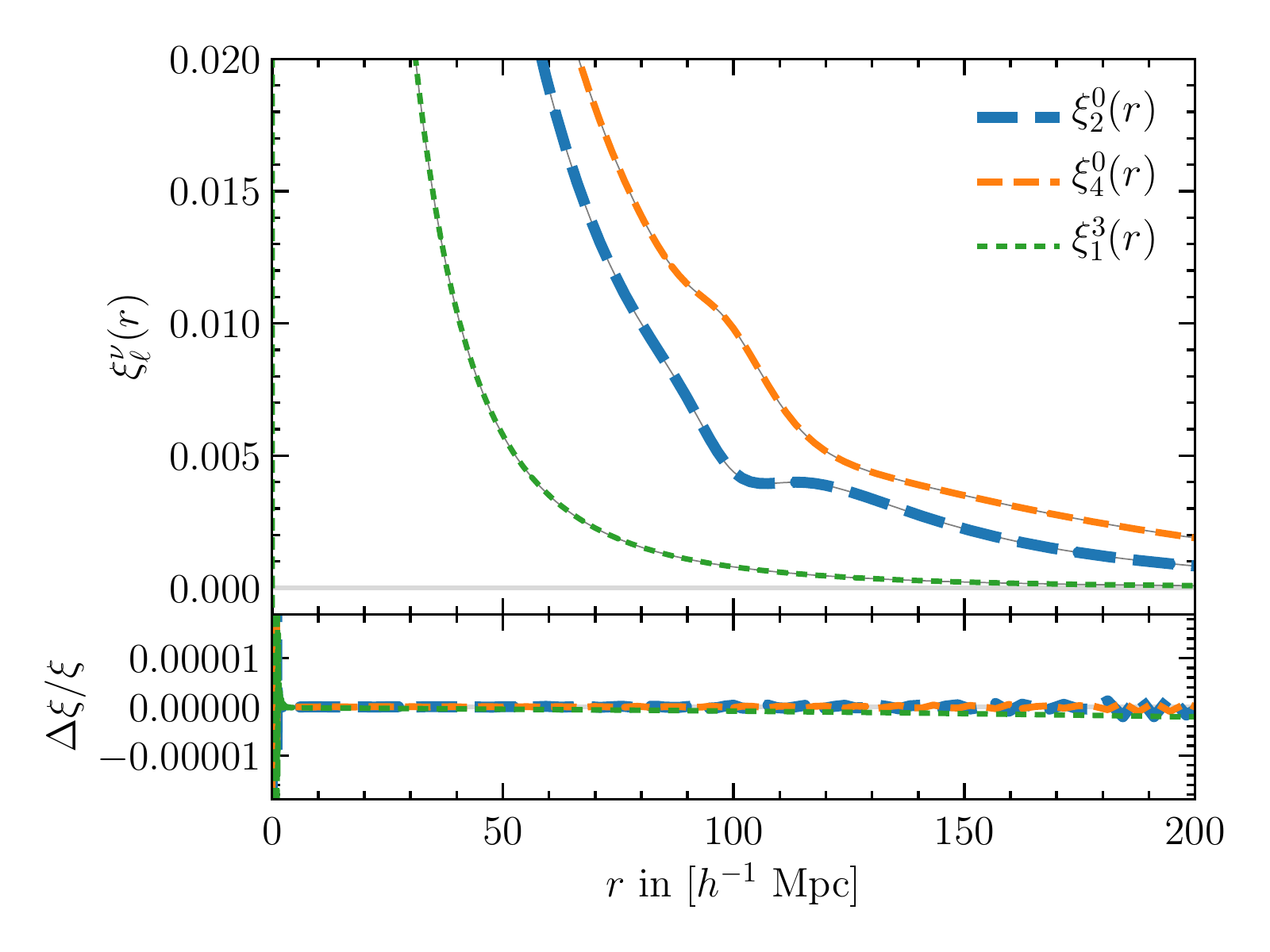}
    \caption{Accuracy comparison for different implementations of
		$\xi_\ell^\nu(r)$. Left: The upper panel shows the
    real-space correlation function $\xi^0_0(r)$ and its first and second
    derivatives calculated with the 2-FAST algorithm (dashed colored lines) and
    with the \texttt{quadosc} algorithm (solid gray lines). To calculate the first
    and second derivatives with the 2-FAST algorithm we use \refeqs{xid}{xidd}.
    For the \texttt{quadosc} algorithm we take the derivatives by creating a
    fifth-order spline of $\xi^0_0(r)$, and taking derivatives of the
    spline. The lower panel shows the relative difference between the 2-FAST
    results and the \texttt{quadosc} results. The difference is generally
    less than \SI{\sim0.05}{\percent}, except at zero crossings, and
    at very small and large separations $r$. The differences at
    $r\gtrsim\SI{150}{\per\h\mega\parsec}$ are likely due to pathologies in the
    \texttt{quadosc} algorithm, as closer inspection reveals unnatural
    oscillations in the \texttt{quadosc} curve, see \reffig{unnatural}.
    Right: The same as on the
    left, except for $\xi^0_2(r)$, $\xi^0_4(r)$, and $\xi^3_1(r)$. For the
    2-FAST algorithm we used $N=1024$, $k_\min=\SI{e-5}{\h\per\mega\parsec}$, and
    $k_\max=\SI{e3}{\h\per\mega\parsec}$.
    Increasing $N$ leads to better agreement between 2-FAST and \texttt{quadosc}.
    }
    \label{fig:xi}
\end{figure*}
\begin{figure}
    \includegraphics[width=0.49\textwidth]{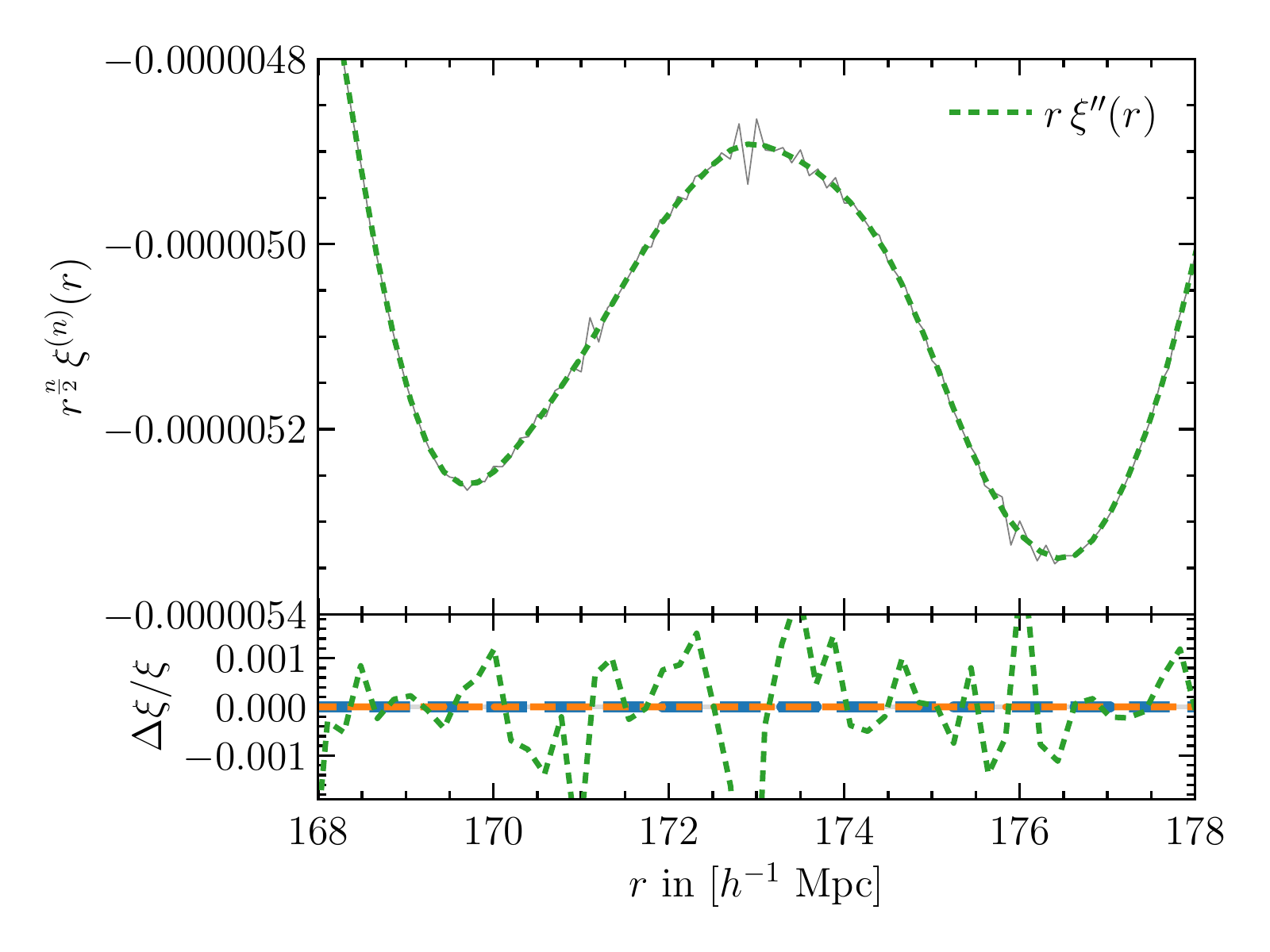}
    \caption{A zoom-in of \reffig{xi} at large $r$. Here we
    chose $N=16384$ for the 2-FAST algorithm (green dashed curve) to get a more dense
    sampling in $r$. The \texttt{quadosc} curve (solid gray) shows unnatural
    erratic behavior. Since the second derivative $\xi''(r)$ is expected to be
    smooth, it is likely that this erratic behavior is due to limitations of
    the \texttt{quadosc} algorithm.}
    \label{fig:unnatural}
\end{figure}

We assess the accuracy of the 2-FAST algorithm by comparing the result
with a slow, but accurate benchmark algorithm. The \texttt{quadosc}
\citep{NR3rd}
algorithm can integrate oscillatory functions accurately over an infinite
interval.
The \texttt{quadosc} algorithm works by integrating between successive zeros of
the integrand using Gauss-Kronrod quadrature, and then using a series
convergence acceleration to sum up the terms effectively out to infinity.
For the  convergence acceleration we use the Levin $u$-transform as described
in \cite{NR3rd}. \changed{In addition, we also verified that our results agree
with the results from FFTLog \citep{hamilton:2000} to within the accuracy
achievable with \texttt{quadosc}.}

\reffig{xi} compares the result from 2-FAST with the result from
\texttt{quadosc}.
The left panel shows the configuration-space two-point correlation function
[$\xi(r)=\xi^0_0(r)$, thick blue dashed line] and its first [$r^{1/2}\xi'(r)$,
orange dashed lines] and second [$r\xi''(r)$, green dashed line] derivatives.
Using the identities for the spherical Bessel function,
$(2\ell+1)\,j'_\ell(x)=\ell\,j_{\ell-1}(x) - (\ell+1)\,j_{\ell+1}(x)$ and
$(2\ell+1)j_\ell(x)/x = j_{\ell-1}(x) + j_{\ell+1}(x)$, the first and second
derivatives of $\xi(r)$ can also be calculated using the 2-FAST algorithm:
\ba
\label{eq:xid}
\xi'(r) &= - \frac{1}{r}\xi^{-1}_1(r) \\
\xi''(r) &= \frac{1}{r^2}\big[\xi^{-2}_2(r) - \xi^{-1}_1(r)\big]
\label{eq:xidd}
\ea
The right panel of \reffig{xi} shows the results for $\xi^0_2(r)$ (blue dashed
line), $\xi^0_4(r)$ (orange dashed line), and $\xi^3_1(r)$ (green dashed
lines). For all cases, we show the corresponding results of the \texttt{quadosc}
algorithm as solid gray lines. For the 2-FAST calculation, we used
$N=1024$, $k_\min=\SI{e-5}{\h\per\mega\parsec}$,
$k_\max=\SI{e3}{\h\per\mega\parsec}$, and $r_0=\SI{e-3}{\per\h\mega\parsec}$.

To facilitate the comparison better, in the lower panels of \reffig{xi}, we
show the fractional difference between the derivatives calculated from the two
methods (\texttt{quadosc} and its numerical derivatives and 2-FAST). For
$\xi_0^0$, $\xi_2^0$ and $\xi_4^0$, the difference between the two methods is
smaller than \SI{\sim0.05}{\percent}, and the accuracy improves when increasing
the sampling size $N$. The derivatives are less accurate, in particular, at
very small $r$, at zero crossings, and at
$r\gtrsim\SI{150}{\per\h\mega\parsec}$.
\changed{The accuracy of the $\xi_2^{-2}(r)$ term is worst, since more negative
$\nu$ puts more weight on small scale structure.}

The residuals are particularly large for the second derivative. We find that
the oscillatory features in the residuals of the second derivative
on small scales $r<\SI{100}{\per\h\mega\parsec}$ are improved when increasing the
sampling frequency $N$. However, on large scales
($r>\SI{100}{\per\h\mega\parsec}$) we find that the discrepancies
reflect limitations of the \texttt{quadosc} algorithm.
In \reffig{unnatural} we show a zoom-in of \reffig{xi} for the second
derivative $\xi''(r)$, where for the 2-FAST algorithm we used $N=16384$ in
order to get a dense sampling in $r$ space. The gray curve, again, shows the
result from using the \texttt{quadosc} algorithm to calculate $\xi(r)$, taking a
fifth-order spline, and then using the second derivative of the spline
function. The \texttt{quadosc} curve shows unnatural erratic behavior. Since the
second derivative should be a smooth function, it is likely that the
differences between the 2-FAST and the \texttt{quadosc} results are due to
limitations of our implementation of the \texttt{quadosc} algorithm and are not a
limitation of the 2-FAST method.

\subsection{Results: Performance}
\begin{figure}
    \includegraphics[width=0.48\textwidth]{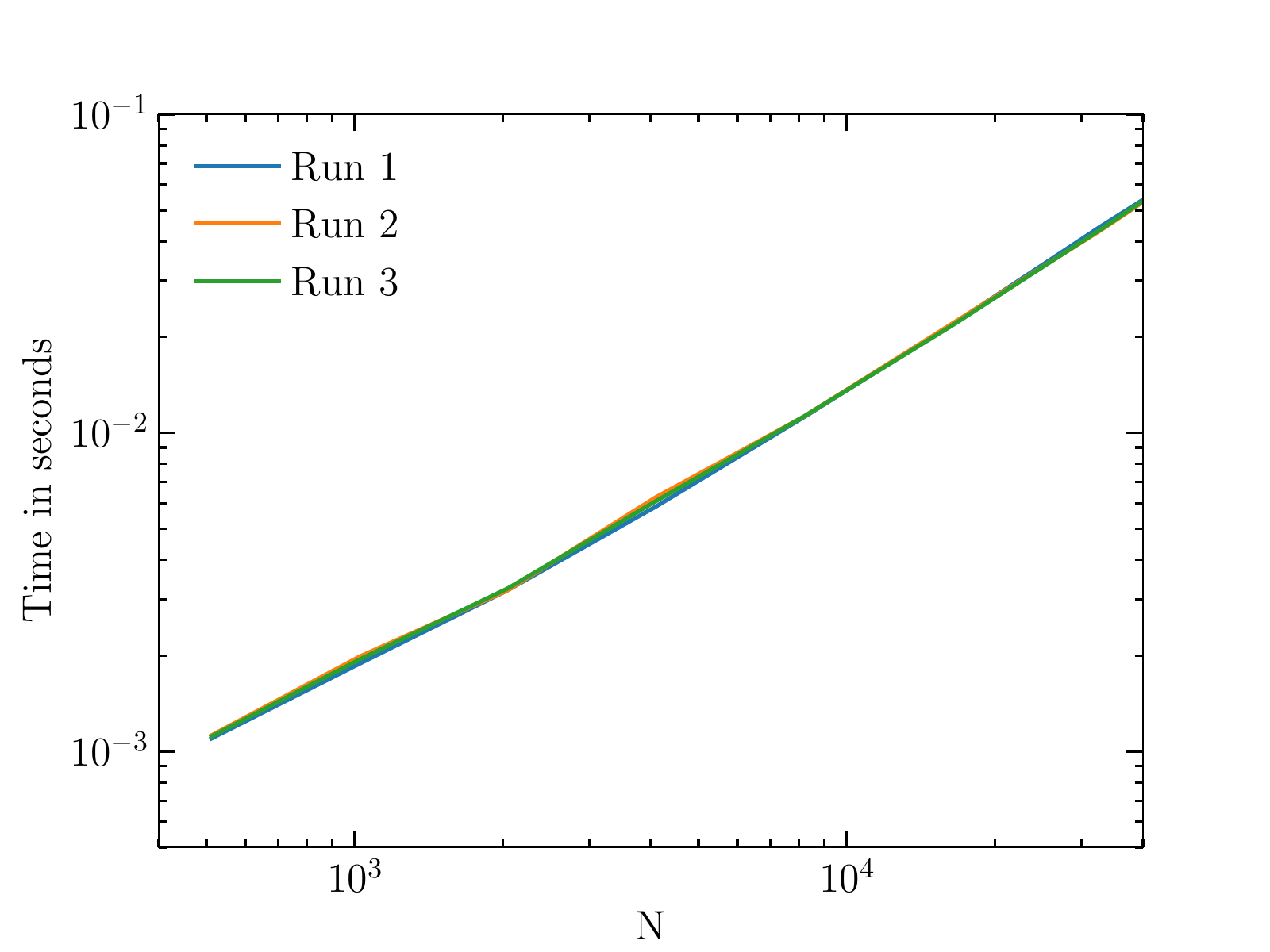}
    \caption{Performance of the 2-FAST algorithm as a function of sampling
	points $N$. We show the timings of three runs at each $N$, Run 1 to 3,
	\changed{each run having} equal settings. The algorithm
    achieves $\SI{\sim1.3}{\milli\second}$ performance for $N=1024$.}
    \label{fig:xispeed}
\end{figure}
In \reffig{xispeed} we show the performance of the 2-FAST algorithm as a
function of the number of sampling points $N$. The figure shows that with the
2-FAST algorithm we can calculate each curve in \reffig{xi} in
$\SI{\sim1.3}{\milli\second}$ on the test laptop that we describe at the
end of \refsec{intro}.
The majority of the time is spent in the FFT, for which we use the Fastest
Fourier Transform in the West\footnote{\url{http://fftw.org}} (FFTW) package.
Note that the memory allocation time of \texttt{Julia} adds to the run time, so further
optimization using a lower-level language is still possible, if necessary.
\changed{The FFTLog software \citep{hamilton:2000} has similar performance,
though a different feature set.}

\section{Projection onto spherical harmonic space}
\label{sec:well}
We now turn to the case with two spherical Bessel functions, \refeq{well}.
With the biasing parameter $q$, \refeq{well} becomes
\ba
w_{\ell\ell'}(\chi,\chi')
&= \frac{2}{\pi} \int_0^\infty \dd{}k\,k^{2-q}\,P(k)\,k^qj_\ell(k\chi)\,j_{\ell'}(k\chi')\,.
\label{eq:well2}
\ea
As in \refsec{xi}, we shall turn
\refeq{well2} into a convolution integral
by introducing logarithmic variables
$\kappa$, $\rho$, and the ratio $R=\chi'/\chi$ which are defined as
\ba
\frac{k}{k_0} &= e^\kappa
&  \frac{\chi}{\chi_0} &= e^\rho
&  \chi' &= R\chi = R\chi_0e^\rho\,,
\label{eq:kappa,rho,R}
\ea
for some pivot wave number $k_0$ and pivot distance $\chi_0$.
Then, \refeq{well2} becomes
\ba
w_{\ell\ell'}(\chi,R)
&= \frac{2k_0^3}{\pi}\,e^{-q\rho} \int_{-\infty}^\infty \dd{}\kappa
\,e^{(3-q)\kappa}\,P(k_0e^\kappa)
\vs&\qquad\quad\times
e^{q(\kappa+\rho)}j_\ell(\alpha e^{\kappa+\rho})\,
j_{\ell'}(R \alpha e^{\kappa+\rho})\,.
\label{eq:welllog2}
\ea
where $\alpha=k_0\chi_0$.
Parallel to \refsec{xi},
we introduce the Fourier transformation of the multiplication of two spherical
Bessel functions $M_{\ell\ell'}^{q}(t,R)$ as
\ba
e^{q\sigma}\,j_\ell(\alpha e^\sigma)\,j_{\ell'}(\beta e^\sigma)
&=
\int\frac{\dd{}t}{2\pi}\,e^{it\sigma}\,M_{\ell\ell'}^q(t,R)\,.
\label{eq:Mellell'2}
\ea
We can now rewrite \refeq{welllog2} by using $\phi^q$ and $M_{\ell\ell'}^q$
as
\ba
w_{\ell\ell'}(\chi,R)
&= 4k_0^3\,e^{-q\rho}
\int\frac{\dd{}t}{2\pi}\,
e^{it\rho}\,
\phi^q(t)\,
M_{\ell\ell'}^q(t,R)\,,
\label{eq:well1d}
\ea
\changed{where $\chi$ is related to $\rho$ by \refeq{kappa,rho,R}.}
\refeq{well1d} is the core of the 2-FAST algorithm for calculating the
angular power spectrum in harmonic space. We show the
discrete version of \refeq{well1d} that we use for the implementation in
\refapp{well1d-discrete}.
We have already discussed the FFTLog transformation in \refsec{phi}, and the key
to evaluate $w_{\ell\ell'}(\chi,R)$ is computing $M_{\ell\ell'}^q(t,R)$, which
we shall turn to next.

\subsection{Fourier transform of two spherical Bessel functions}
\begin{figure}
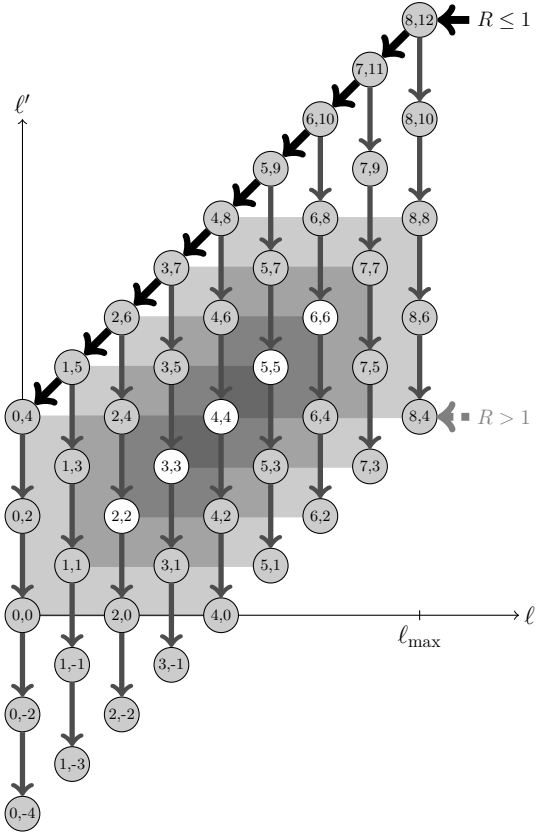

    \input flow-ell-ladder2.tex
    \caption{
    In order to calculate the Fourier transform of two spherical Bessel
    functions $M^q_{\ell\ell'}(t,R)$, we employ recursions along the paths in
    $\ell\ell'$ space shown in this figure for $\ell_\max=8$. Each node shows
    its $(\ell,\ell')$ coordinates. For $R\leq1$ we start at $\ell_\max$ with
    $\Delta\ell=\ell'-\ell=4$, and proceed down along the path $\ell'=\ell+4$
    until $\ell=0$. At each $\ell$, we then proceed with a recursion
    $\Delta\ell\rightarrow\Delta\ell-2$ until $\Delta\ell=-4$. The gray
    underlying squares centered on the white nodes indicate the values of
    $w_{\ell\pm2,\ell\pm2}$ which are needed to calculate the $C_\ell$ with
    linear redshift-space distortion (see \refapp{cl}). In gray we also
    indicate the start of the recursion for $R>1$, which is stable along the
    paths $\ell'=\ell-4$ and $\Delta\ell\rightarrow\Delta\ell+2$ (not shown).
    The recurrence relations and their stability properties are derived in
    \refapp{hyp2f1} and \refapp{Mellell-recurrence}. }
    \label{fig:ellladder}
\end{figure}
\begin{figure*}
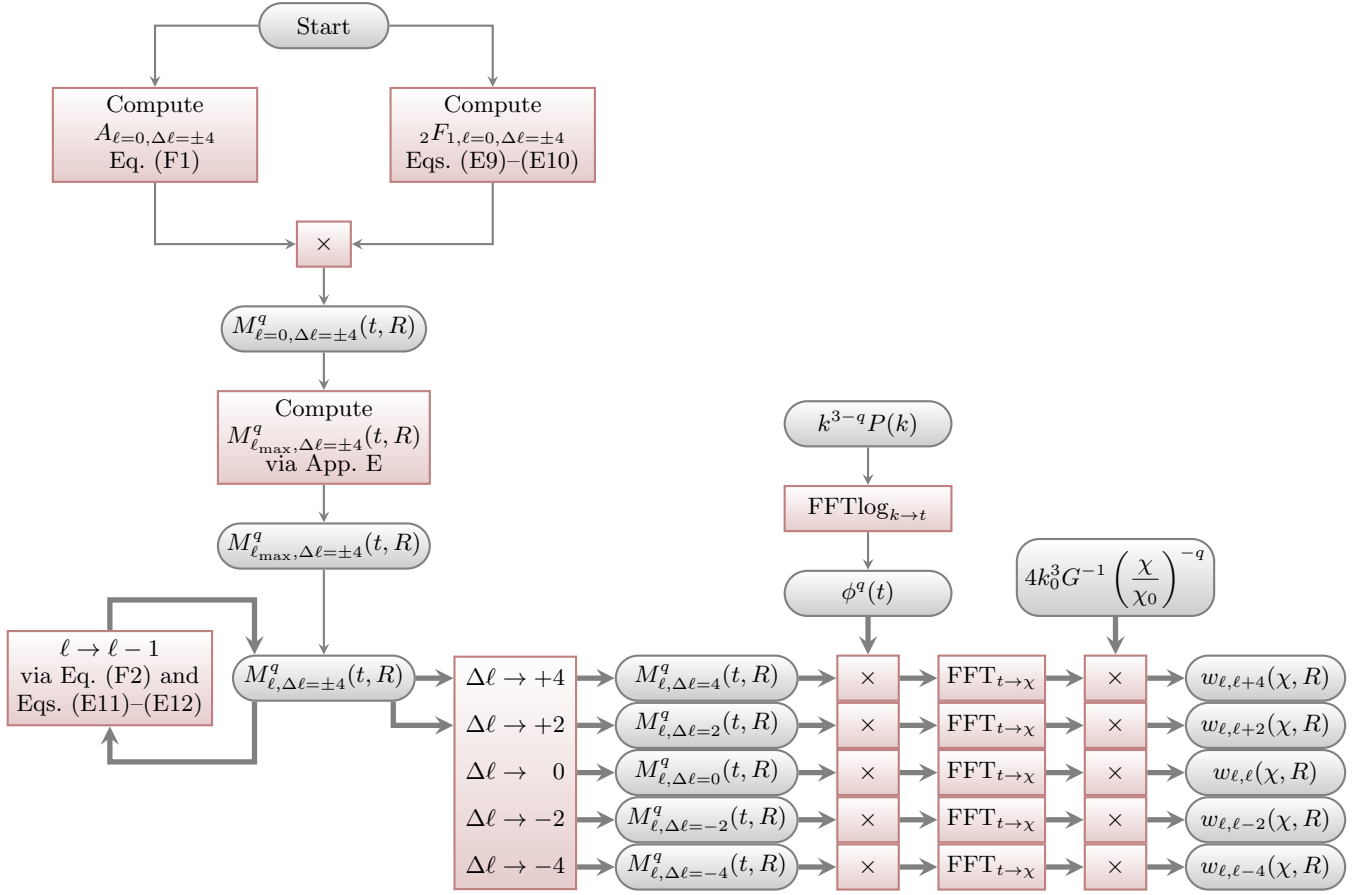

    \input flowchart_2f1-v2.tex
    \caption{Overview of the 2-FAST algorithm of calculating
		$w_{\ell\ell'}(\chi,\chi')$. Starting at the top left, we
    start by calculating the cosmology-independent part
    $M^q_{\ell\ell}(t,R)=A_{\ell,\Delta\ell}\,\,{}_2F_{1,\ell,\Delta\ell}$ at
    $\ell=0$ and $\Delta\ell=\pm4$. The plus sign is chosen for $R\leq1$, and the
    negative sign for $R>1$. Following \refapp{hyp2f1}
    the cosmology-independent part is calculated for $\ell_\max$ from $\ell=0$.
    Then, the recursion from $\ell_\max\rightarrow0$ is used, and at each
    step the products $w_{\ell,\ell\pm(0,2,4)}(\chi,R)$ are calculated. Note
    that everything to the left of where the Fourier transform $\phi^q(t)$ of the
    power spectrum enters is independent of the
    cosmology and can be precomputed. Thick arrows signify paths that need to
    be taken multiple times as the recursion over $\ell$ progresses.
    Furthermore, since the transformation from $w_{\ell\ell'}(\chi,\chi')$ to
    $w_{\ell,jj'}(\chi,\chi')$ is linear (see \refapp{cl}), that can also be
    done before multiplying by $\phi^q(t)$. However, this does not significantly
    change the method here.}
    \label{fig:flowchart}
\end{figure*}
The $P(k)$-independent part $M_{\ell\ell'}^q(t,R)$ is given by
the Fourier transformation of the product of two spherical Bessel functions:
\ba
M^q_{\ell\ell'}(t,R)
&= \int \dd{}\sigma\,e^{(q-it)\sigma}\,
j_\ell(\alpha e^{\sigma})\,j_{\ell'}(R \alpha e^{\sigma})
\vs
&= \alpha^{-1} \int \dd{}s\,\left(\frac{s}{\alpha}\right)^{q-1-it}\,j_\ell(s)\,j_{\ell'}(Rs)
\vs
&= \alpha^{it-q} \int \dd{}s\,s^{q-1-it}\,j_\ell(s)\,j_{\ell'}(Rs)
\vs
&= \alpha^{it-q}\, U_{\ell\ell'}(R,q-1-it)\,,
\label{eq:Mellell'R}
\ea
where $s=\alpha e^{\sigma}$, or $\sigma=\ln(s/\alpha)$, and
$U_{\ell\ell'}(R,n)$ is given in terms of the Gauss hypergeometric function
${}_2F_1$ as
\ba
&U_{\ell\ell'}(R,n) \vs
&= 2^{n-2} R^{\ell'}\pi\,
\frac{\Gamma\big[(1+\ell+\ell'+n)/2\big]}{\Gamma\big[(2+\ell-\ell'-n)/2\big] \Gamma\big[\frac32 + \ell'\big]}
\vs&\quad\times
{}_2F_1\bigg(
\frac{-\ell+\ell'+n}{2},
\frac{1 + \ell + \ell' + n}{2};
\frac32 + \ell'
;R^2
\biggl)\,,
\label{eq:Uellellp}
\ea
\changed{which we obtained from \texttt{Mathematica} \citep{mathematica11}}.
Here, $n = q-1-it$. Note that the general expression \refeq{Uellellp} is valid
for $|R|<1$, $\Re(n)<2$ and $\Re(\ell+\ell'+n)>-1$. Furthermore, for $R=1$, the
Gauss hypergeometric function converges only if $\Re(1-n)>0$.
These conditions put constraints on the choice of the biasing parameter $q$:
\ba
-\ell-\ell' < q < 2\,.
\label{eq:wl_q}
\ea
The method for calculating the function $U_{\ell\ell'}(R,n)$, however, may put
further constraints on $q$. For example, when evaluating $U_{\ell\ell'}(R,n)$
by recursion (see below for the details of the recursion), we need to know
$U_{\ell\ell'}$ at $\ell=\ell'=0$.
In that case, the Gamma function in the numerator
of \refeq{Uellellp} becomes infinite when $n$ is a negative odd integer, which
happens for $t=0$ and nonpositive even integer values of $q$. Hence, we have
the further constraint
\ba
q &\neq -2m \qquad\text{for }m=0,1,\cdots\,.
\ea
Furthermore, $q\neq1$ is required for our implementation of the case $R=1$; see
\refapp{special}. For the calculation of $w_{\ell\ell'}(\chi,\chi')$, we find
that $q<1.5$ is required to suppress the aliasing effect associated with the
convolution for $\chi, \chi'\gtrsim \SI{10}{\per\h\mega\parsec}$ (see
\refapp{xiq}).

The $R=\chi'/\chi>1$ cases can also be obtained from \refeq{Uellellp}
which is valid only for $|R|<1$, because by simply
changing the integration variable from $s$ to $s'=Rs$, \refeq{Mellell'R}
becomes
\ba
M_{\ell\ell'}^q(t,R)
&= (R\alpha)^{it-q} \int \dd{}s'\,s'^{q-1-it}\,j_\ell(R^{-1}s')\,j_{\ell'}(s') \vs
&= (R\alpha)^{it-q} \, U_{\ell'\ell}(R^{-1},q-1-it)\,.
\label{eq:Mellellsymmetry}
\ea
Note that $M_{\ell\ell'}$ is now proportional to $U_{\ell'\ell}$.
We use \refeq{Mellellsymmetry} when calculating for $R>1$ cases.

Now, the efficiency and accuracy of the $2$-FAST algorithm depends on our
ability to calculate the Gauss hypergeometric function ${}_2F_1$ in
\refeq{Uellellp}. Here, we use a set of recurrence
relations based on contiguous relations for the Gauss hypergeometric function
that we list in \refeqs{2F1rA}{2F1rH}. We describe the details of the
recursion in \refapp{hyp2f1} and \refapp{Mellell-recurrence}, and outline the
key procedure here. In particular, our implementation is based upon the
following three properties of ${}_2F_1$ in \refeq{Uellellp}:
(A) the backward recursion $\ell\to\ell-1$ is stable in all cases of interest;
(B) the recursion $\Delta\ell=4 \to\Delta\ell=-4$ is stable for $R<1$ cases;
(C) the recursion $\Delta\ell=-4 \to\Delta\ell=4$ is stable for $R>1$ cases,
for $\Delta\ell\equiv\ell-\ell'$.
Note that from \refeq{Mellellsymmetry}, (B) implies (C).
Here, we call a recursion stable when the error decays as the recursion
proceeds.

Miller's algorithm \cite{bickley+1952} exploits the property (A) and runs the
recursion backwards for a fixed $\Delta\ell$.
Here, we extend Miller's algorithm by using all three properties as
follows. First, we calculate the backward recursion from high
$\ell_\mathrm{seed}$ down to $\ell=0$ for the fixed
$\Delta\ell=4$ (when $R<1$) and $\Delta \ell=-4$ (when $R>1$) cases.
We then run recursions through the $\Delta\ell$ direction to
complete the calculation.  The recursion paths in $\ell\ell'$ space are shown
in \reffig{ellladder}.

In order to run the recursion backward, we need to set up the \emph{initial}
condition at some large multipole moment $\ell_\mathrm{seed}$.
We then run the backward recursion down to $\ell=0$ where we can fix the
normalization by using the analytical expression of ${}_2F_1$ at $\ell=0$.
Because the backward recursion is stable, the only requirement is that we must
choose $\ell_\mathrm{seed}$ sufficiently larger than $\ell_\mathrm{max}$
(maximum $\ell$ desired) so that any inaccuracy in the initial condition
decays sufficiently at $\ell_\mathrm{max}$. We ensure that by requiring that
the ${}_2F_1$ at $\ell_\mathrm{max}$ for different starting $\ell_\mathrm{seed}$
values converge within a fractional error of $10^{-10}$ [see
\refeq{conv_criteria}].
As the error decays throughout the backward recursion, initial conditions
do not have to be exact. The closer the initial conditions are to
the true ${}_2F_1$, however, the more efficient the algorithm is, since
a smaller $\ell_\mathrm{seed}$ would be sufficient. For the $R\ll1$ case, we
use the asymptotic behavior of the recurrence relation in the limit
$\ell\to\infty$ to set the initial conditions.
For the $R\sim1$ case, it turns out that, albeit noisy, the forward recursion
provides a reasonable initial condition at $\ell_\mathrm{seed}$.
We, therefore, set up the initial condition by running the forward recursion
to $\ell_\mathrm{seed}$, and then we apply the backward recursion.

For the $R=1$ case, we use an analytical expression for the hypergeometric
function ${}_2F_1$,
\ba
&{}_2F_1\bigg(
\frac{-\ell+\ell'+n}{2},
\frac{1 + \ell + \ell' + n}{2};
\frac32 + \ell'
;1
\biggl)
\vs
&=
\frac{\Gamma\big(\frac32+\ell'\big)\,\Gamma\big(1-n\big)}
{\Gamma\big[\frac12\big(3+\ell'+\ell-n\big)\big]\,
\Gamma\big[\frac12\big(2+\ell'-\ell-n\big)\big]}\,,
\ea
which we use to initialize the recursion at $\ell_\max$.

\subsection{From \texorpdfstring{$M_{\ell\ell'}^q(t,R)$}{Mll(t,R)} to angular power spectra}
\label{sec:ClfromMll}
\begin{figure*}
	\includegraphics[width=0.49\textwidth]{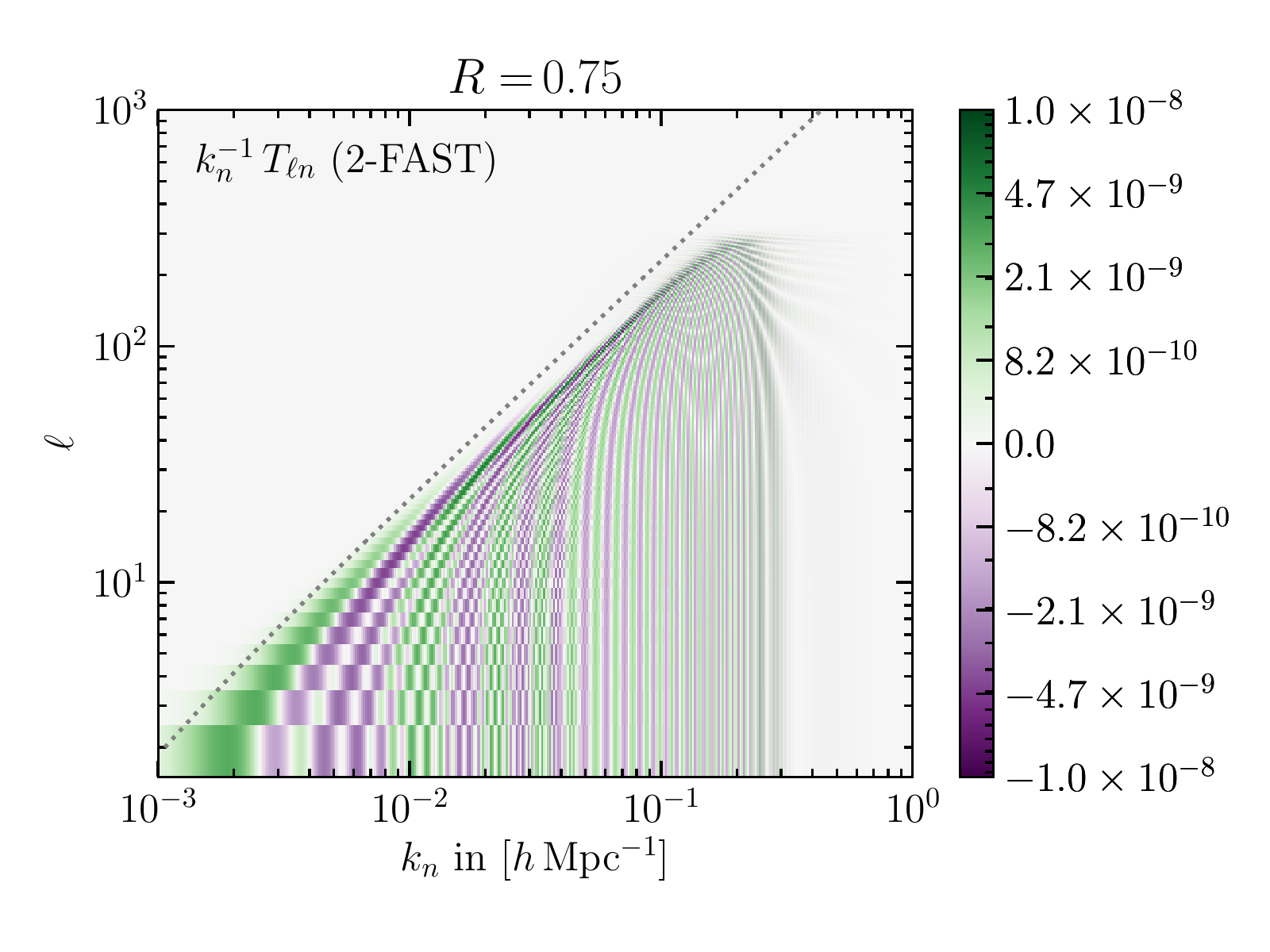}
	\includegraphics[width=0.49\textwidth]{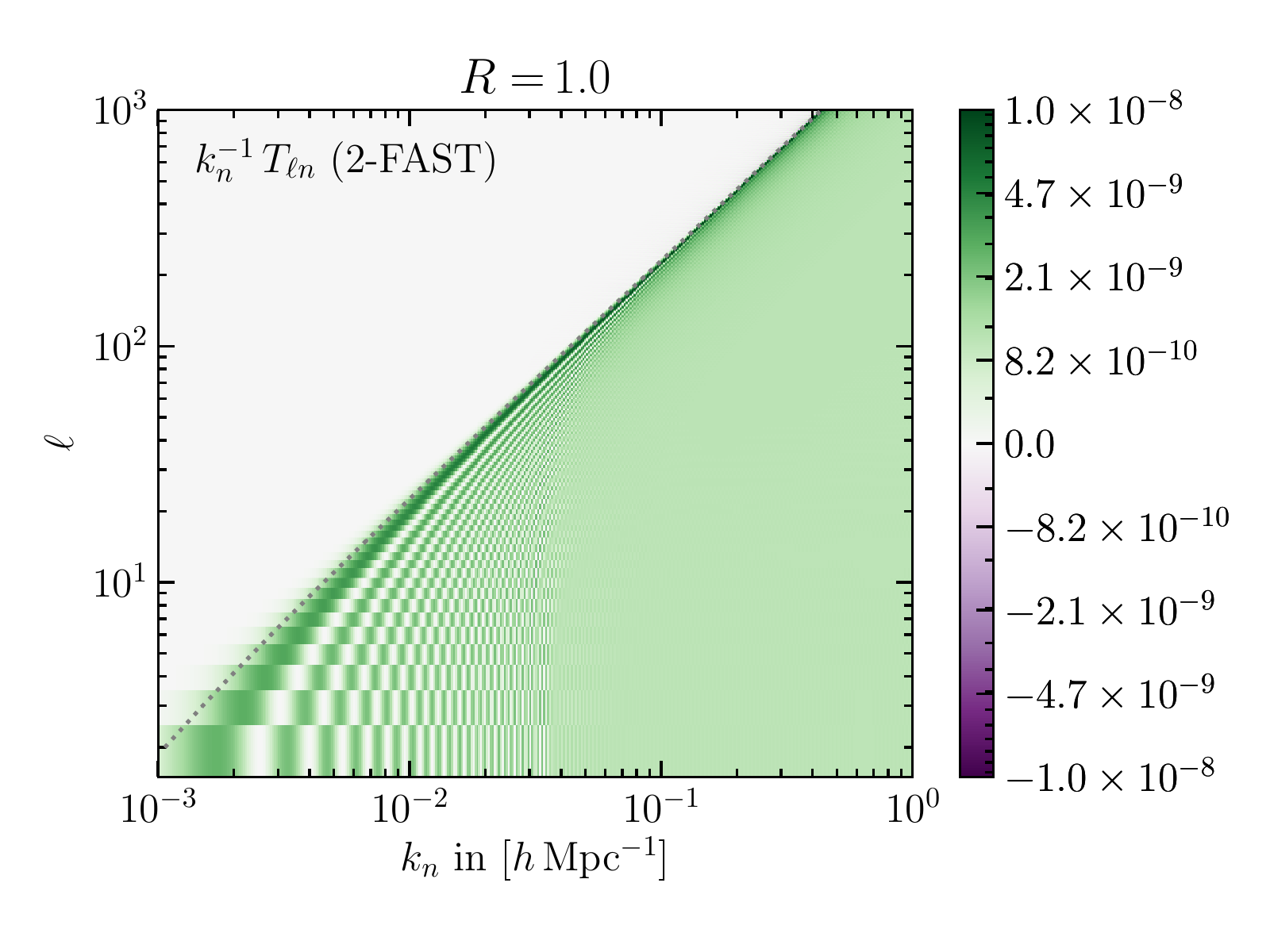}
        \caption{The transformation matrix
        $T_{\ell n}$ given in \refeq{transmat} multiplied by $k_n^{-1}$ for a
        Dirac-delta window function centered around
        $\chi_{n'}=\SI{2370}{\per\h\mega\parsec}$ for $R=0.75$ (left) and $R=1$ (right).
        The color bar shows the value of $k_n^{-1}T_{\ell n}$
        on a nonlinear $\sinh^{-1}$ scale. The minor tick marks in the color
        bars are linearly spaced.
        At high $k$, the transformation matrix is smooth, since for
        each $k_n$, $T_{\ell n}$ represents an integral over multiple
        oscillations of the spherical Bessel functions.
        At high $\ell$, the transformation matrix is significant only along
        $k_n\sim\sqrt{\ell(\ell+1)}/\chi_{n'}$ (gray dotted line), which shows
        that Limber's approximation should work well in this regime.
        }
	\label{fig:transmat}
\end{figure*}

\refeqs{well1d}{Uellellp} describe our method of
computing an integral over two spherical Bessel functions. A discrete version
is given in \refapp{well1d-discrete}. We give an overview of the method in
\reffig{flowchart}.

In linear theory, all harmonic-space power spectra are linear in $P(k)$, and
they can be calculated as a linear combination of $w_{\ell\ell'}(\chi,\chi')$.
One such example is the linear, redshift-space, galaxy power spectrum $C_\ell$
that we present in \refapp{cl}. For these cases, we can carry out the
$P(k)$-independent part of the calculation with $M_{\ell\ell'}^q$ before the
power spectrum enters the calculation. That is, for a given set of spherical
observables that are linearly related to $P(k)$, the 2-FAST method naturally
breaks down the calculation into the $P(k)$-dependent $\phi^q$ and the
$P(k)$-independent part which can be precalculated.

As an example, consider cross-correlating two linear galaxy density fields
spread over redshift ranges centered around, respectively, $z_1$ and $z_2$ with
the survey radial window functions, respectively, $\cW_1(\chi)$ and
$\cW_2(\chi)$. The angular power spectrum in this case is given by
\be
C_\ell =
b_1^2 \int \dd\chi_1 \int \dd \chi_2\; \cW_1(\chi_1)\cW_2(\chi_2)
w_{\ell\ell}(\chi_1,\chi_2),
\ee
which can be calculated by using \refeq{well1d}
\ba
C_\ell &= 4k_0^3 b_1^2 \int \dd\chi_1
\int \dd \chi_2\; \cW_1(\chi_1)\cW_2(\chi_2)
\vs&\times
e^{-q\rho}
\int \frac{dt}{2\pi} \, e^{it\rho}\phi^q(t)M^q_{\ell\ell'}(t,R)
\vs
&= 8k_0^3 \chi_0^2 b_1^2
\int \frac{dt}{2\pi} \,
\phi^q(t)
\vs&\times
\int \dd\rho
\int_0^1 \dd R\, e^{(2-q+it)\rho} \, \cW_1(e^\rho) \, \cW_2(Re^\rho) \,
M^q_{\ell\ell'}(t,R)\,.
\label{eq:Cl_2FAST}
\ea
The second line of the integration is independent from the power spectrum, and
could be calculated for given radial window functions. Since the radial
window functions $\cW_i(\chi)$ are cosmology dependent, encompassing the linear
growth factor, redshift-distance relation and so on, we do not further
investigate this approach in this paper. Upon the quantification of these
radial dependences, rearranging the integrals as in
\refeq{Cl_2FAST} should result in a fast and accurate calculation.

Alternatively, we can also define the transformation matrix between the
Fourier-space power spectrum $P(k)$ and angular power spectrum $w_{\ell\ell'}$.
We write \refeq{well1d} with \refeq{phiq} as
\ba
w_{\ell\ell'}(\chi,R)
&=
\int \frac{\dd k}{k}\,
P(k)\,\bigg[
\frac{2}{\pi}\,
k^3
e^{-q(\kappa+\rho)}\,
\vs&\quad\times
\int \frac{dt}{2\pi} \,
e^{it(\kappa+\rho)}\,
M^q_{\ell\ell'}(t,R)
\bigg]\,,
\ea
If we evaluate the integrals as written using the definition of
$M^q_{\ell\ell'}(t,R)$ in \refeq{Mellell'2}, we recover \refeq{well}.
However, for implementation on a computer, the integrals are approximated as
sums over discrete $k_n$ and $t_m$. Using the discrete versions of our
algorithm in \refapp{discrete}, the term in brackets and the measure
become
\ba
\label{eq:transmat}
T^{\ell',q}_{\ell n}(\chi_{n'},R)
&=
\frac{2}{\pi}\,k_n^3\,\left(\frac{k_n\chi_{n'}}{k_0\chi_0}\right)^{-q}
W_1(k_n)
\vs&\quad\times
\frac{1}{N}
\sum_m
e^{i2\pi (n+n')m/N}\,
W_2(t_m)\,
\vs&\quad\times
M^q_{\ell\ell'}(t_m,R)\,,
\ea
where we included the Fourier-space window functions $W_i(x)$
defined in \refeq{windowfn} to reduce ringing, and $t_m$ is defined in
\refeq{tm}. Then, we can calculate the harmonic-space power spectrum
$w_{\ell\ell'}$ by a matrix multiplication:
\ba
w_{\ell\ell'}(\chi_{n'},R) &= \sum_n T^{\ell',q}_{\ell n}(\chi_{n'},R)\,P(k_n)\,.
\label{eq:Pk_to_Cl}
\ea
In \reffig{transmat}, we show the transformation matrix $T_{\ell
n}^{\ell',q}(\chi_{n'},R)$ for $\chi_{n'}=\SI{2303}{\per\h\mega\parsec}$,
$\Delta\ell=0$, integration limits $k_\min=\SI{e-4}{\h\per\mega\parsec}$,
$k_\max=\SI{e4}{\h\per\mega\parsec}$, and $q=1.1$ for $R=0.75$ and $R=1$. The
figure shows that most of the power comes from a narrow band around
$k_n\simeq\sqrt{\ell(\ell+1)}/\chi_{n'}\simeq(\ell+0.5)/\chi_{n'}$ (gray dotted
line in the figure), which gets narrower towards higher $\ell$. This trend is
consistent with the Limber approximation that maps the Fourier space and the
harmonic space by $k\simeq (\ell+0.5)/\chi$ \citep{loverde/afshordi:2008} and
$R=1$. As shown in \reffig{transmat}, the Limber approximation is accurate only
at large $\ell$. While the transformation matrix is non-negative for $R=1$ (it
is proportional to the square of the spherical Bessel function), the $R=0.75$
case shows the \emph{beat} between the two Bessel functions with
different \emph{frequencies}.

To understand the 2-FAST algorithm better, we compare with a more traditional
approximation of \refeq{well}:
\be
\label{eq:traditional}
w_{\ell\ell'}(\chi,R)
=
\sum_n
\bigg[\frac{2}{\pi}
\Delta k_n\,
k_n^2\,j_{\ell}(k_n\chi)\,j_{\ell'}(k_nR\chi)
\bigg]
P(k_n)\,.
\ee
To use this traditional method, the sampling of $k_n$ needs to be very dense at
high $k$ so as to capture the oscillations of the spherical Bessel functions.
The 2-FAST method avoids the need for a dense sampling in $k$ by calculating
$M_{\ell\ell'}^q(t,R)$ analytically. Then, the linear transformation matrix
between $P(k)$ and $w_{\ell\ell'}(\chi,R)$ effectively averages out the
high-$k$ oscillation of the spherical Bessel functions.

Using the transformation matrix \refeq{transmat} is useful to gain some
insight into the spherical harmonic projection of the power spectrum.
For example, we can easily see the response of the angular power spectrum
(observables) to the changing cosmological parameters that alter the
three-dimensional power spectrum. That calculation is particularly useful
for a Fisher matrix analysis.
For calculating the harmonic-space power spectrum in practice, however,
following the 2-FAST algorithm \refeq{well1d} is faster.
This is because, from a given set of $M_{\ell\ell'}^q(t,R)$, the
matrix multiplication operation in \refeq{Pk_to_Cl} takes
\changed{$\orderof(N_\ell N_R N^2)$ time, where $N_\ell$ is the number of
$\ell$-values, $N_R$ the number of $R$-values, and $N$ the number of $k$ and
$\chi$ values}, while the 2-FAST algorithm in \refeq{well1d} only takes
\changed{$\orderof(N_\ell N_R N\log N)$} time thanks to the fast Fourier
Transformation.

\subsection{Results: Accuracy}
\begin{figure*}
    \includegraphics[width=0.49\textwidth]{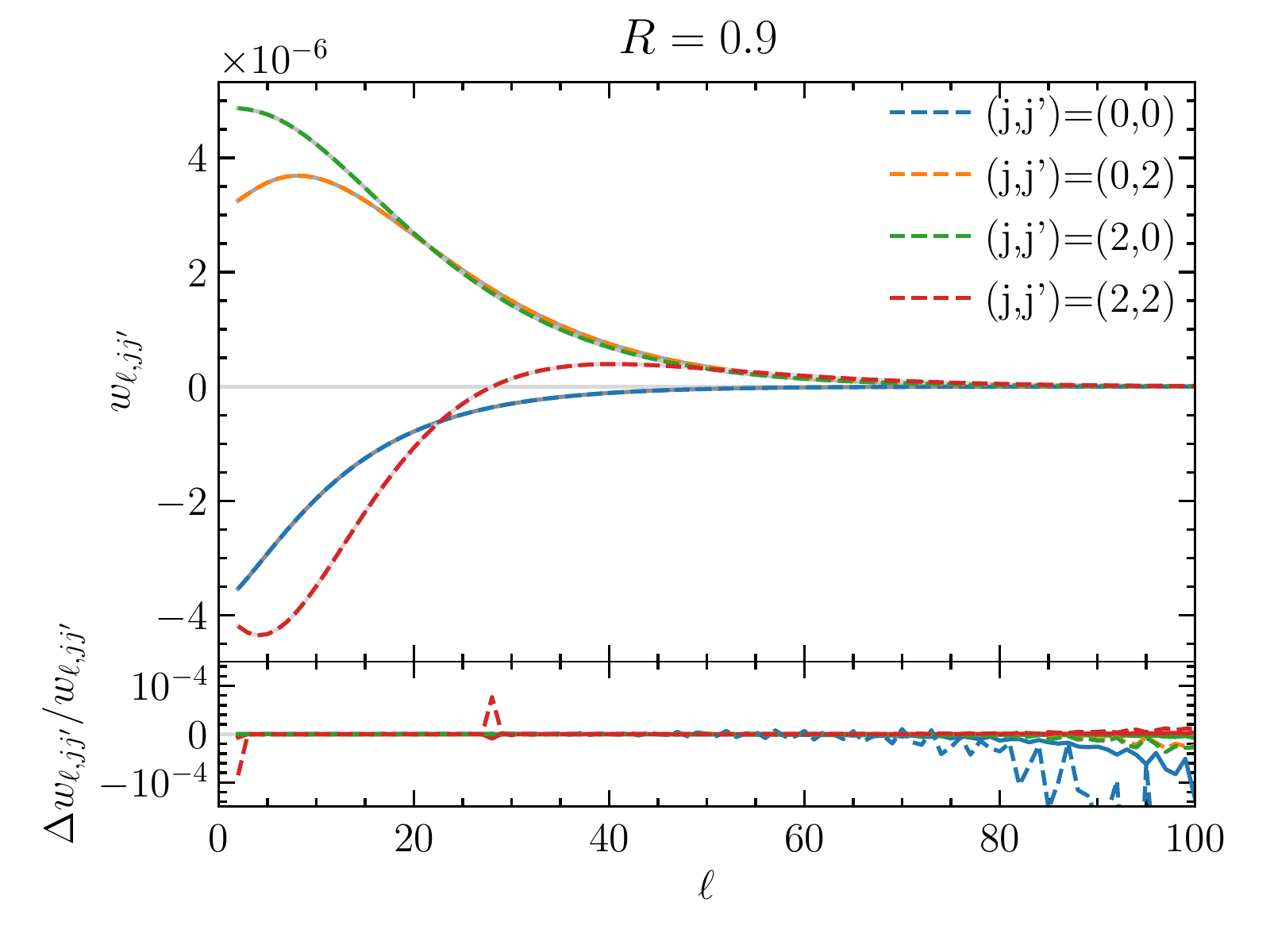}
    \includegraphics[width=0.49\textwidth]{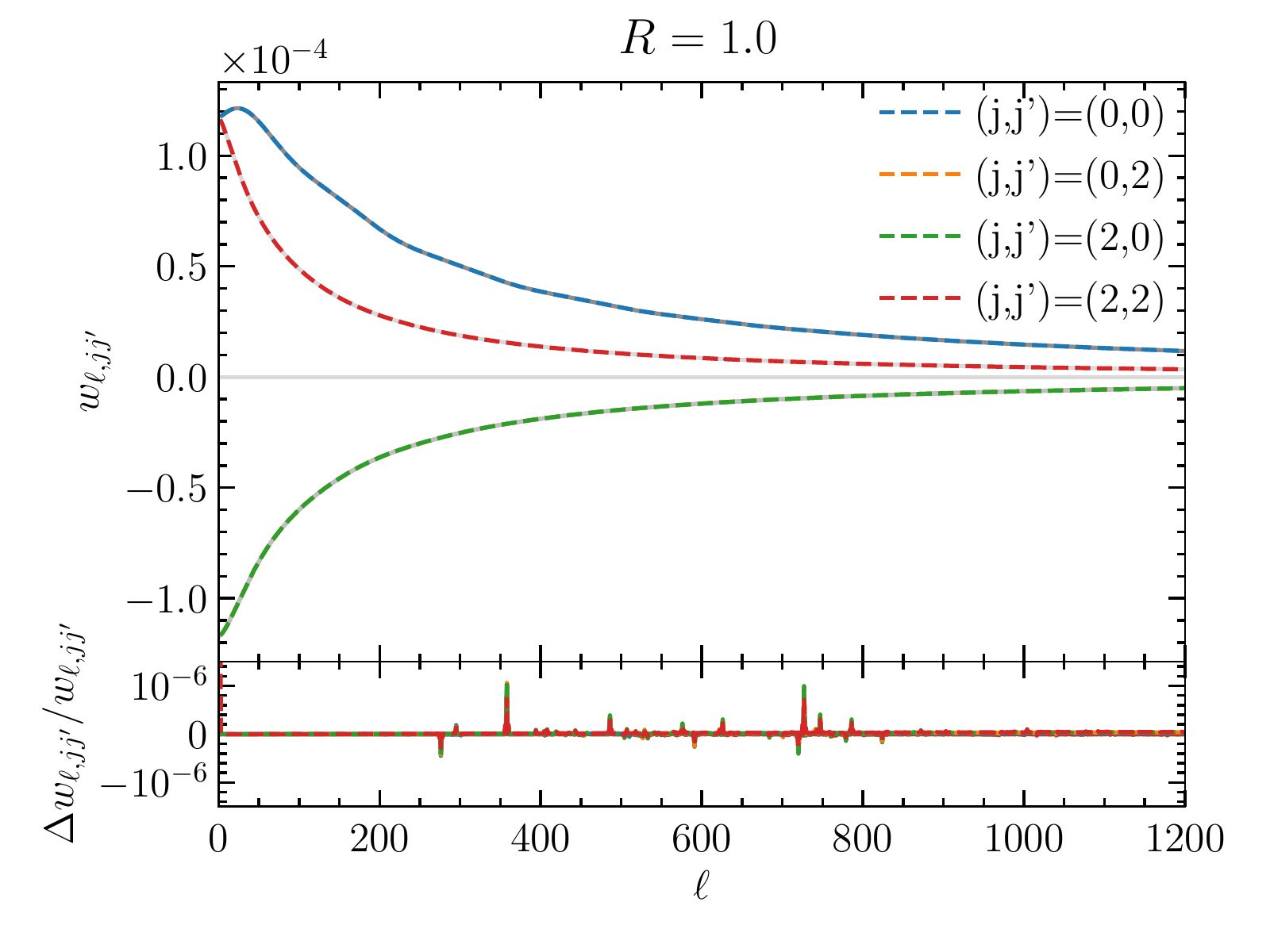}
    \caption{Comparison between the 2-FAST algorithm and the \cite{lucas1995}
    algorithm for the quantity $w_{\ell,jj'}$ [see \refeq{wljj}].
    Left: The top panel shows the value of $w_{\ell,jj'}$ for $R=0.9$. By eye, no differences between the two algorithms are apparent. The
    bottom panel shows the relative difference. Typical differences are on the
    order of one part in \num{e4}.
    Right: The same for $R=1.0$.
    Differences are on the order of one part in \num{e6}, except for $\ell=2$
    and $(j,j')=(2,2)$. In that case the relative difference is $\sim\num{e-4}$.
    Here we chose $N=1600$, $k_\min=\SI{e-5}{\h\per\mega\parsec}$,
    $k_\max=\SI{e5}{\h\per\mega\parsec}$,
    $\chi=\SI{2370}{\per\h\mega\parsec}$, $q=1.1$.
    The glitches in the residuals for $R=1$ are likely due to inaccuracies in
    our implementation of the Lucas algorithm.
    For $R=0.9$, the differences at large $\ell$ can be reduced by increasing
    the number of sample points on the power spectrum, e.g. to $N=4096$.
    The differences at small $\ell$ are due to aliasing, and are reduced by
    increasing the width of the integration interval, e.g. by decreasing the lower
    bound to $k_\min=\SI{e-6}{\h\per\mega\parsec}$. The bottom residual panels
		show the result of both these changes as the colored solid lines.}
    \label{fig:cl-R1}
\end{figure*}
\begin{figure*}
    \includegraphics[width=0.73\textwidth]{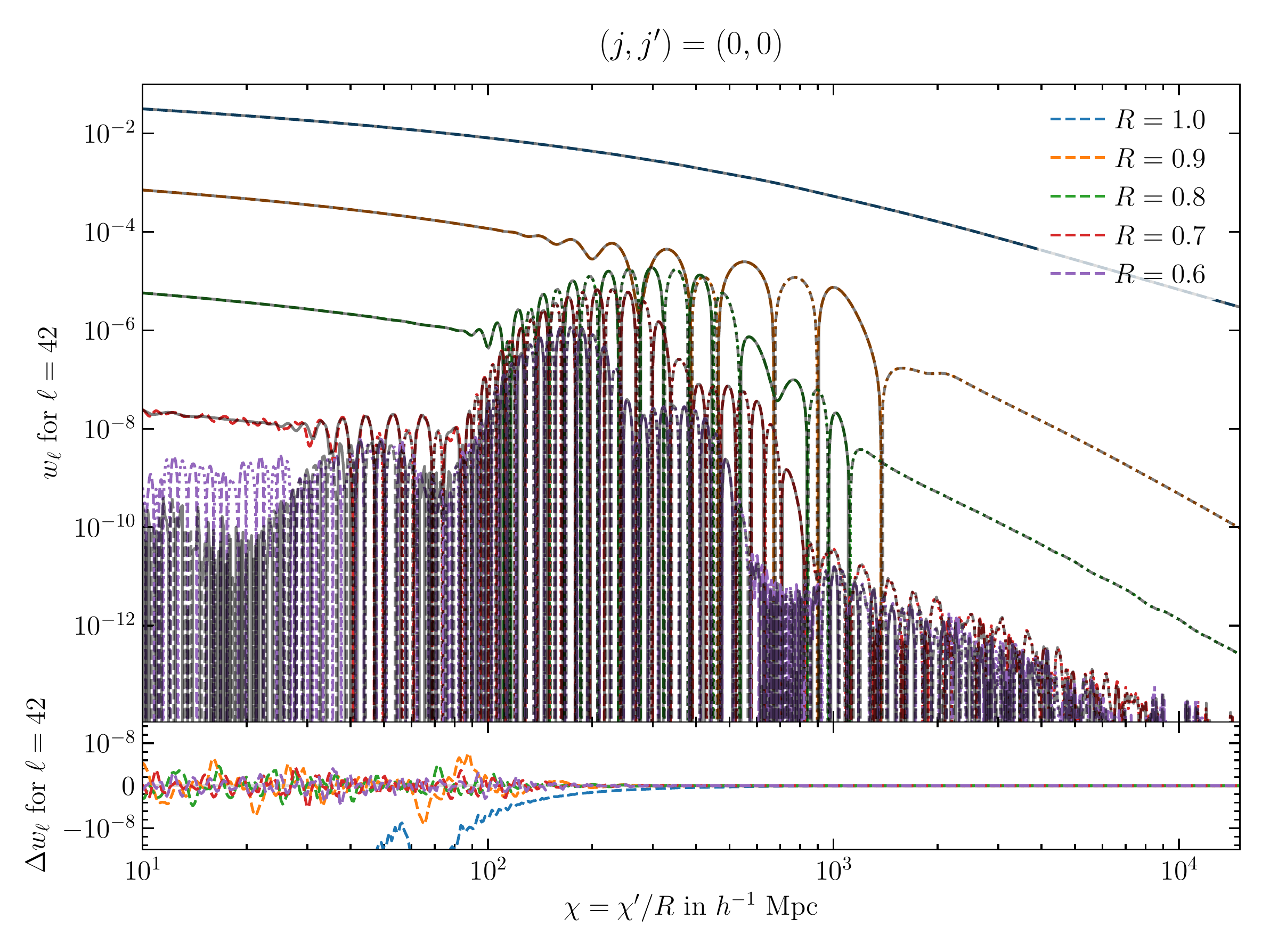}
    \includegraphics[width=0.73\textwidth]{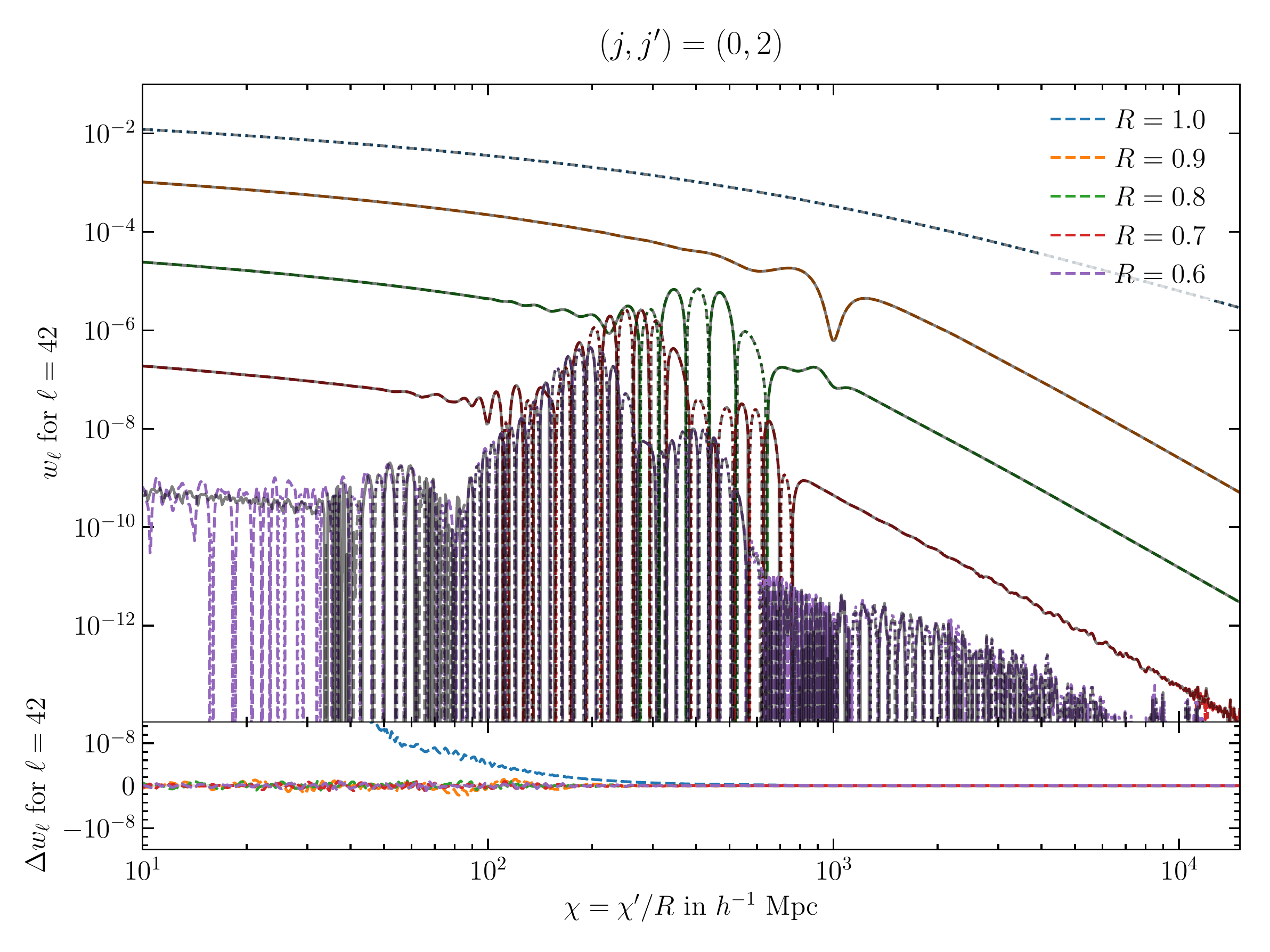}
    \caption{Comparison of the 2-FAST algorithm (dashed lines) with the
    Lucas algorithm (gray solid lines) for $\ell=42$ and $(j,j')=(0,0)$ on the
    top and $(j,j')=(0,2)$ on the bottom for a range of ratios $R=\chi'/\chi$.
    The top panel in each plot shows the value of $w_{\ell,jj'}$ [defined in \refeq{wljj}].
    Dotted colored lines for the 2-FAST algorithm and dashed gray lines for the
    Lucas algorithm indicate negative values. The bottom panels show the
    difference from the results of the Lucas algorithm. The largest differences occur for the $R=1$
    lines at $\chi\lesssim\SI{e2}{\per\h\mega\parsec}$. Closer
    inspection reveals that the relative differences in these cases are less
    than 1 part in \num{e5} throughout the figure. This difference can be
    reduced by sampling more densely, and decreasing $k_\min$ and increasing
    $k_\max$. For this
    plot we used $N=\num{4096}$, and $k_\min=\SI{e-5}{\h\per\mega\parsec}$,
    $k_\max=\SI{e5}{\h\per\mega\parsec}$, $q=1.1$,
    $\chi_0=\SI{1}{\per\h\mega\parsec}$.}
    \label{fig:comparison-ell42}
\end{figure*}

We test the accuracy of our implementation of the 2-FAST algorithm
by calculating
\ba
\label{eq:wljj}
w_{\ell,jj'}(\chi,\chi')
&= \frac{2}{\pi}\int_0^\infty \dd{}k\,k^2\,P(k)\,
j_\ell^{(j)}(k\chi)\,j_{\ell}^{(j')}(k\chi')\,,
\ea
where $j$ and $j'$ denote the number of derivatives on the spherical Bessel
functions. The functions $w_{\ell,jj'}$ appear in the calculation of the
angular power spectrum of galaxies in redshift space.
In \refapp{cl} we present the full expression for the angular power spectrum
of the redshift-space galaxy distribution, and derive $w_{\ell,jj'}$ in terms of
$w_{\ell\ell'}(\chi,\chi')$.
We then compare the 2-FAST result with a slow, but accurate computation
using the Lucas algorithm \cite{lucas1995} that we summarize
in \refapp{lucas1995}.

In \reffig{cl-R1} we show the comparison with the Lucas algorithm for all
values of $(j,j')$ needed for linear redshift-space distortion for the $R=1$
($\chi'=\chi$, right panel) and $R=0.9$ ($\chi'=0.9\chi$, left panel) cases.
The two algorithms
agree well and the curves (color curves for 2-FAST, gray curves for Lucas) lie
on top of each other at all $\ell$ shown here. The bottom panels of
\reffig{cl-R1} show that the fractional residuals are $\lesssim\!\num{e-6}$ in
the case of $R=1$ and $\lesssim\!\num{e-4}$ in the case of $R=0.9$ for all
$(j,j')$ pairs relevant for calculating the linear redshift-space galaxy power
spectrum. The one exception is for the $(j,j')=(2,2)$ case at the multipole
$\ell=2$, where the error is as large as \SI{0.1}{\percent}. A larger sampling
number $N$ results in a better match. The differences at small $\ell$ are
due to aliasing and can be reduced by choosing a wider integration interval
or choosing a different biasing parameter $q$ (see \refapp{xiq}).
In \reffig{cl-R1} we show the effect of a larger $N$ and a wider integration
interval on the residuals as colored solid lines.
Some of the glitches in the residuals are likely due to inaccuracy in our
implementation of the Lucas algorithm, which we discuss briefly in
\refapp{lucas1995}.

We show a comparison for $w_{\ell=42,jj'}(\chi,R\chi)$ as a function of the
comoving distance ($\chi$) for different values of $R=1$, $0.9$, $0.8$, $0.7$,
$0.6$ in \reffig{comparison-ell42}. The curves for the Lucas algorithm are in
solid gray for positive values and dashed gray for negative values. The 2-FAST
curves are positive for colored dashed lines, and negative for colored dotted
lines. The top plot shows the result for $(j,j')=(0,0)$ and the bottom
plot shows it for $(j,j')=(0,2)$. For both plots, we show corresponding
residuals between the 2-FAST and Lucas algorithms in the lower panels. Note
that here we show the absolute error instead of the relative error because the
function $w_{\ell=42,jj'}$ frequently crosses zero when $R\neq1$. The absolute
error is generally less than \num{e-8}. The exception is when $R=1.0$ (blue
dashed line). However, in that case the relative error is still $\num{<e-5}$.
This can be improved by choosing a wider integration interval $G$ or adopting
a smaller biasing parameter $q$ (see \refapp{xiq}).

\subsection{Results: Performance}
\begin{table*}
    \newcommand{\msec}[1]{\SI{#1}{\milli\second}}
    \newcommand{\secs}[1]{\SI{#1}{\second}}
    \begin{threeparttable}[b]
        \caption{Performance results.}
        \begin{tabular}{c|c|c|c||c|c|c||c|c}
            $N$\tnote{a}
                & $N_\chi$\tnote{b}\phantom{.}
                & $N_R$\tnote{c}\phantom{.}
                & $\ell_\max$
                & ${}_2F_{1,\ell_\max}$
                & $M^q_{\ell\ell'}$
                & $C_\ell$
                & Total\tnote{d}\phantom{.}
                & IO\tnote{e}
            \\
            \hline
            1600 & 1     &  1 &   500
                & \msec{326} & \msec{215} & \msec{28}
                & \msec{569} & \msec{68} \\
            1600 & 1     &  1 &  1200
                & \msec{393} & \msec{446} & \msec{60}
                & \msec{899} & \msec{142} \\
            1600 & 1600  &  1 &  1200
                & \msec{404} & \msec{453} & \msec{69}
                & \msec{926} & \msec{163}\tnote{f}\phantom{.} \\
            3200 & 3200  &  5 &  1200
                & \secs{3.85} & \secs{3.44} & \secs{0.45}
                & \secs{7.74} & \secs{1.10} \\
        \end{tabular}
        \begin{tablenotes}
        \item[a] Number of sample points on the power spectrum $P(k)$
        \item[b] Number of redshifts, or number of $\chi$
        \item[c] Number of ratios $R=\chi'/\chi$
        \item[d] Sum of the three preceding times
        \item[e] Time spent reading and writing to the disk
        \item[f] Since we are only interested in compute times here, we did not
            save all 1600 values to the disk in this case.
        \end{tablenotes}
        \label{tab:speedy}
    \end{threeparttable}
\end{table*}

We test the performance of our implementation of the 2-FAST algorithm
by measuring the time it takes to calculate the angular
power spectra in \reffig{cl-R1} and \reffig{comparison-ell42}, and
variations thereof. The result is summarized in \reftab{speedy} for computing
four different scenarios. In the table, $N$ is the number of sampling points on
the power spectrum. It defines the size of the FFT
array. $N_\chi$ is the number of redshifts (comoving radii) we are interested
in, $N_R$ the number of ratios $R=\chi'/\chi$, and $\ell_\max$ the maximum
multipole moment.

The first two scenarios show that the performance scales roughly proportional to
$\ell_\max$, which is the total number of multipole moments. The second and third
scenarios show that the performance is almost independent of the number
of redshifts $N_\chi$. The 2-FAST algorithm always calculates the
$w_{\ell\ell'}(\chi,R\chi)$ at different comoving radii, even when only one
redshift is desired, and the FFT takes only a marginal fraction of the total
time. This is one of the strengths of the 2-FAST algorithm: one automatically gets
$w_{\ell\ell'}(\chi,R\chi)$ for \emph{all} $\chi$ at once.

Finally, the last test scenario demonstrates that the time scales
proportionally to the number $N_R$ of ratios $R$ and roughly proportionally to
the number of sampling points $N$. Hence, it is feasible to create a dense
grid of $R$-values to cover a large fraction of the $\chi$-$\chi'$ plane.
This will be useful, for example, when calculating the angular power spectrum
for surveys with a broad radial window function or for weak gravitational
lensing convergence.

Note that when we need to calculate the angular harmonic projections of
several different power spectra, then the cosmology-independent
${}_2F_{1,\ell_\max}$ and $M^q_{\ell\ell'}$ can be precalculated and cached
as described in \refsec{ClfromMll}.
In that case, only the timing from the ``$C_\ell$'' column is relevant.

The 2-FAST method scales with the number of sample points $N$, the number of
ratios $N_R$, and the number of multipole moments $N_\ell$ desired, which we
here set as $\ell_\max$ (i.e. no binning in multipoles). That is, the time $T$
to take for the calculation scales as
\changed{\ba
T &\propto N_\ell \cdot N_R \cdot N\cdot\log{N}\,.
\ea}
\changed{In our tests,} the time for the FFT is negligible compared to other
operations \changed{that scale with $N$}.

We recommend caching the \emph{initial} value of ${}_2F_{1,\ell_\max}$ at
$\ell_\mathrm{max}$. While caching $M_{\ell\ell'}^q$ may make sense in some
cases, the $M_{\ell\ell'}^q$ cache may demand very large disk space.

\section{Applications}
\label{sec:applications}
In this section we consider three applications of the 2-FAST algorithm. First,
we study the radial BAO signal, then the lensing potential power spectrum, and
finally the lensing-convergence-galaxy cross-correlation. These three test
cases demonstrate that we can apply the 2-FAST algorithm for calculating the
cross-correlation between two widely separated redshift bins as well as
angular autocorrelation and cross-correlation of widely spread-out density fields.

\subsection{Radial baryon acoustic oscillations}
\label{sec:baonobao}

\begin{figure*}
    \includegraphics[width=0.49\textwidth]{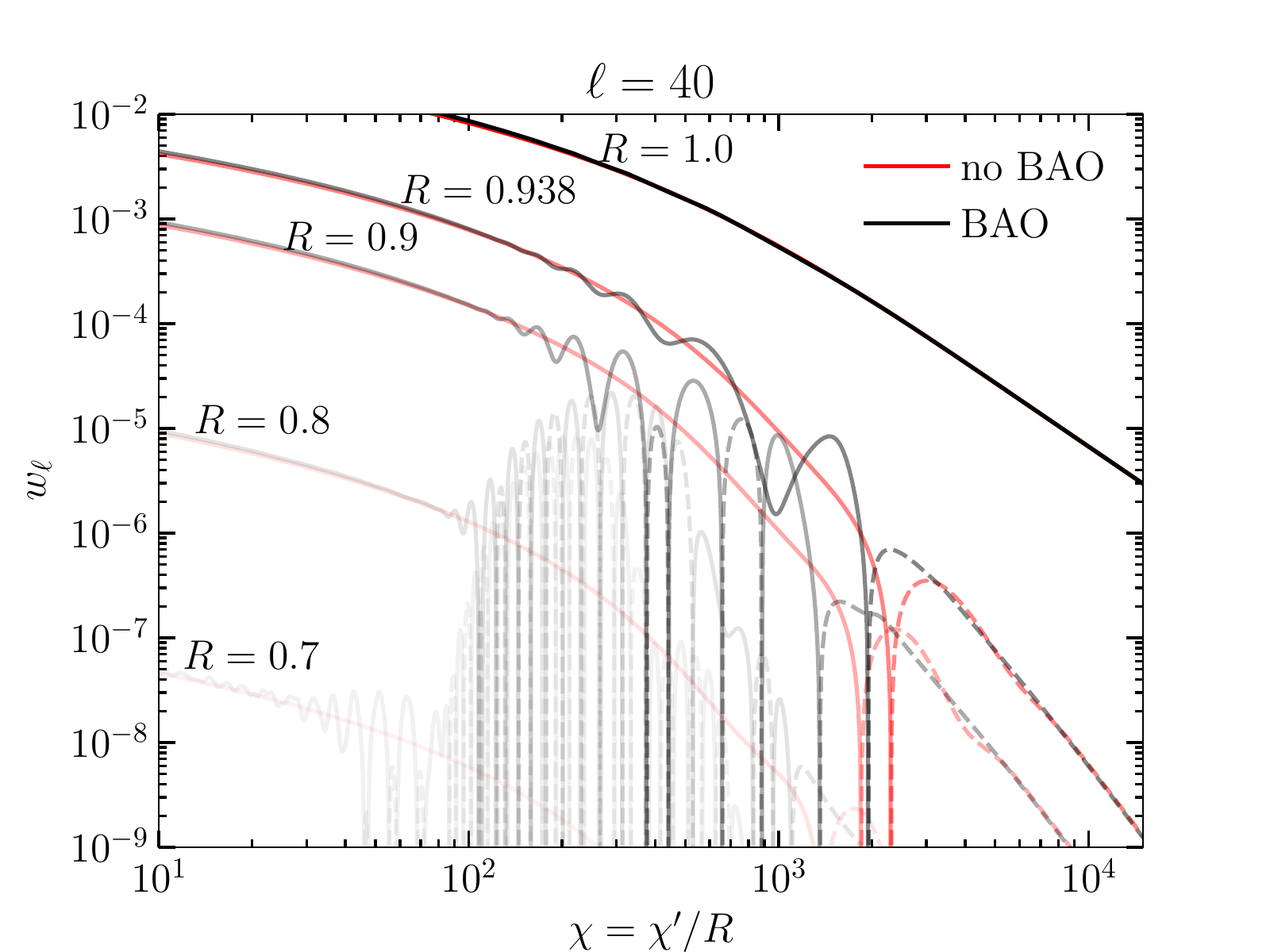}
    \includegraphics[width=0.49\textwidth]{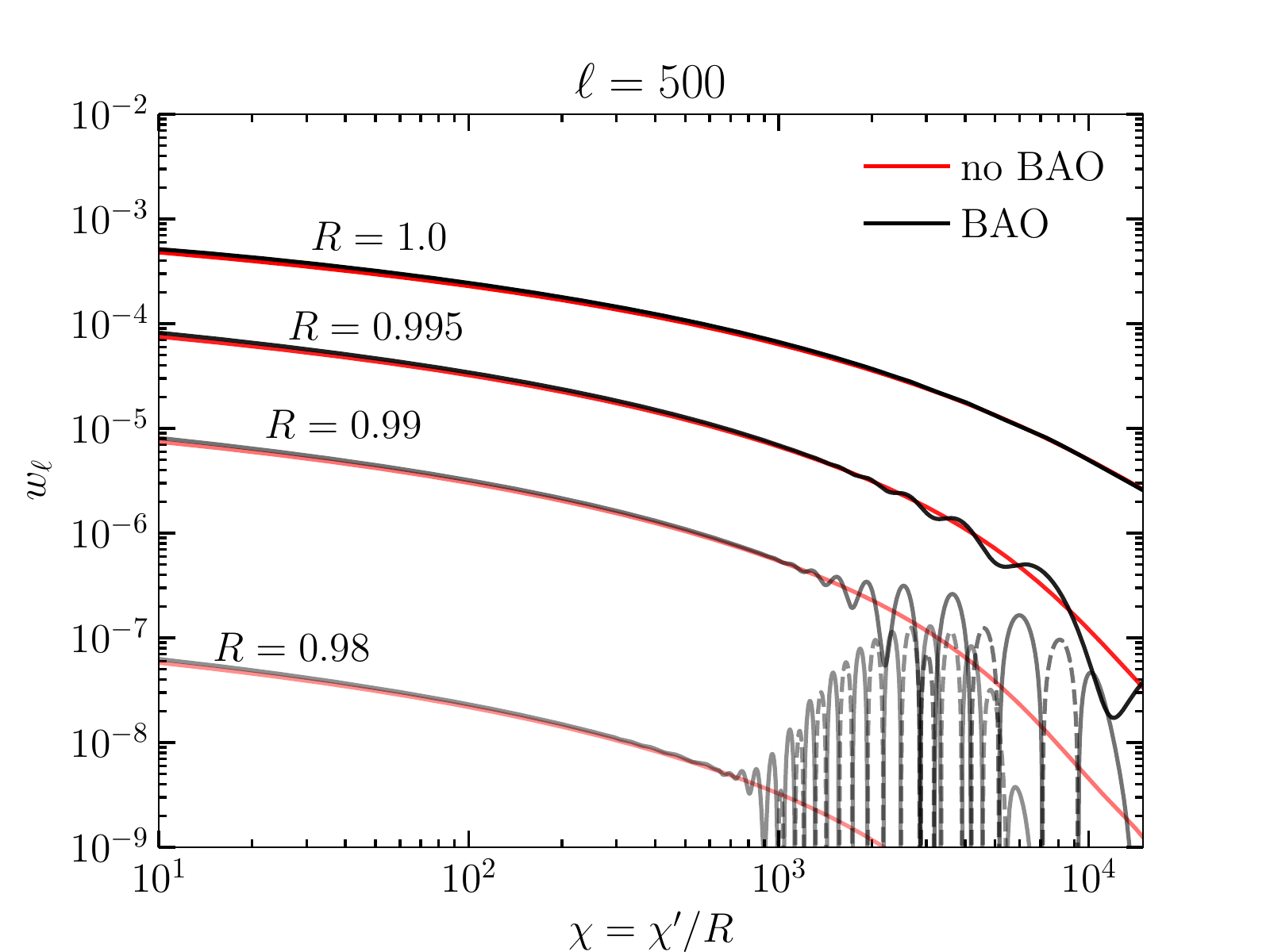}
    \caption{Projected power spectra with and without BAO.
		Left: For $\ell=40$. Right:
    For $\ell=500$. In gray is the BAO power spectrum, in red the power
    spectrum without BAO; dashed lines indicate negative values.
    The BAO appears as
    wiggles for $R\neq1$, which for larger $\ell$ begin at larger distances.}
    \label{fig:wiggles}
\end{figure*}


With an accurate and efficient implementation of the 2-FAST algorithm, we
study the radial BAO appearing in the harmonic-space correlation function,
$w_{\ell}(\chi,R\chi)$:
\be
w_{\ell}(\chi,R)
=
\frac{2}{\pi}\int_0^\infty \dd k k^2 P(k) j_{\ell}(k\chi) j_{\ell}(kR\chi)\,.
\ee
Here we study the harmonic-space correlation function by itself.
Summing $w_\ell(\chi,\chi')$ over $\ell$ corresponds to the real-space two-point
correlation function \citep{lepori/etal:2017} with the wide-angle formula.

In order to highlight the BAO feature, we compare the angular power spectrum
$w_\ell$ with the $P(k)$ from the \texttt{CAMB} output (with BAO)
to the $w_\ell$ with the $P(k)$ from the no-BAO fitting formula given
by \citet{eisenstein/hu:1998}. We study the radial BAO signature by
fixing the multipole moments $\ell$ and the ratio $R$ and plotting
$w_{\ell}(\chi,R\chi)$ as a function of comoving radial distance $\chi$.

\reffig{wiggles} shows the comparison for the $w_{\ell=40}$ (left panel)
and $w_{\ell=500}$ (right panel) cases. In both panels, the black curves and
the red curves show, respectively, the power spectrum with BAO and without BAO.
The radial BAO feature is most prominent for $R\neq1$ cases.
Because the acoustic scale of $d_\mathrm{BAO}\simeq\SI{106}{\per\h\mega\parsec}$ is
fixed, when fixing the ratio $R$ between two radii, the radial BAO features
appear at larger (smaller) radius for larger (smaller) multipole moments that
correspond to the smaller (larger) angular scales. That is, for a
standard ruler of size $d_\mathrm{BAO}$ where radial distance to each end is
$\chi$, $\chi'=R\chi$, the angle subtended by the ruler is
$\cos\theta = [1+R^2 - (d_\mathrm{BAO}/\chi)^2]/(2R)$. For small angles
one can approximate $\theta\simeq\pi/\ell$, so that
$\chi_\mathrm{BAO}\simeq d_\mathrm{BAO}/\sqrt{(1-R)^2 + R(\pi/\ell)^2}$.

For a randomly oriented ruler of BAO size, the viewing-angle-average
projected length is $(\pi/4)d_\mathrm{BAO}$, from which we estimate the
characteristic radius at which the radial BAO appears as
\be
\chi_\mathrm{BAO} \simeq \frac{\ell}{4} d_\mathrm{BAO}.
\ee
The corresponding $R$ is
\ba
R_\mathrm{BAO}
&\sim
1 \pm \frac{\sqrt{16-\pi^2}}{\ell}\simeq 1\pm \frac{2.48}{\ell}
\ea
to first order in $1/\ell$.
For a fiducial $\Lambda$CDM cosmology, we find
$\chi_\mathrm{BAO}=\SI{1060}{\per\h\mega\parsec}$, $R_\mathrm{BAO}=0.938$ for $\ell=40$ and
$\chi_\mathrm{BAO}=\SI{13250}{\per\h\mega\parsec}$, $R_\mathrm{BAO}=0.995$ for $\ell=500$,
which are consistent with \reffig{wiggles}.

The result shows that the BAO feature in the angular power spectrum is spread
over many multipole moments and distance ratios $R$. As we have shown before,
this is because the BAO is a sharp feature defined in the configuration
space with a fixed distance scale. Therefore, although we show the radial BAO
here as a performance test for the 2-FAST algorithm, the best method of
detecting BAO would be to detect in configuration space.
After all, one does not need a spherical projection for the BAO, as long as
the BAO scale is much smaller than the radial distances to the survey.

\subsection{Lensing potential power spectrum}
\label{sec:lenspotential}
We now turn to the case for the angular power spectrum of a widely
spread density distribution using the 2-FAST algorithm. As an
example, we calculate the lensing potential power spectrum $C^{\psi\psi}_\ell$
for the cosmic microwave background (CMB) lensing, where the source plane is at the CMB's
last-scattering surface ($z_\star\simeq1089$). We denote the comoving angular
diameter distance to the surface of last scattering as
$\chi_\star\equiv \chi(z_\star)$.

The lensing potential for the CMB lensing is \citep{lewis/challinor:2006}
\ba
\psi(\nhat)
&= -2\int_0^{\chi_\star} \dd\chi\,\frac{\chi_\star-\chi}{\chi_\star\chi}\,
\Phi(\chi\nhat)\,,
\label{eq:lens-potential}
\ea
where the gravitational potential $\Phi(\chi\nhat)$ is related to the
density contrast by Poisson's equation:
\ba
k^2\Phi(\bfk,a)
=&\, 4\pi G a^2 \bar\rho_{m}(a)\,\delta_m(\bfk,a)
\vs
=&\, \frac32 a^2H^2 \Omega_{m}(a)\,\delta_m(\bfk,a)\,.
\ea
The angular power spectrum of the lensing potential is then given by
\ba
C^{\psi\psi}_{\ell}
&=
\int_{0}^{\chi_\star} \dd\chi\,\varphi(\chi)\,
\int_{0}^{\chi_\star} \dd\chi'\,\varphi(\chi')\,
\vs&\qquad\times
\frac2\pi \int_0^\infty \dd k\,k^{-2}\,P(k)\,
j_{\ell}(k\chi)\,
j_{\ell}(k\chi')
\\
&=
\int_{0}^{\chi_\star} \dd\chi\,\varphi(\chi)\,
\int_{0}^{\chi_\star} \dd\chi'\,\varphi(\chi')\,
w_{\ell\ell}^p(\chi,\chi')\,,
\ea
where the index $p=-4$ reminds us that in order to utilize the 2-FAST algorithm
\changed{as described in \refsec{well}, we need to replace the function $P(k) \to
k^{-4}P(k) \propto P_\psi(k)$.} We also defined the radial weighting function
$\varphi(\chi)$ as
\ba
\varphi(\chi)
&= \frac{\chi_\star - \chi}{\chi_\star \chi}\,(1+z)\,D(\chi)\,,
\ea
where $D(\chi)$ is the linear growth factor.
Furthermore, we introduce $R=\chi'/\chi$, and we use the symmetry
$w_{\ell\ell'}(\chi,\chi')=w_{\ell'\ell}(\chi',\chi)$ to find that
\ba
C^{\psi\psi}_{\ell}
&=
\int_{0}^{1} \dd R\,
\int_{0}^{\chi_\star} \dd\ln\chi\,
\big[2\,
\chi^2\,\varphi(\chi)\,
\varphi(R\chi)\,
\big]
w_{\ell\ell}^p(\chi,R)\,.
\ea
We first calculate $w_{\ell\ell}^p(\chi,R)$ by using the 2-FAST algorithm
and perform the integration over $\ln\chi$ and $R$ using the trapezoidal
method \cite{NR3rd}.
The sampling in $\ln\chi$ is given by $N$ that we use for FFT in 2-FAST.
As shown in \reffig{cl-R1}, for $\ell\lesssim100$ the
$w_{\ell\ell}(\chi,R)$ are a slowly varying function of
$R$, whereas for high $\ell$ they are narrowly peaked around $R\sim1$. Hence,
for $R$ we choose different samplings for $\ell\leq100$ and $\ell>100$.
Specifically, we choose 51 evenly spaced sampling points between $R=1$ and
$R=0.9$ for $\ell>100$, and 51 sampling points between
$R=1$ and $R=0$ for $\ell\le100$.
Finally, because the power spectrum is divided by $k^4$ compared to the
matter density power spectrum, the biasing parameter $q$ also needs to be
adjusted to $q\sim-2.5$ (see \refapp{xiq}).

\begin{figure}
    \includegraphics[width=0.48\textwidth]{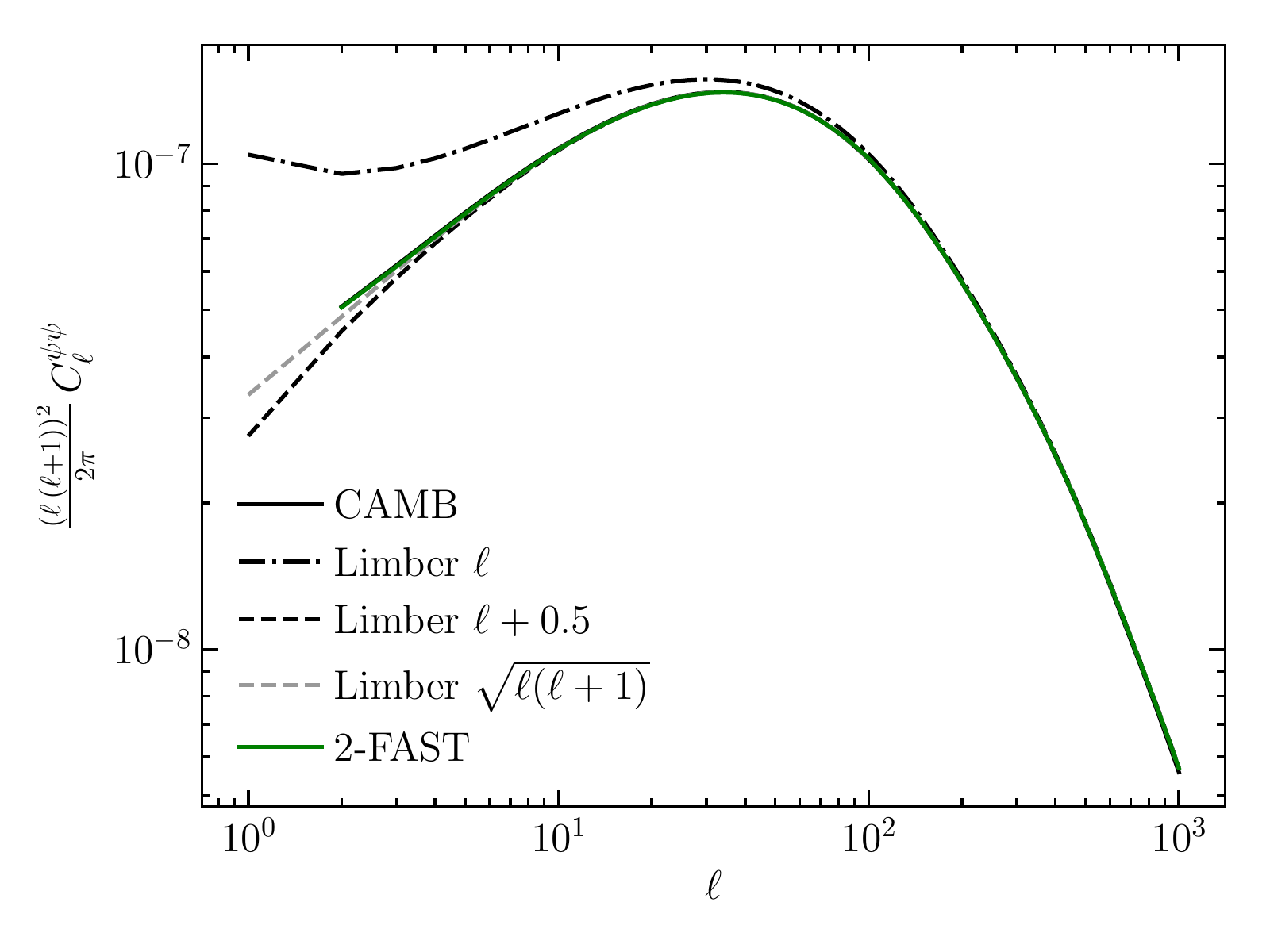}
    \caption{The lensing potential angular power spectrum
    $C_\ell^{\psi\psi}$. In solid black is the result from \texttt{CAMB}, in dashed-dotted black
    the Limber approximation with $\nu=\ell$ [see \refeq{limber}], in dashed
    black the Limber approximation with $\nu=\ell+0.5$, in dashed gray
    Limber's approximation with $\nu=\sqrt{\ell(\ell+1)}$, and in solid green
    the 2-FAST method presented in this paper. The multiplication by
    $[\ell(\ell+1)]^2$ amplifies the error of the $\nu=\ell$ Limber
    approximation. Our 2-FAST result is too close to the result from \texttt{CAMB} to be
    distinguished in this graph.}
    \label{fig:psipsi}
\end{figure}
The resulting lensing potential power spectrum is shown in \reffig{psipsi}
as a solid green line, which lies on top of the \texttt{CAMB} output (black solid line).

\subsubsection{Limber's approximation}
\label{sec:psipsi-limber}
We compare the result with Limber's approximation
(see, for example \citep{loverde/afshordi:2008}), where the spherical Bessel
integration is approximated as
\ba
w_{\ell\ell}(\chi,\chi')
&= \frac{2}{\pi}\int \dd k\,k^2\,P(k)\,j_\ell(\chi k)\,j_{\ell}(\chi'k)
\\
&\approx
\frac{\delta^D(\chi-\chi')}{\chi^2}\,P\left(\frac{\nu}{\chi}\right)\,,
\ea
with $\nu=\ell+\frac12$.
Using Limber's approximation, the lensing potential $C_\ell$ becomes
\ba
C^{\psi\psi}_{\ell}
&\approx
\int_{0}^{\chi_\star} \dd\chi\,
\frac{\varphi^2(\chi)}{\chi^2}\,
P\left(\frac{\nu}{\chi}\right)\,,
\label{eq:limber}
\ea
which we integrate using Gauss-Kronrod integration. In the
literature, the numerator in the argument to the power spectrum is often
approximated as $\nu=\ell$ instead of $\nu=\ell+0.5$. In \reffig{psipsi} we
show both for comparison, as well as the exact calculation from the 2-FAST
algorithm.

We note that the Limber approximation reproduces the exact calculation for
larger multipole moments $\ell\gtrsim100$, but the result deviates from the
exact calculation for larger angular scales. In particular, the ``old'' Limber
approximation with $\nu=\ell$ shows the largest deviation, whereas the proper
Limber approximation with $\nu=\ell+0.5$ as derived in
\cite{loverde/afshordi:2008} follows the correct value to the larger scales
$\ell\simeq 10$. We note that a further improvement can be achieved by using
$\nu=\sqrt{\ell(\ell+1)}$ (gray dashed line), which was already
hinted at in \citep{loverde/afshordi:2008}.

\subsection{Lensing convergence-galaxy cross correlation}
\label{sec:lensing-galaxy-cross-correlation}
\begin{figure*}
	\includegraphics[width=0.49\textwidth]{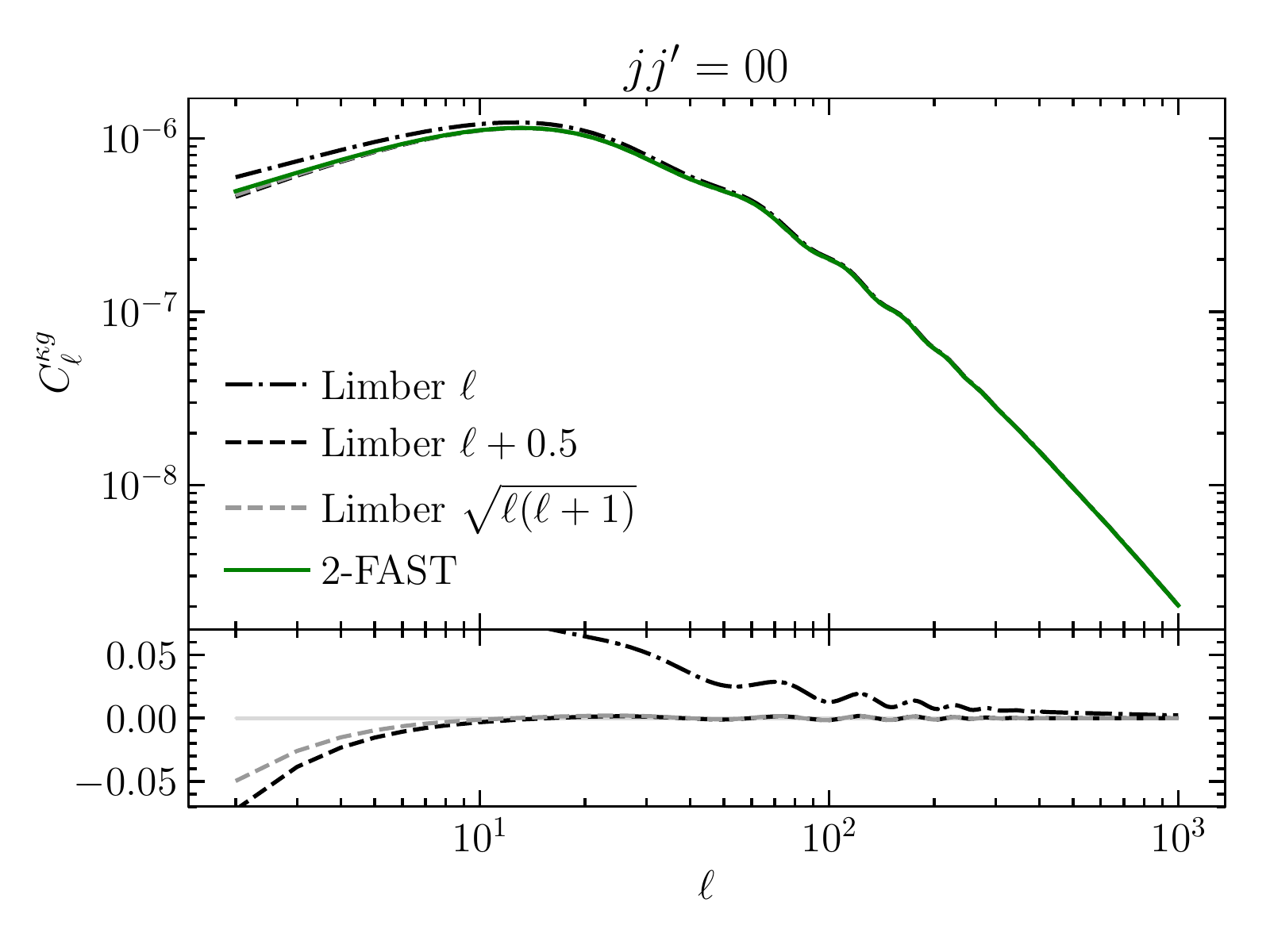}
	\includegraphics[width=0.49\textwidth]{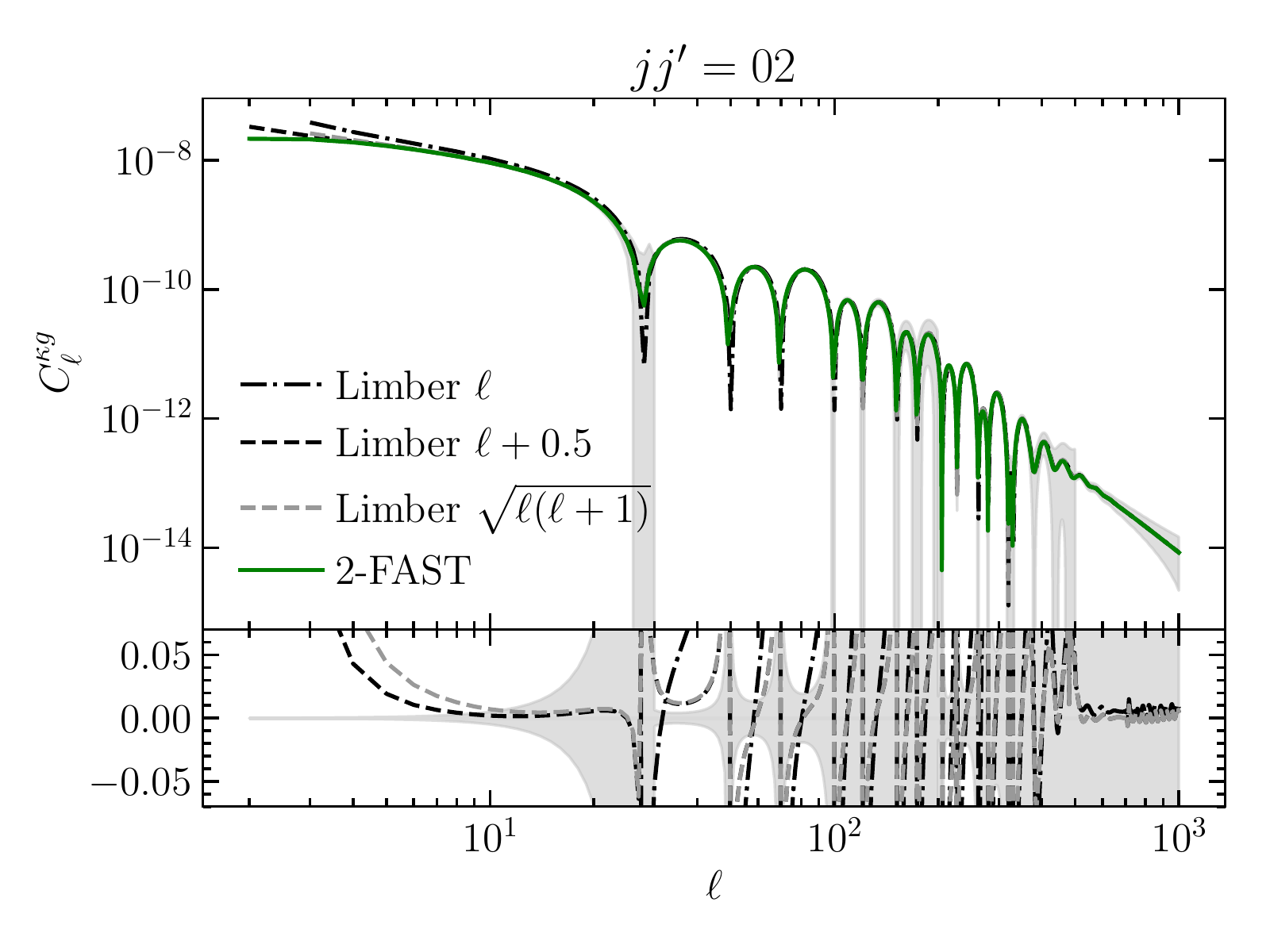}
	\caption{The two terms contributing to \refeq{cl2-kappa-g-chi}:
	$jj'=00$ on the left and $jj'=02$ on the right. In addition to
	our 2-FAST method, we show three versions of the Limber approximation,
	which are defined as in \reffig{psipsi}.
	The gray bands show the estimated error due to the discrete sampling of
	$R'=\chi/\chi'$.
	For large $\ell$, the Limber
	approximation agrees well with the more exact 2-FAST calculation.
	However, at $\ell\lesssim10$, none of the Limber approximations achieves
	better than percent-level precision. }
	\label{fig:kappa-g-jj}
\end{figure*}
As a final test case, we calculate the cross-correlation $C^{\kappa g}_\ell$
between foreground galaxies at comoving distance $\chi'$ (redshift $z'$) and
the lensing convergence field $\kappa$ reconstructed from the source galaxies
at distance $\chi_\star$ (redshift $z_\star$).
Such a cross-correlation dominates the cross-correlation between galaxies
widely separated in redshift, because the lensing magnification traces
the line-of-sight directional convergence.

Besides the relativistic corrections (see \cite{jeong/schmidt:2015} for a
review), the dominant components of the observed galaxy density contrast
$\delta_g$ of galaxies are given by
\ba
\delta_g(\bfk) &= [b_g + f(\nhat\cdot\khat)^2]\,\delta_m(\bfk) + 2(\Q-1)\kappa\,,
\ea
where $b_g$ is the galaxy bias, $f$ the linear growth rate
$f\equiv\dd \ln D/\dd\ln a$, $\delta_m(\bfk)$ the matter density contrast,
$\Q$ the slope of the luminosity function at the
survey limit, and $\kappa$ is the lensing convergence.
Here, we neglect the factor $2(\Q-1)$ \changed{and we set $b_g=1$, as those
factors are specific to the galaxies and survey.}

The lensing convergence is given by
\ba
\kappa(\chi\nhat)
&= -\tfrac12 \nabla_\theta^2 \psi(\nhat)\,,
\ea
where the lensing potential $\psi(\nhat)$ is given in \refeq{lens-potential}.
Then, the cross-correlation between lensing convergence for the sources at
distance $\chi_\star$ and galaxies at distance $\chi'$ is given by
\cite{jeong/komatsu/jain:2009}
\ba
C^{\kappa g}_\ell(\chi_\star,\chi')
&=
\tfrac32\Omega_m H_0^2\,
\ell(\ell+1)
\int_0^{\chi_\star} \frac{d\chi}{\chi}\,\frac{\chi_\star-\chi}{\chi_\star}
\vs&\qquad\times
\frac{D(z)D(z')}{a}
\vs&\qquad\times
\big[
    b'w^p_{\ell,00}(\chi,\chi')
    - f'w^p_{\ell,02}(\chi,\chi')
\big]\,,
\label{eq:cl2-kappa-g-chi}
\ea
where we attach the suffix $p=-2$ to $w_{\ell,jj'}$ to signify that the biased
power spectrum $k^{-2}P(k)$ is to be used.
To use the 2-FAST algorithm, it is advantageous to exploit the symmetry
$w_{\ell,jj'}(\chi,\chi')=w_{\ell,j'j}(\chi',\chi)$ and introduce
$R'=\chi/\chi'$. That is,
\ba
w^p_{\ell,jj'}(\chi,\chi')
&= w^p_{\ell,j'j}(\chi',R'\chi')\,.
\ea
This way we can keep $\chi'=\mathrm{const}$ while performing the integral over
$R'$. With $\ln\chi=\ln{R'}+\ln{\chi'}$ we get
\ba
C^{\kappa g}_\ell(\chi_\star,\chi')
&=
\tfrac32\Omega_m H_0^2\,
\ell(\ell+1)
\int_{0}^{\chi_\star/\chi'} d\ln{R'}\,
\frac{\chi_\star-\chi}{\chi_\star}
\vs&\qquad\times
\frac{D(z)D(z')}{a}
\vs&\qquad\times
\big[
	b'w^p_{\ell,00}(\chi',R'\chi')
    - f'w^p_{\ell,20}(\chi',R'\chi')
\big]\,.
\label{eq:cl2-kappa-g}
\ea
We partition the range in $\ell$ into four intervals. In each interval we
choose a different sampling for $\ln{R'}$. This is needed, since the integrand
is broad at low $\ell$, but becomes a narrowly peaked function at high $\ell$.
Specifically, for our test case $z'=0.3$
($\chi'=\SI{835}{\per\h\mega\parsec}$), $z_\star=2.2$
($\chi_\star=\SI{3796}{\per\h\mega\parsec}$) we found the following choices to
work well for the integer values $m$:
\begin{center}
	\begin{tabular}{r@{$\,\rightarrow\,$}r|l|c}
		\multicolumn{2}{c|}{$\ell$-interval}
		& \multicolumn{1}{c|}{$R'$}
		& $m$ \\
		\hline
		500 & 1000$\,$ & $\,e^{0.0001\,m}$ & $-400,\ldots,400$ \\
		200 & 499$\,$  & $\,e^{0.0005\,m}$ & $-200,\ldots,200$ \\
		30  & 199$\,$  & $\,e^{0.002\,m}$  & $-400,\ldots,200$ \\
		2   & 29$\,$   & $\,e^{0.01\,m}$   & $-300,\ldots,151$
	\end{tabular}
\end{center}
These choices ensure an accurate calculation of all the terms in
\refeq{cl2-kappa-g}. However, since \refeq{cl2-kappa-g} is dominated by the
$(j,j')=(0,0)$ term, a less dense grid in $R'$ may be sufficient for many
applications. We then integrate using the trapezoidal method. Here, we
use $q=0.5$ (\refapp{xiq}).

When either $z'$ or $z_\star$ differs from its values investigated here, then
the values for $R$ in the table above may be used as a starting point, and for
each $\ell$ region one may increase the interval in $R$, as well as the number
of sampling points in $R$ until convergence is reached.

Alternatively, one can avoid the somewhat \emph{ad hoc} choice of $R$-sampling by
integrating $M_{\ell\ell'}(t,R)$ over the radial window function as shown in
\refeq{Cl_2FAST}. While we have not further investigated here,
\citet{assassi/etal:2017} have shown that the integration can be done with
the hypergeometric function ${}_3F_2$ if the radial function can be
approximated by a sum of polynomials. We shall further study this in
future publications.

\subsubsection{Comparison}
\begin{figure}
	\includegraphics[width=0.48\textwidth]{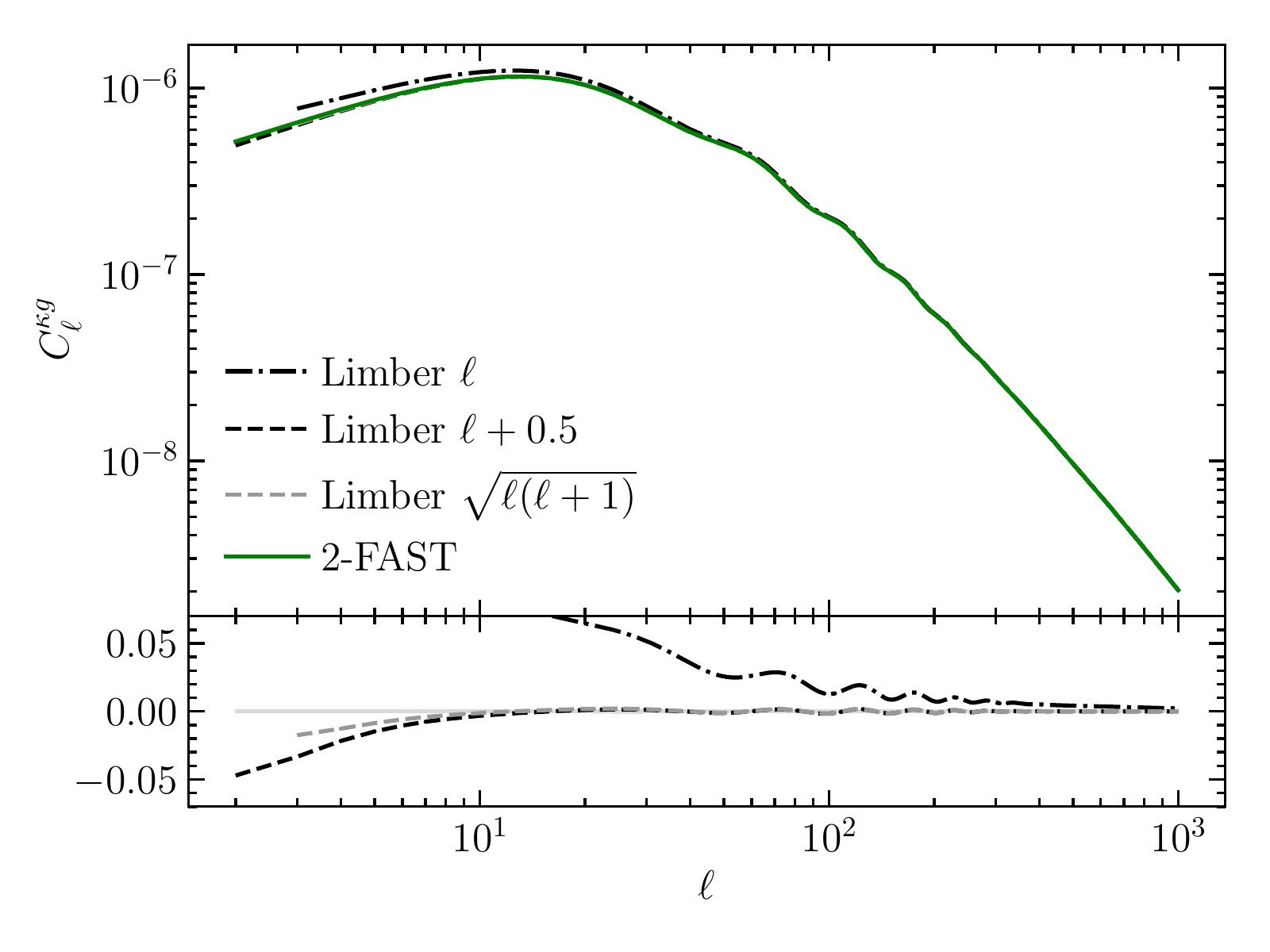}
	\caption{The lensing-convergence-galaxy
	cross-correlation \refeq{cl2-kappa-g-chi}, the sum of the two
	plots in \reffig{kappa-g-jj}. All labels have the same meaning as in
	that figure. The $jj'=00$ term dominates the cross-correlation.}
	\label{fig:kappa-g}
\end{figure}

In this section we compare our results with several versions of Limber's
approximation. To be applicable to the $w_{\ell,02}$ term in
\refeq{cl2-kappa-g-chi}, we extend Limber's approximation to include the
cases $\ell\neq\ell'$ by using the results from \citet{loverde/afshordi:2008}.
We defer the details of this derivation to \refapp{general_limber}.

\reffig{kappa-g-jj} and \reffig{kappa-g} show the comparison of the 2-FAST
calculation with three versions of Limber's approximation.
\reffig{kappa-g-jj} shows the two terms in \refeq{cl2-kappa-g-chi} separately
in the left and right panels, whereas \reffig{kappa-g} shows the full
lensing-convergence-galaxy cross-correlation power spectrum.
In both figures the estimated error due to discretization of the $R'$-integral
is shown as a gray band around the 2-FAST line. Also shown in the lower panels
of the figures is the relative difference of Limber's approximation to the
2-FAST algorithm calculation.

Again, Limber's approximation is accurate for larger multipole moments, but
deviates from the exact calculation from 2-FAST on small multipoles (larger
angular scales). However, we note that as in \refsec{psipsi-limber}, the
$\nu=\sqrt{\ell(\ell+1)}$ version of Limber's approximation agrees quite
well with the 2-FAST results. Percent-level precision is achieved with
this version of Limber's approximation for $\ell\gtrsim10$.

\section{Conclusion}
\label{sec:conclusion}
In this paper, we have presented the 2-FAST algorithm for
projecting the three-dimensional power spectrum onto two-point correlation
functions in configuration space as well as in spherical harmonic space.
Based on the FFTLog method in \cite{talman:1978,siegman:1977,hamilton:2000}, we
generalize to the case $\ell\neq\ell'$.

By decomposing the power spectrum with FFTLog basis functions and the
coefficients $\phi^q$, the infinite-range integrations in the 2-FAST
algorithm are done as gamma functions and Gauss hypergeometric functions
${}_2F_1$ for,
respectively, calculating $\xi_\ell^\nu(r)$ and $w_{\ell\ell'}(\chi,\chi')$.
Therefore, 2-FAST bypasses the difficulties in dealing with oscillatory
spherical Bessel functions with large arguments.
At the core of the 2-FAST algorithm is a recursion algorithm for computing the
Gauss hypergeometric function. In particular, the stable backward recursion
enables a fast, high-precision calculation of $w_{\ell\ell'}(\chi,\chi')$.

Using the fast Fourier transformation, the 2-FAST algorithm calculates the
$w_{\ell\ell'}(\chi,R\chi)$ for multiple values of $\chi$ (regularly spaced
in the logarithmic interval) by one operation of FFT.
In addition, an efficient recursion along the $\ell$-direction provides
$w_{\ell\ell'}(\chi,R\chi)$ essentially to arbitrary $\ell$ and $\ell'$
values (although the current implementation is done for $\Delta \ell=\pm4$,
the extension is trivial, if necessary).
Furthermore, we can then easily map out $w_{\ell\ell'}(\chi,\chi')$ in the
$\chi$-$\chi'$ plane by repeating the procedure for different values of $R$.

From the transformation matrix from $P(k)$ to $C_\ell$ in \refsec{ClfromMll}
and the two examples in \refsec{lenspotential} and
\refsec{lensing-galaxy-cross-correlation}, we have demonstrated that
Limber's approximation with the identification $k = \sqrt{\ell(\ell+1)}/\chi$
performs far better than the traditional prescription of $k=\ell/\chi$ or
$k=(\ell+\frac12)/\chi$, and it reaches down to $\ell=10$ for the convergence-galaxy
cross-correlation power spectrum. For small-angle galaxy surveys, using this
improved Limber approximation may suffice.
For the future galaxy surveys with large angular footprints, however, we need
to use the full calculation, and the 2-FAST algorithm will make it fast and accurate.

There are two directions in which we can extend the 2-FAST algorithm.
First, we can boost the speed of integration over the survey radial window
function by approximating the window function as a sum over polynomials.
In this case, the integration over a polynomial window function can be
done as a hypergeometric function ${}_3F_2$ \cite{assassi/etal:2017}, and
a detailed study of recursion can accelerate the calculation faster than
the current 2-FAST method that integrates over the precalculated
$w_{\ell\ell'}(\chi,\chi')$ on the $\chi-\chi'$ space.
Second, for the $n$th-order angular polyspectra, we need to calculate the
overlapping integration of $n$ spherical Bessel functions with some polynomial.
We surmise that the Fourier-based method that we presented here can aid
greatly in this type of calculation as well. These two issues must be
addressed to fully exploit the large angular scale galaxy clustering signatures
from future galaxy surveys.

The \texttt{Julia} version of the 2-FAST implementation can be obtained
\changed{from \url{https://github.com/hsgg/twoFAST}}. The authors have a plan
to implement the \texttt{FORTRAN} and \texttt{C} versions in the future.

\begin{acknowledgments}
\changed{The authors would like to thank Zachary~Slepian and the anonymous
referee for helpful comments in  improving the paper.}
H.~G. and D.~J. acknowledge support from the National Science Foundation Grant No. AST-1517363.
\end{acknowledgments}

\bibliography{reference}

\appendix   
\begin{widetext}
\section{Discrete versions of the algorithm}
\label{app:discrete}
In the main text, we discuss the 2-FAST algorithm based on the integration of
continuous functions over an infinite range. In this appendix, we shall give
discrete versions of the relevant equations that we used when
implementing the algorithm.
First, we present the equations for the FFTLog transformation of the biased
power spectrum that we have implemented through the fast Fourier Transformation
of the array spaced with a constant logarithmic interval
(\refapp{phi-discrete}).
We then show the discrete version for calculation of $\xi_\ell^\nu(r)$
(\refapp{xifftlog-discrete}) and $w_{\ell\ell'}(\chi,\chi')$
(\refapp{well1d-discrete}).
Finally, we shall present some basic equations appearing commonly for all cases
and clarify the relation between variables.

\subsection{FFT of biased power spectrum}
\label{app:phi-discrete}
Here we derive the discrete version of \refeq{phiq}.
We define
\ba
\label{eq:xm-kappam}
x_m &= m \frac{L}{N}  &  \kappa_n &= n\,\frac{2\pi}{L}
\ea
where $N$ is the number of sample points, $L$ is the size of the interval, and
$n,m=0,\ldots,N-1$.
Then \refeq{phiq} becomes
\ba
\label{eq:phi-discrete-simple}
\phi^{q}(x_m)
&= \frac{1}{L}\sum_n e^{i2\pi nm/N}\,e^{(3-q)\kappa_n}\,P(k_0e^{\kappa_n}) \\
&= \frac{1}{L} \, \Big\{\mathrm{RFFT}\big[e^{(3-q)\kappa_n}\,
    P(k_0e^{\kappa_n})\big]\Big\}^*
\label{eq:phi-discrete}
\ea
where ``RFFT'' denotes the fast Fourier Transform specialized for a real
function, and the ${}^*$ symbol denotes complex conjugation.

To reduce ringing we avoid sharp edges at the interval boundaries by applying
the same window function as Eq.~(C.1) in \cite{McEwen/etal:2016}. We repeat it
here for completeness:
\ba
\label{eq:windowfn}
W(x)
 &=
\begin{cases}
    \frac{x-x_\min}{x_\text{left}-x_\min}
    - \frac{1}{2\pi}\sin\left(2\pi\,\frac{x-x_\min}{x_\text{left}-x_\min}\right),
    & x < x_\text{left}
    \\
    1,  &  x_\text{left} < x < x_\text{right}
    \\
    \frac{x_\max-x}{x_\max-x_\text{right}}
    - \frac{1}{2\pi}\sin\left(2\pi\,\frac{x_\max-x}{x_\max-x_\text{right}}\right),
    & x > x_\text{right}
\end{cases}
\ea
We apply this window function to the biased power spectrum both before and
after Fourier transforming.

\subsection{Discrete version of single Bessel function transform}
\label{app:xifftlog-discrete}
Here we give a discrete version of \refeq{xifftlog}. Let $G$ be the size of the
logarithmic integration interval, and $N$ the number of sample points.
\refeq{xifftlog} with $t_m=2\pi m/G$ and $\rho_n=nG/N$ then becomes
\ba
\xi^\nu_\ell(r_n)
&= \frac{k_0^3e^{-(q_\nu+\nu)\rho_n}}{\pi\alpha^\nu G}
\sum_m
e^{i2\pi mn/N}\,
\phi^{q_\nu+\nu}(t_m)\,
M^{q_\nu}_\ell(t_m)
\\
&= \frac{k_0^3e^{-(q_\nu+\nu)\rho_n}}{\pi\alpha^\nu G}
\,\mathrm{BRFFT}\big[
    \phi^{q_\nu+\nu}(t_m)\,
    M^{q_\nu}_\ell(t_m)
\big]
\label{eq:xi-discrete}
\ea
where $\mathrm{BRFFT}[\tilde f(x_m)]=G\times\mathrm{IRFFT}[\tilde f(x_m)]$ is the inverse
transform of a real function $f(x)$ without dividing by $G$.

\subsection{Discrete version of two Bessel function transform}
\label{app:well1d-discrete}
Here we give a discrete version of \refeq{well1d}. With $t_m=2\pi
m/G$ for $N$ samples over an interval $G$, and $\rho_n=nG/N$ we get
\ba
w_{\ell\ell'}(\chi_n,R)
&= 4k_0^3\,\left(\frac{\chi_n}{\chi_0}\right)^{-q}
\frac{1}{G}
\sum_m
e^{i2\pi nm/M}\,
\phi^q(t_m)\,
M^q_{\ell\ell'}(t_m,R)
\vs
&= \frac{4k_0^3}{G}\,\left(\frac{\chi_n}{\chi_0}\right)^{-q}
\,\mathrm{BRFFT}\big[
\phi^q(t_m)\,
M^q_{\ell\ell'}(t_m,R)
\big]
\label{eq:well1d-discrete}
\ea
where $\mathrm{BRFFT}[\tilde f(x_m)]=G\times\mathrm{IRFFT}[\tilde f(x_m)]$ is the inverse
transform of a real function $f(x)$ without dividing by $G$.

\subsection{Relation between variables}
Since the $t_m$ arguments to $\phi^q$ and $M^{\nu,q}_\ell$ [\refeq{xifftlog}], or $\phi^q$
and $M^q_{\ell\ell'}$ [\refeq{well1d}] are identical, we have
\ba
x_m &= m\,\frac{L}{N} = t_m = m\,\frac{2\pi}{G}
\label{eq:tm}
\ea
Thus,
\ba
\label{eq:G}
G &= \frac{2\pi N}{L} = \ln\bigg(\frac{k_\max}{k_\min}\bigg)
\ea
where the last equality follows from the choice $k_0=k_\min$, using the second
of \refeq{xm-kappam}, and by writing the first of \refeq{kappa,rho,R} [or
\refeq{kappa-r}] for $k_\max$ as
\ba
\frac{k_\max}{k_\min} &= e^{2\pi N/L}
\ea
Finally, from our discretization of $\rho_n$ above, \refeq{kappa,rho,R} becomes
\ba
\chi_n &= \chi_0 \, e^{nG/N}
\ea
This also holds for \refeq{kappa-r} when $\chi\rightarrow r$.

\section{On the choice of \texorpdfstring{$N$}{N}, \texorpdfstring{$G$}{G}, and
the biasing parameter \texorpdfstring{$q$}{q}}
\label{app:xiq}
In this appendix, we study the effect of different biasing parameters $q$ and
the resolution $N$ and interval $G$ of the FFTLog transformation.
For the continuous FFTLog transformation, $q$ is only constrained by
\refeq{phi:q}, \refeq{xi:q} and \refeq{wl_q}, which together constitute the requirement that
the integrals converge.
When we implement FFTLog as a discrete Fourier transformation, however, numerical
artifacts affect the accuracy of the result, and the error depends
sensitively on different choices of the parameters $q$, $N$ and $G$.
There are three types of numerical error involved in the Fourier
transformation: (A) ringing, which is the loss of
high-frequency modes, is minimized by avoiding sharp edges in the power
spectrum; (B) aliasing, due to limited sampling, can be reduced by
increasing the number of sampling points $N$; (C) wrap-around, due to
the FFT assuming periodic data so that the high-$k$ modes influence the
low-$k$ modes, is mitigated via zero-padding.

In particular, it turns out that the accuracy of the 2-FAST algorithm depends
sensitively on the choice of the biasing parameter $q$.
We used $q=1.9$ for calculating $\xi_\ell^\nu(r)$ and $q=1.1$ for
calculating $w_{\ell\ell'}(\chi,\chi')$. When integrating the overlapping of
two spherical Bessel functions with $k^{-2}P(k)$ (galaxy-lensing
cross power spectrum, \refsec{lensing-galaxy-cross-correlation}) and
$k^{-4}P(k)$ (lensing potential power spectrum, \refsec{lenspotential}), we
use, respectively, $q=0.5$ and $q=-2.5$.
In this section, we systematically study these choices of the parameter $q$.

We implement \refeq{xifftlog} and \refeq{well1d} by the 2-FAST algorithm
as follows: we first calculate $\phi^q(t)$ by using the discrete FFTLog and
then multiply it by the analytically calculated $M_{\ell}^{q_\nu}(t,R)$ or
$M_{\ell\ell'}^q(t,R)$. We then apply the FFTLog again to calculate
$\xi_\ell^\nu(r)$ or $w_{\ell\ell'}(\chi,R)$.
In this procedure, there are two aliasing effects that worsen the
accuracy of the 2-FAST algorithm: the aliasing effect associated
with the calculation of $\phi^q(t)$, and the aliasing effect associated with
the backward-FFTLog for the convolution.
Of course, we can remedy the aliasing effects by reducing the sampling
intervals $\Delta_k = G/N$ and $\Delta_t=2\pi/G$, respectively, in the
wave number space and its dual space. We find that, however, it is more
efficient to reduce the aliasing effect by choosing an appropriate biasing
parameter $q$. We study the two aliasing effects in the following
subsections.

\subsection{Aliasing effect in FFTLog of \texorpdfstring{$P(k)$}{P(k)}}
\begin{figure}
    \includegraphics[width=0.49\textwidth]{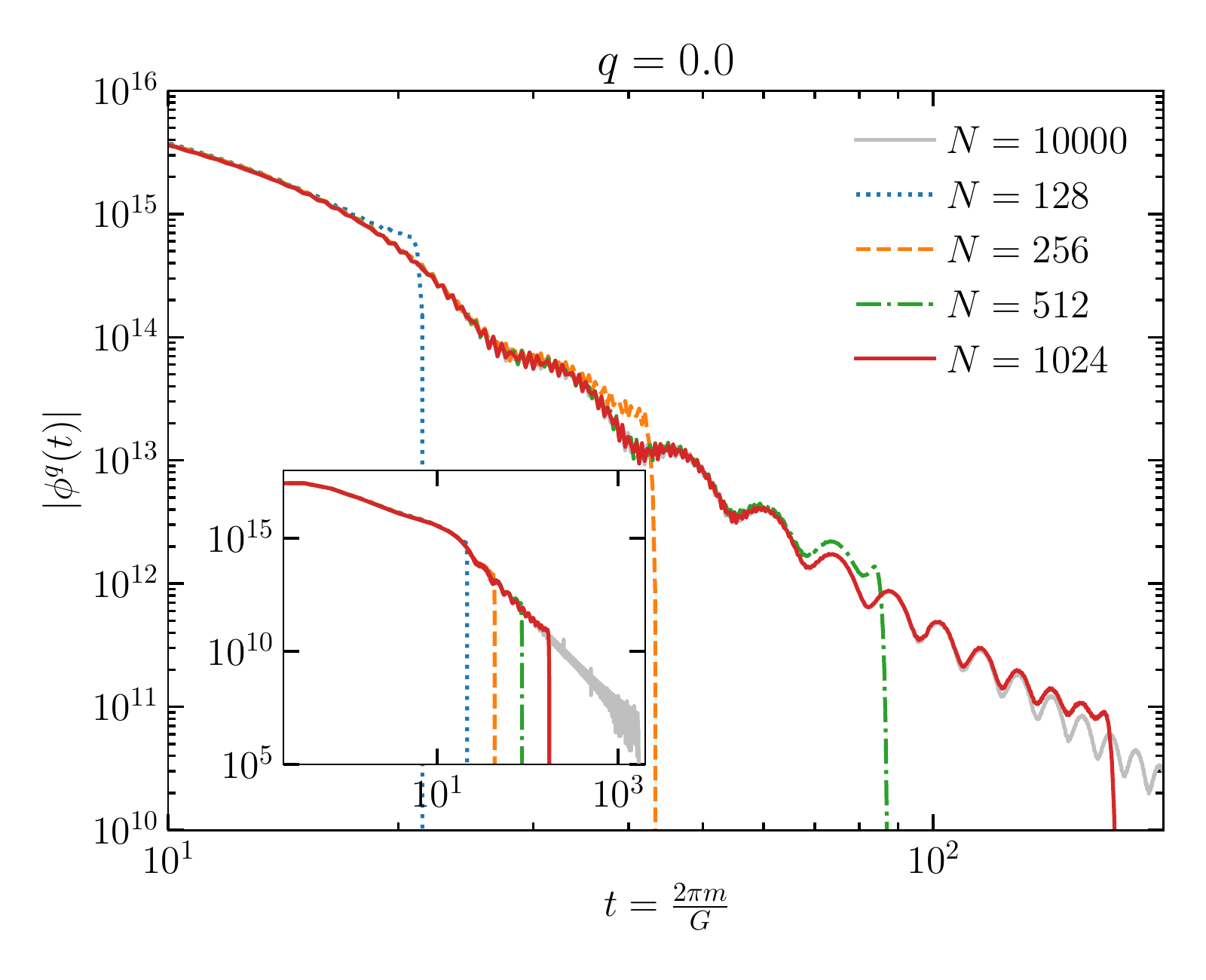}
    \includegraphics[width=0.49\textwidth]{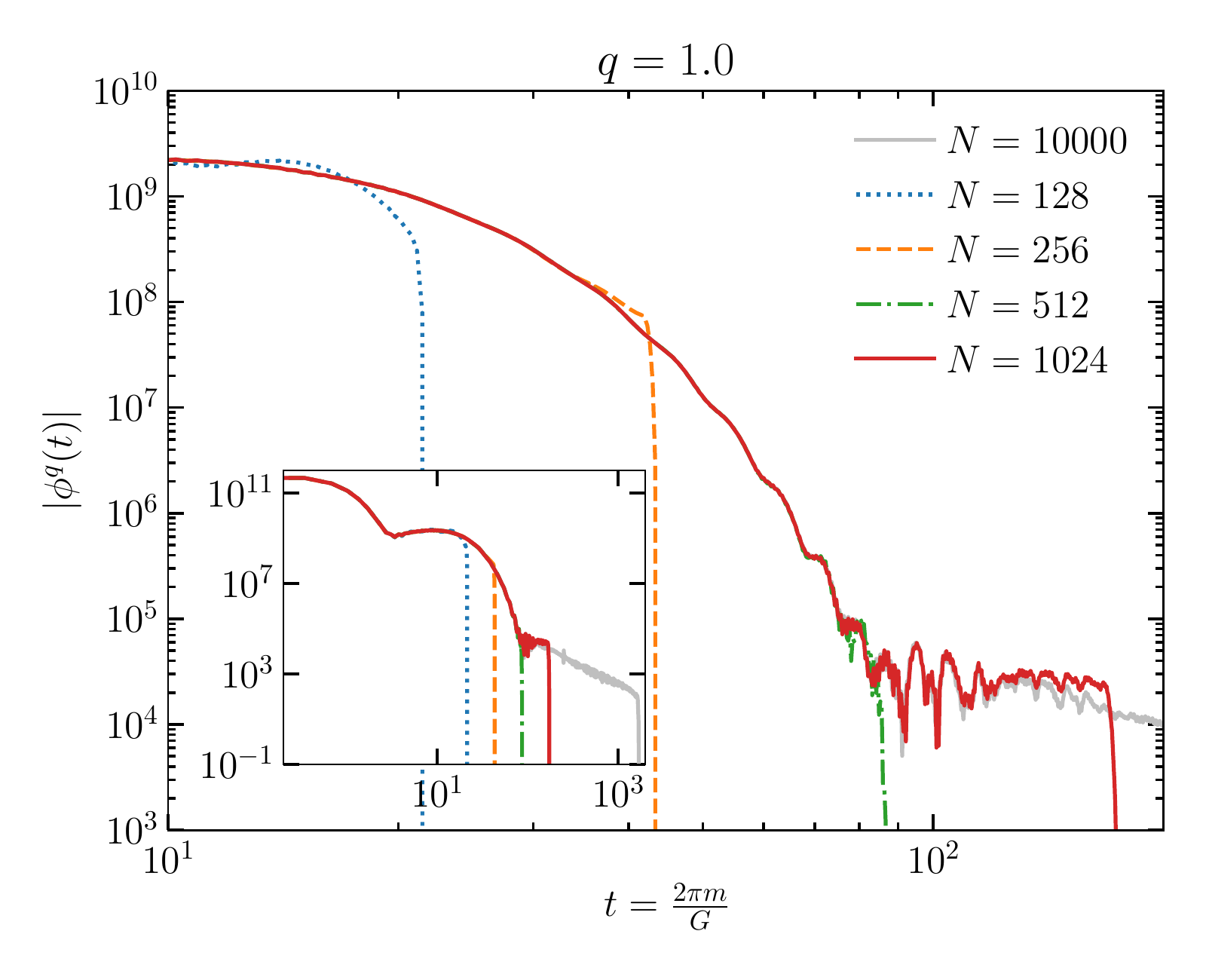}
    \includegraphics[width=0.49\textwidth]{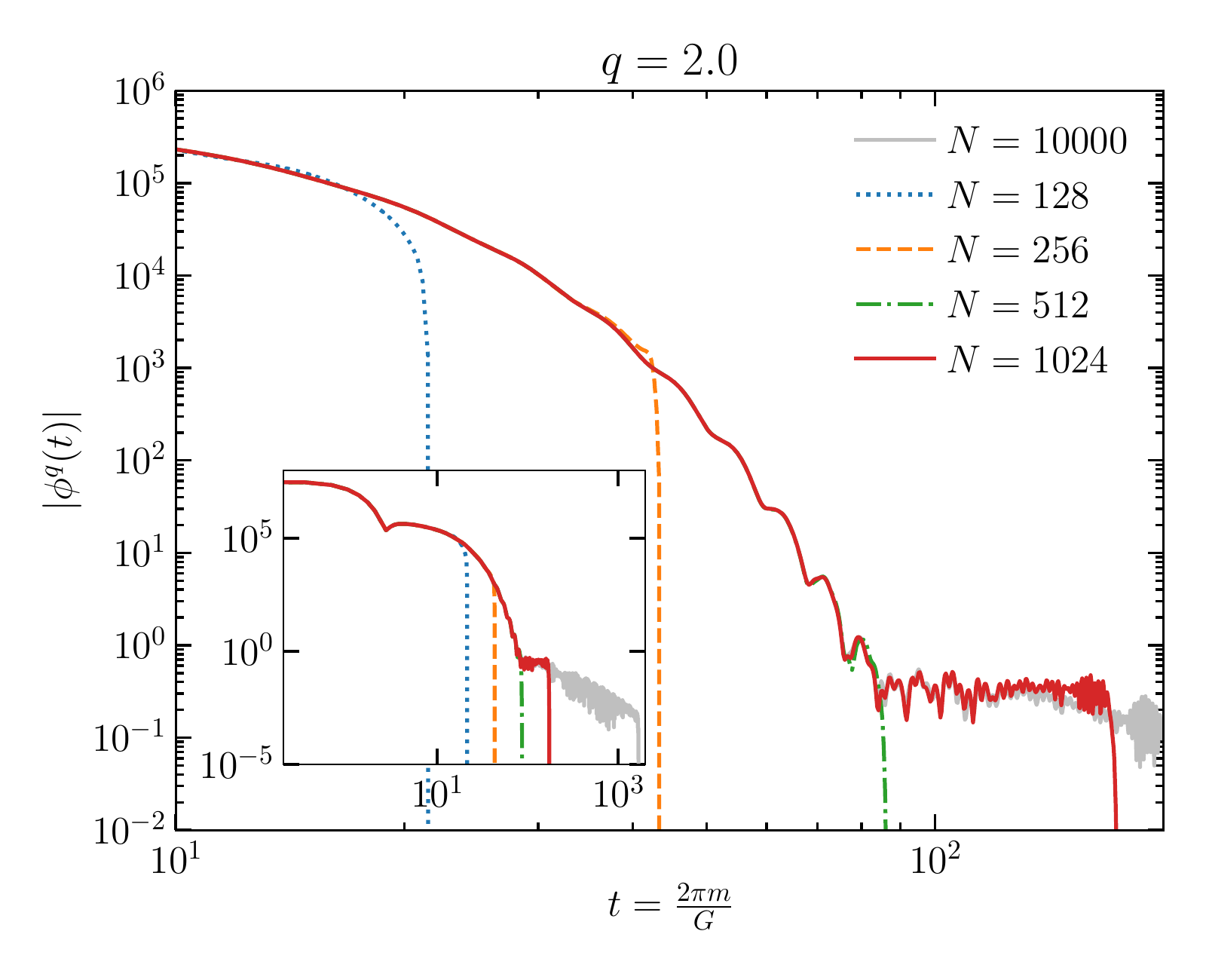}
    \includegraphics[width=0.49\textwidth]{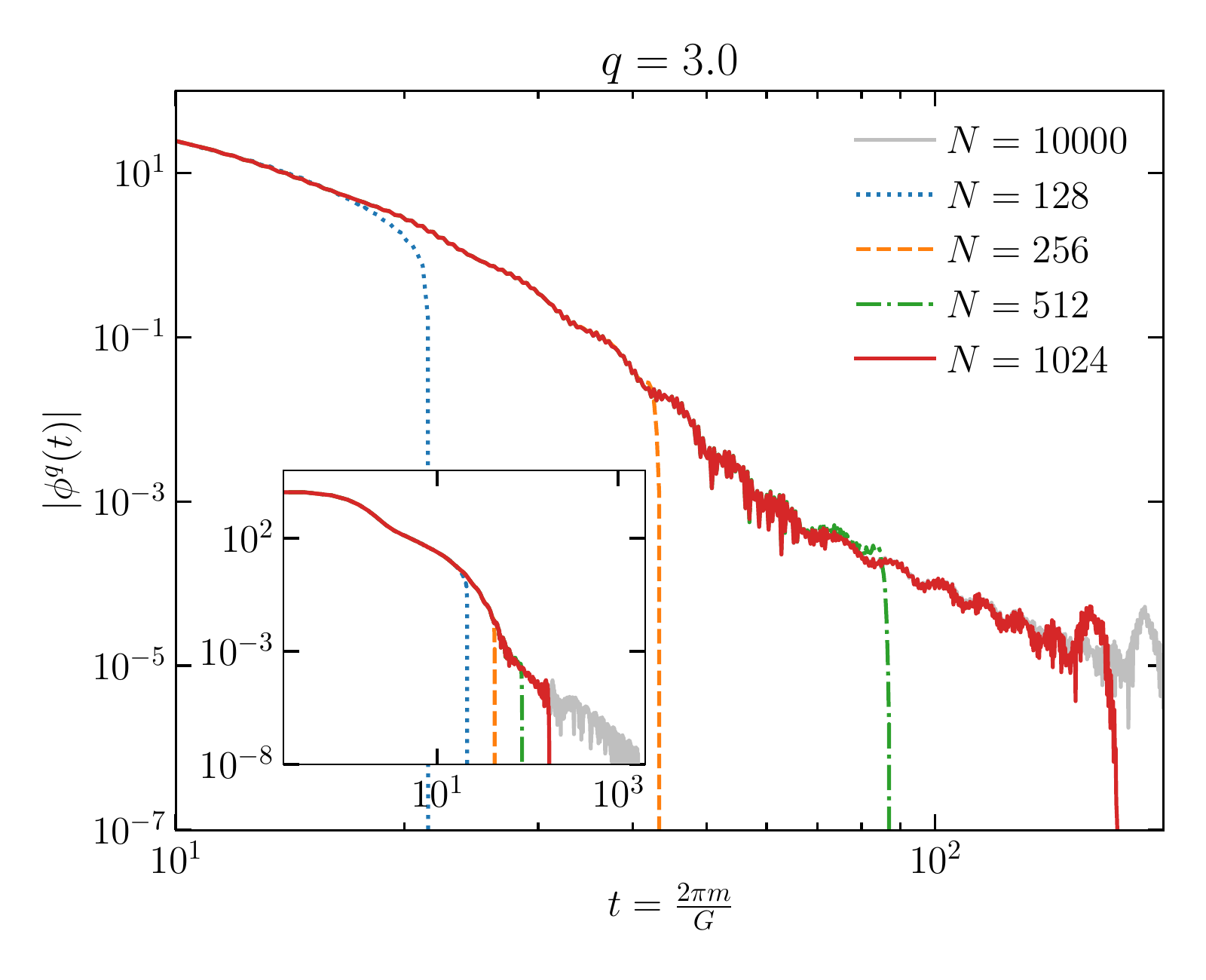}
    \caption{
    The Fourier-space function $|\phi^q(t)|$ for five different numbers of
    samplings $N=128,~256,~512,~1024,~10000$ for
    each biasing paramter $q=0,1,2,3$. The insets show the full spectrum; the
    main panels are zoom-ins. The dropoff of the respective maximum
    $t$ for each $N$ is due to the application of the window function after the
    FFTLog transform. Aliasing is strongest for the $q=0$ case. The features in
    the spectrum are captured by $N\gtrsim512$, though we choose $N=1024$. Here
    we used the interval $k_\min=\SI{e-5}{\h\per\mega\parsec}$ to
    $k_\max=\SI{e3}{\h\per\mega\parsec}$.}
    \label{fig:phiqt_NN}
\end{figure}
\begin{figure*}
    \includegraphics[width=0.49\textwidth]{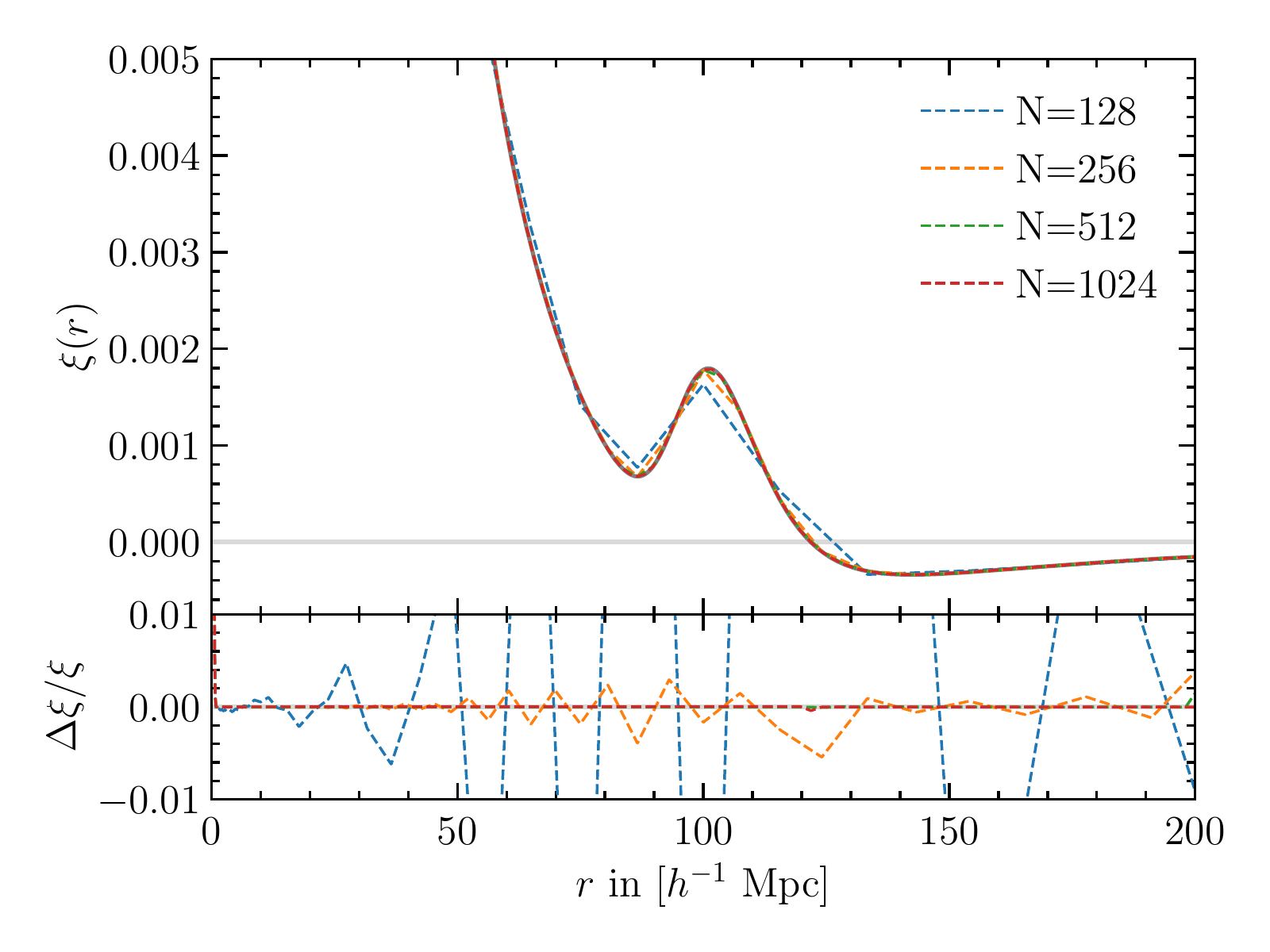}
    \includegraphics[width=0.49\textwidth]{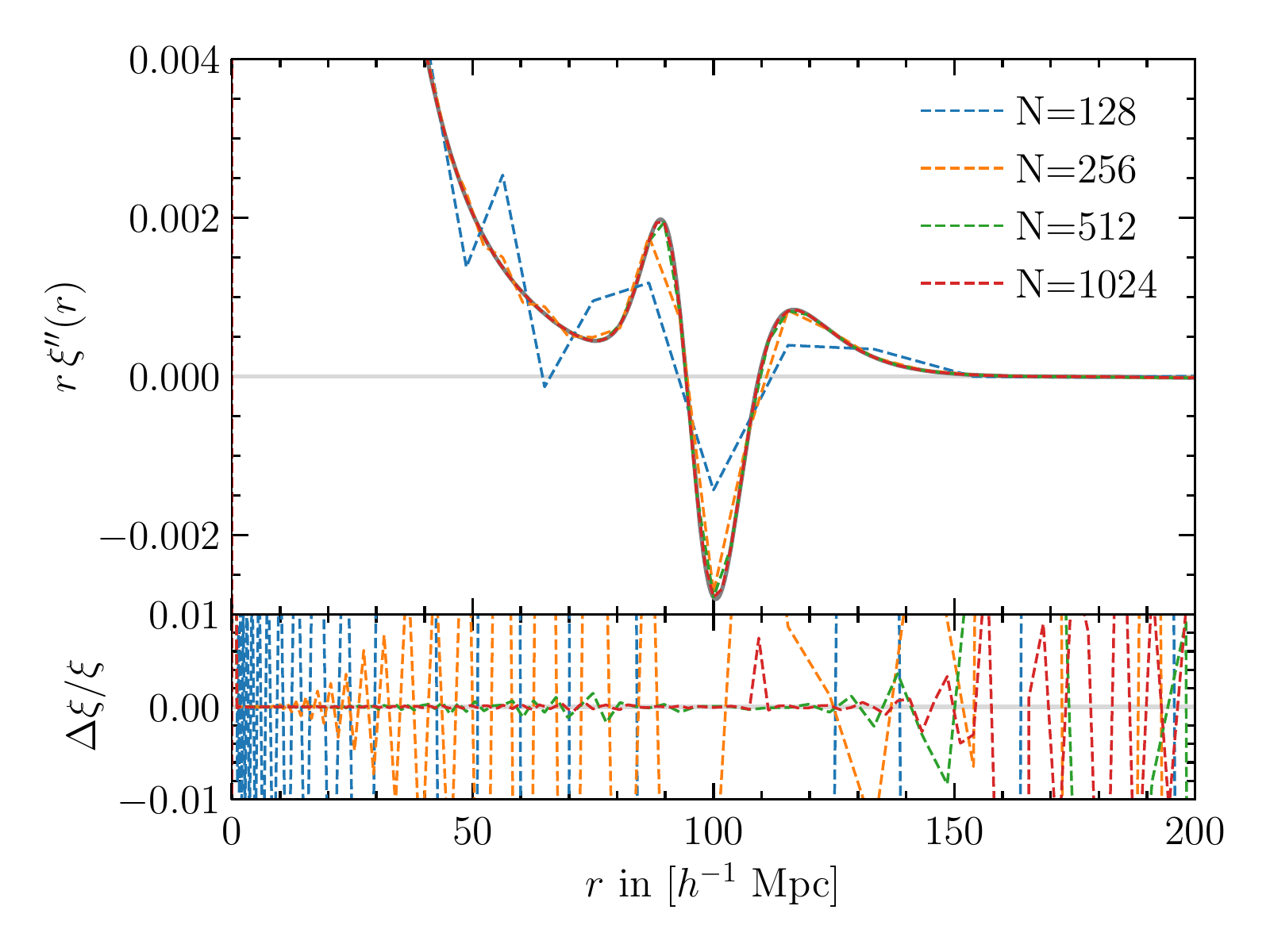}
    \caption{
        The result of the 2-FAST algorithm for $\xi(r)$ (left) and its second
        derivative (right) for several choices of the sampling number $N$. For
        $\xi(r)$ itself, percent-level precision is achieved with $N\geq256$.
        The second derivative is noisier as it depends on smaller structure in
        the power spectrum, and so it achieves percent-level precision with
        $N\geq512$. (The differences at $r\gtrsim\SI{150}{\per\h\mega\parsec}$
        are likely due to pathologies in the \texttt{quadosc} algorithm as shown
        in \reffig{unnatural}.)}
    \label{fig:xiNdependence}
\end{figure*}
\begin{figure}
    \includegraphics[width=0.49\textwidth]{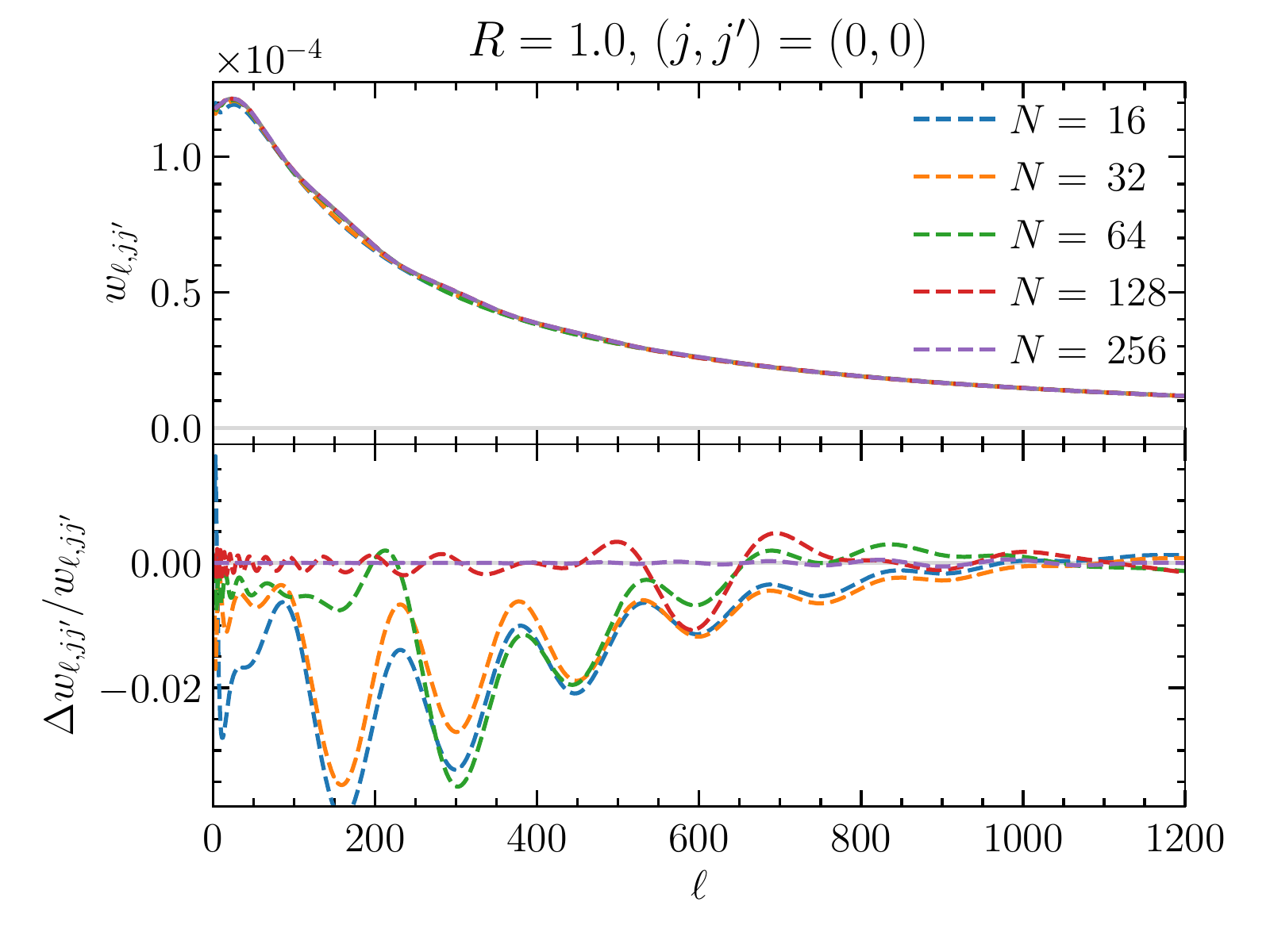}
    \includegraphics[width=0.49\textwidth]{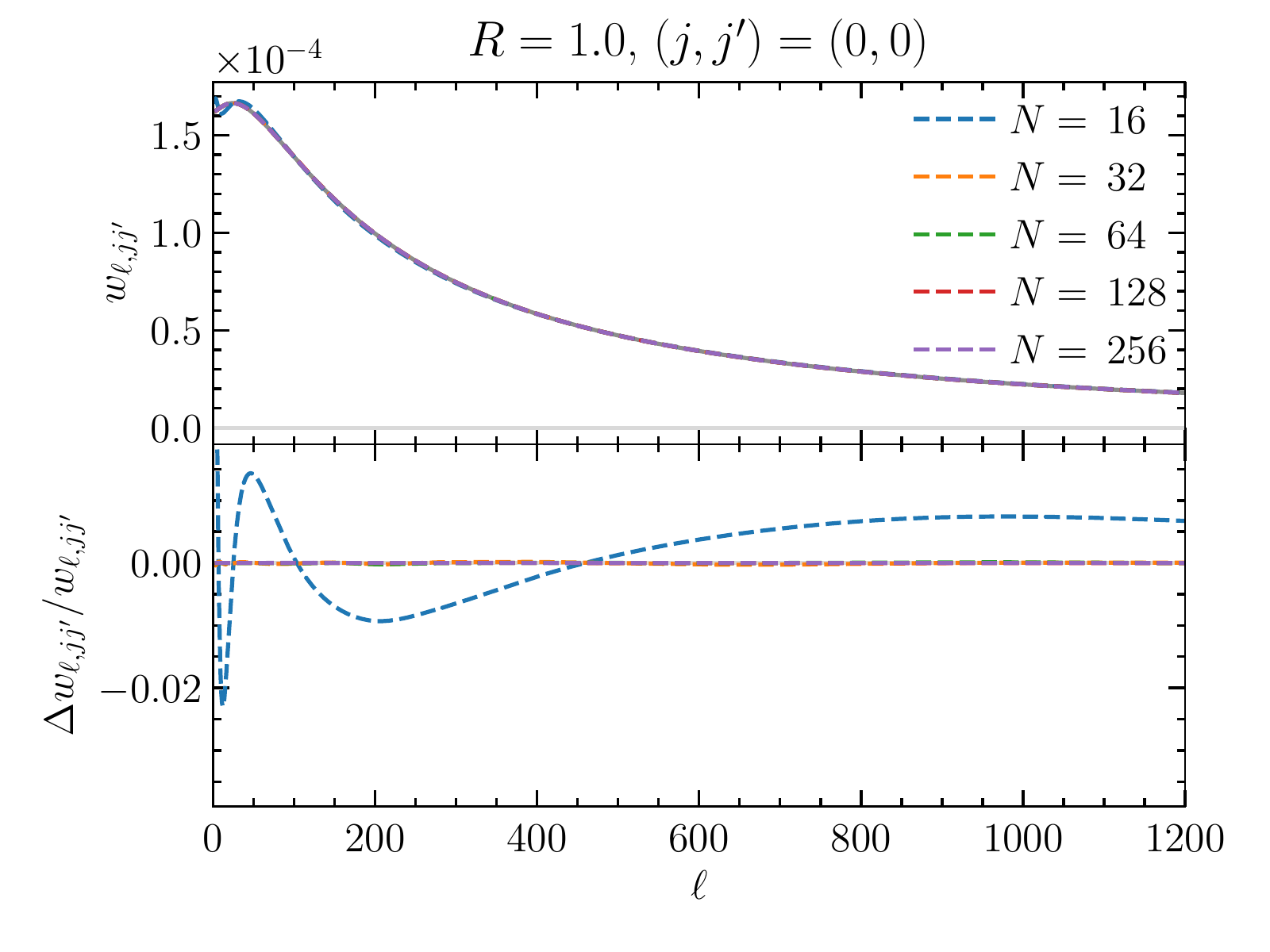}
    \caption{
        Here we show the behavior of the method as it depends on the number of
        sample points $N$ in the interval $k_\min=\SI{e-5}{\h\per\mega\parsec}$
        to $k_\max=\SI{e5}{\h\per\mega\parsec}$ for a power spectrum with BAO (left)
        and a power spectrum without BAO (right) at
        $\chi=\chi'=\SI{2370}{\per\h\mega\parsec}$. In each graph, the top
        panel
        shows the power spectrum, and the bottom panel shows the residual to
        the 2-FAST method with $N=2048$.
        Left: Using a power spectrum
        with BAO. If the number of sampling points is below $N\sim128$, nearly
        all of the BAO features are missed, and show up in the residuals. At
        $N=128$, the BAO feature is missed only at $\ell\gtrsim500$. This is
        due to the logarithmic spacing of the sample points $k_n$ of the
        power spectrum $P(k_n)$, since for larger $\ell$, most of the power
        comes from larger $k$, where the sample points are more sparse.
        Right: For a power spectrum without BAO. In this case even
        $N=32$ sampling points capture much of the structure of the power
        spectrum. }
    \label{fig:wlNdependence}
\end{figure}
\begin{figure*}
    \includegraphics[width=0.49\textwidth]{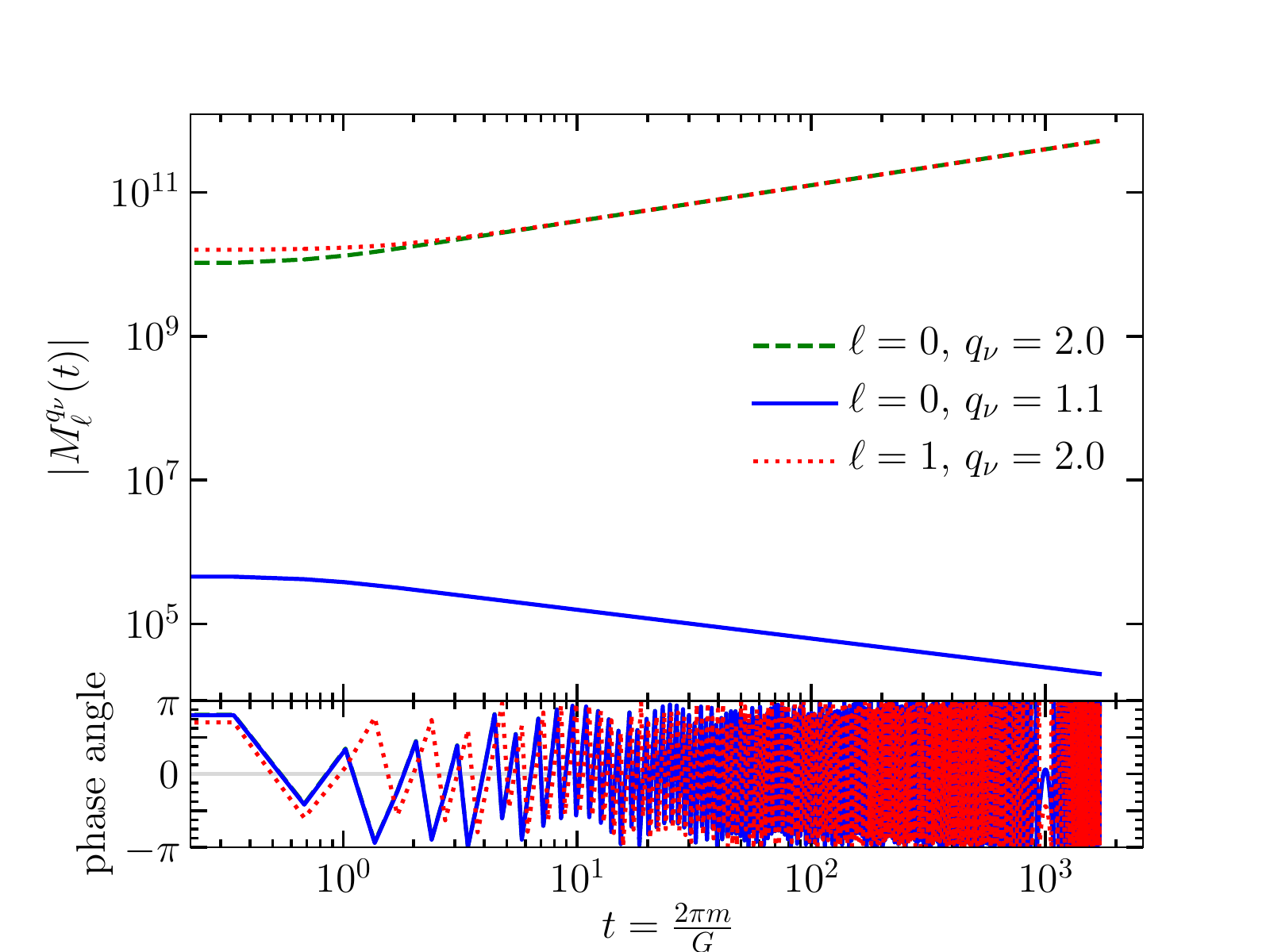}
    \includegraphics[width=0.49\textwidth]{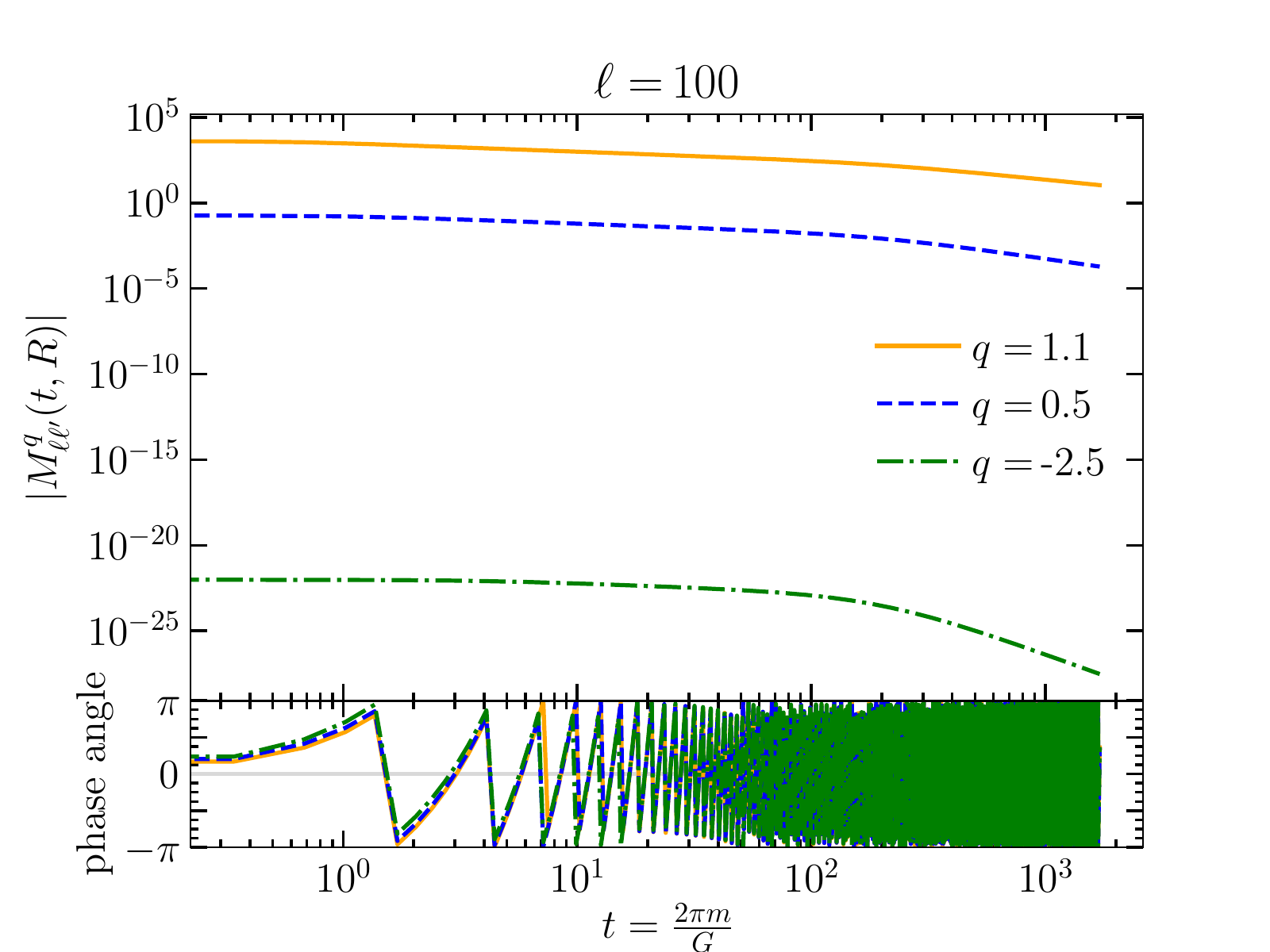}
    \includegraphics[width=0.49\textwidth]{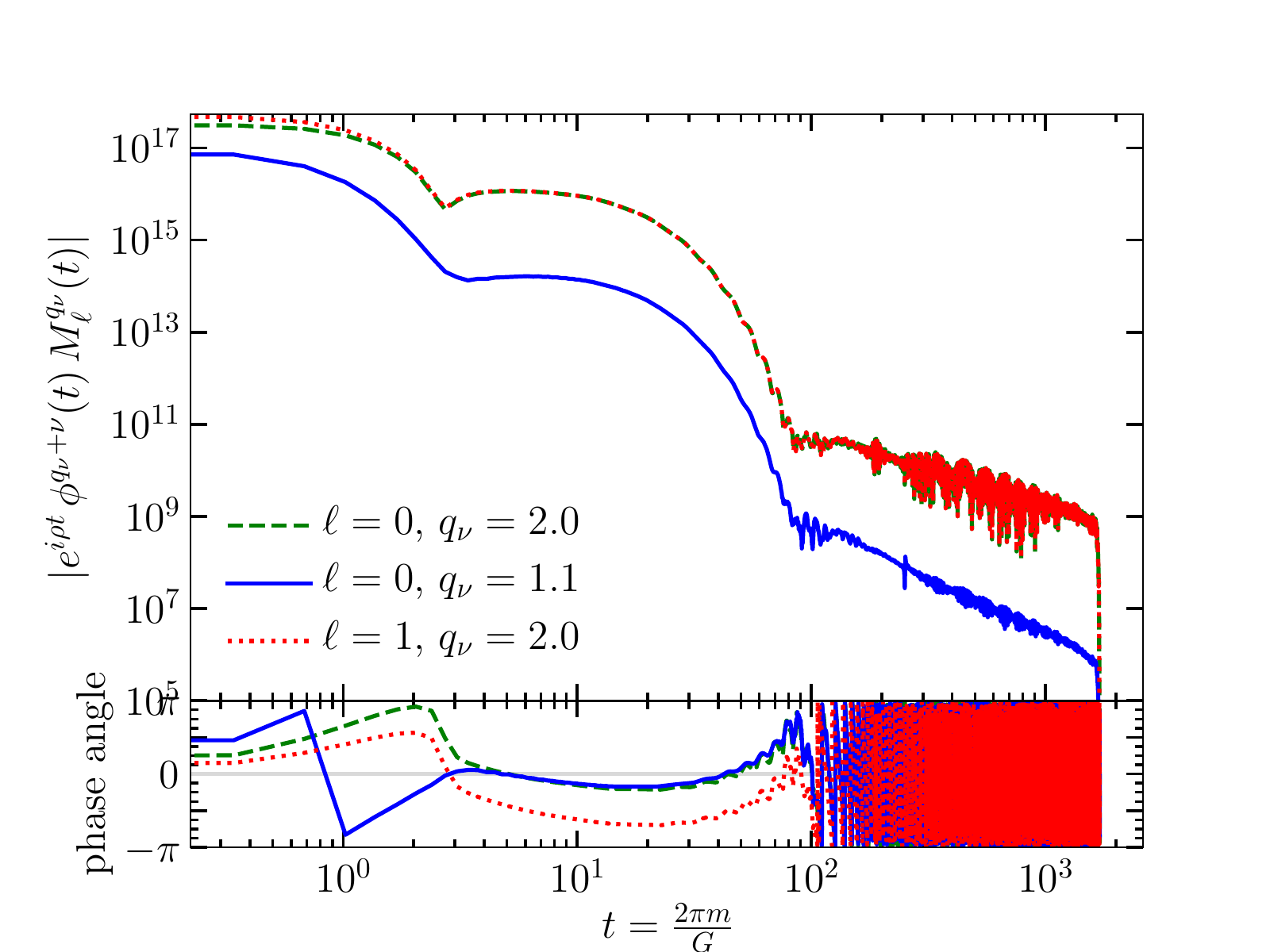}
    \includegraphics[width=0.49\textwidth]{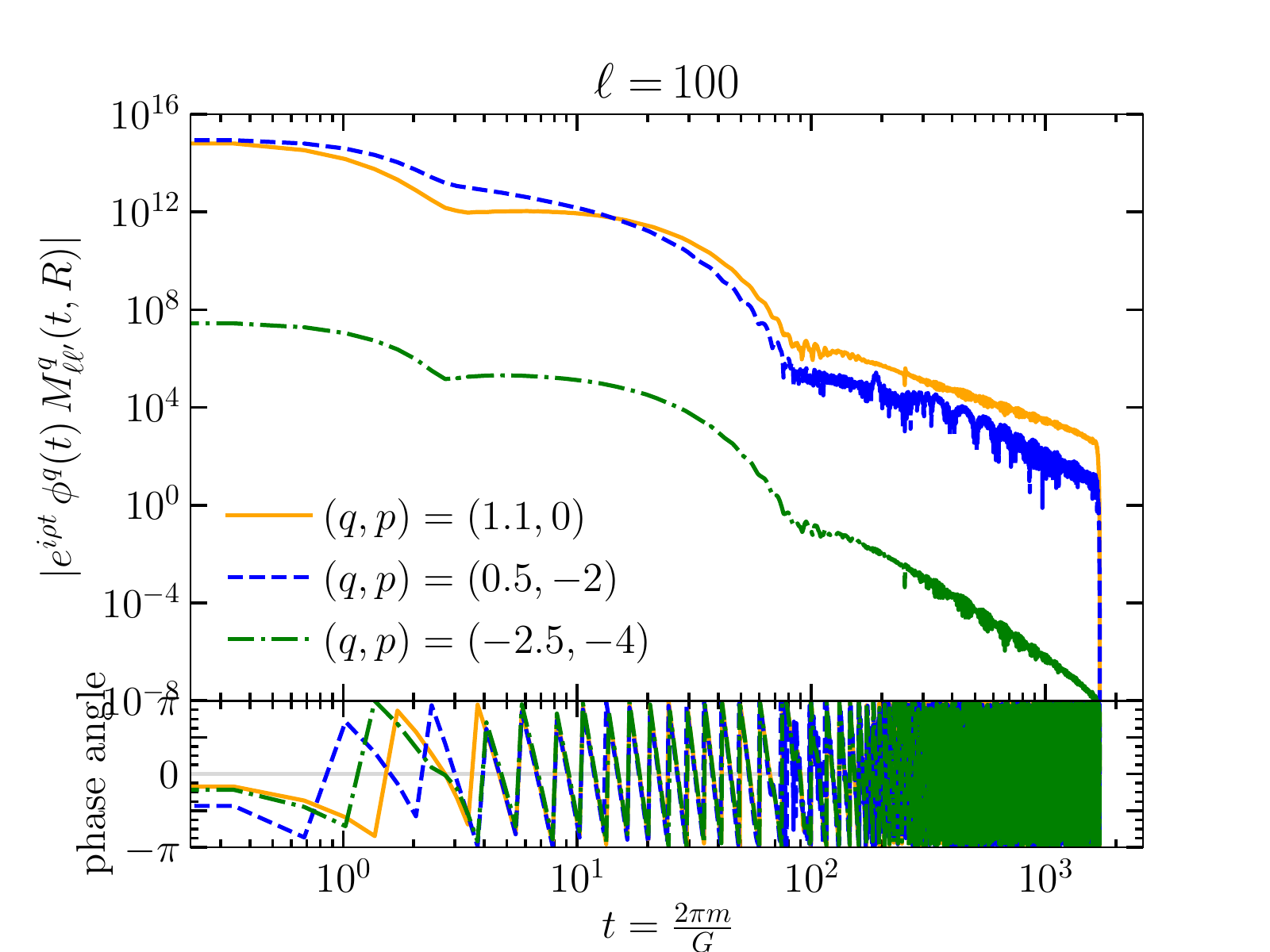}
    \caption{
        Top left: The absolute value (top panel) and the
        complex phase angle (bottom panel) of $M^{q_\nu}_\ell(t)$ as a function
        of the integration parameter $t$. Note that the absolute value is quite
        featureless and merely rises or decays somewhat.
        Top right: The same but for $M^q_{\ell\ell'}(t,R)$ as needed for
        the integral over two spherical Bessel functions.
        Bottom left: Here we show the absolute value and the complex
        phase angle of the full integrand in \refeq{xifftlog} for $\nu=0$.
        Bottom right: The same but for two spherical Bessel functions
        \refeq{well1d}, using the biased power spectrum $k^p\,P(k)$.}
    \label{fig:Mlqnut}
\end{figure*}
\begin{figure*}
    \includegraphics[width=0.49\textwidth]{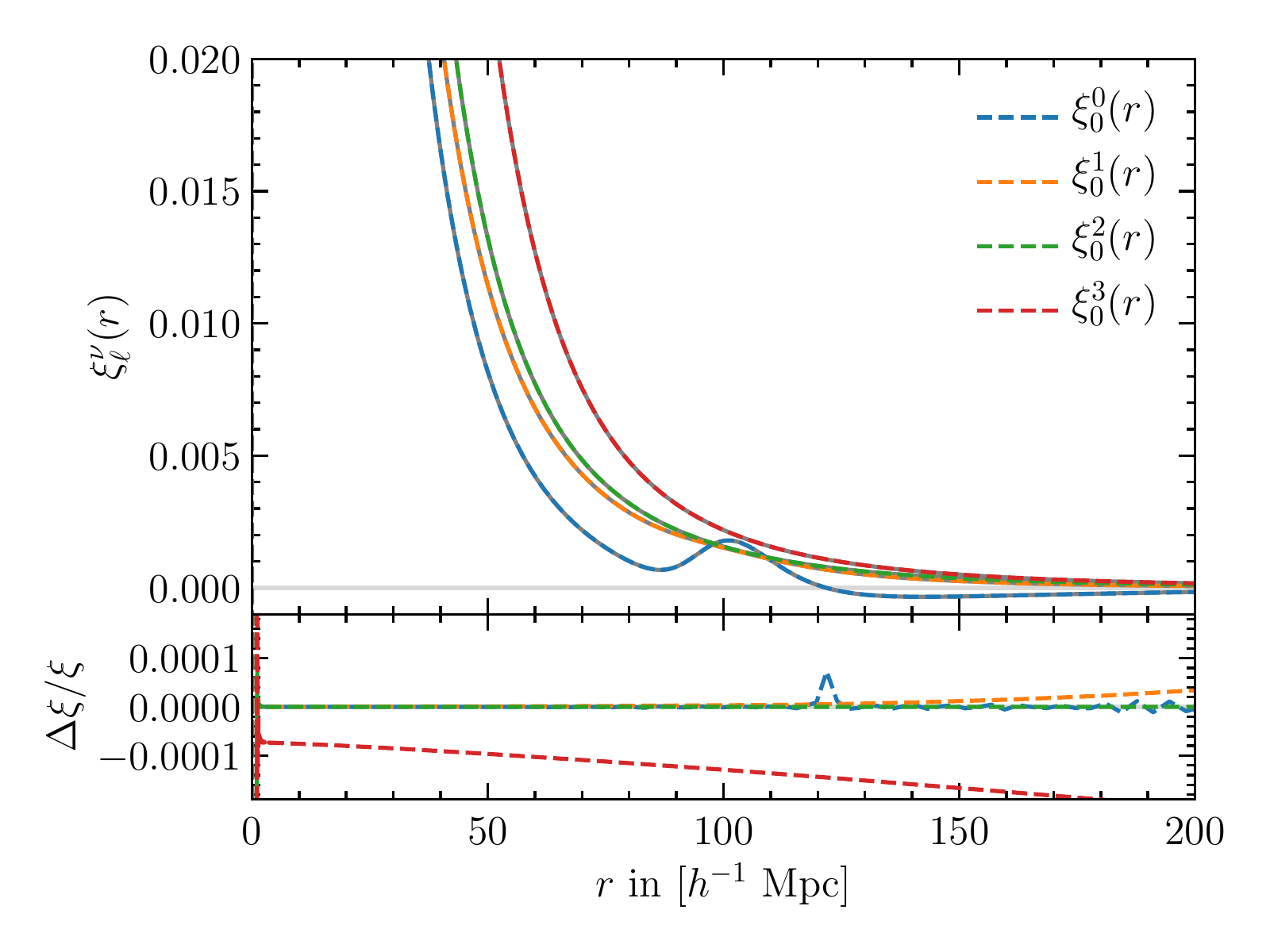}
    \includegraphics[width=0.49\textwidth]{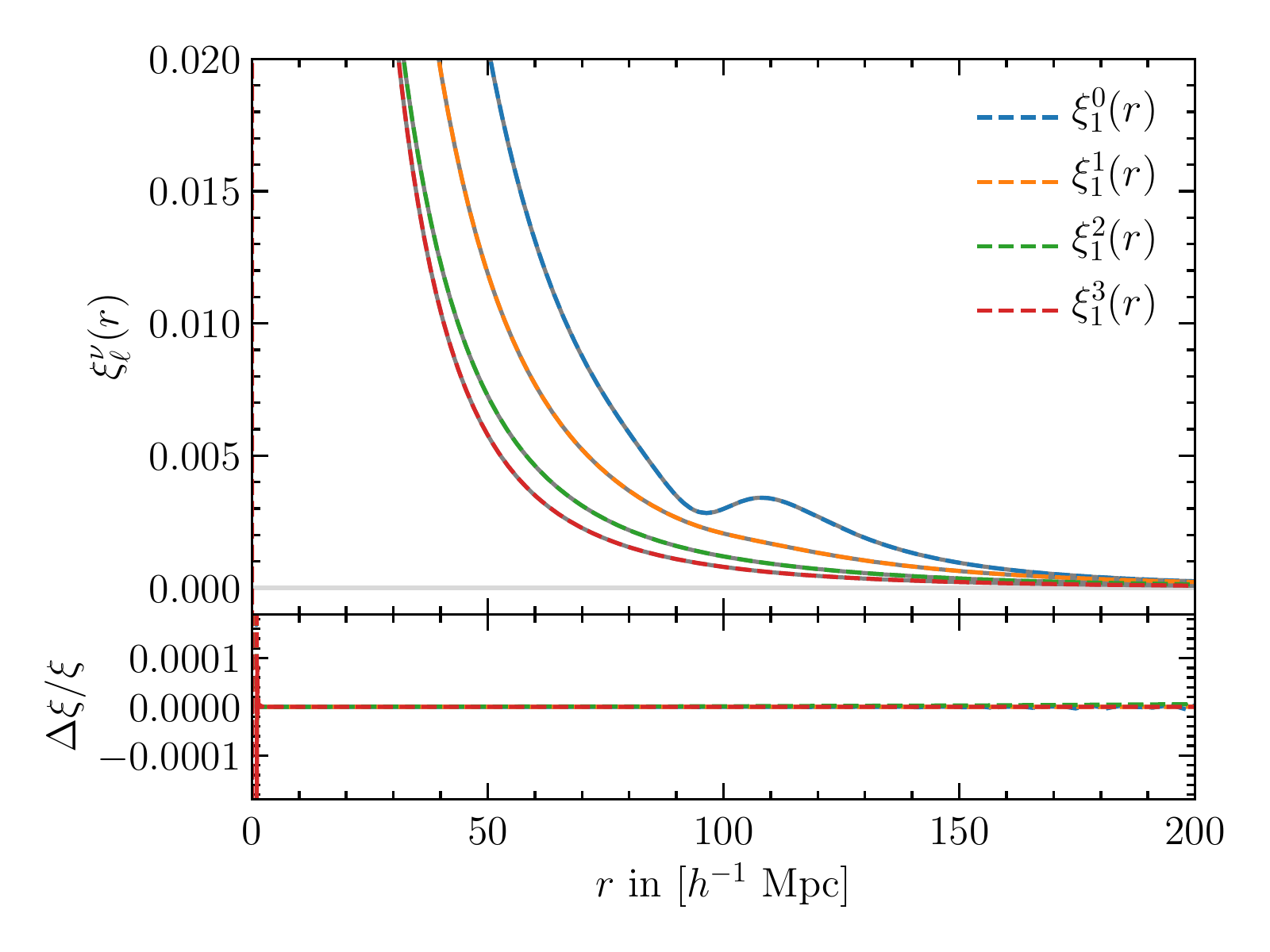}
    \includegraphics[width=0.49\textwidth]{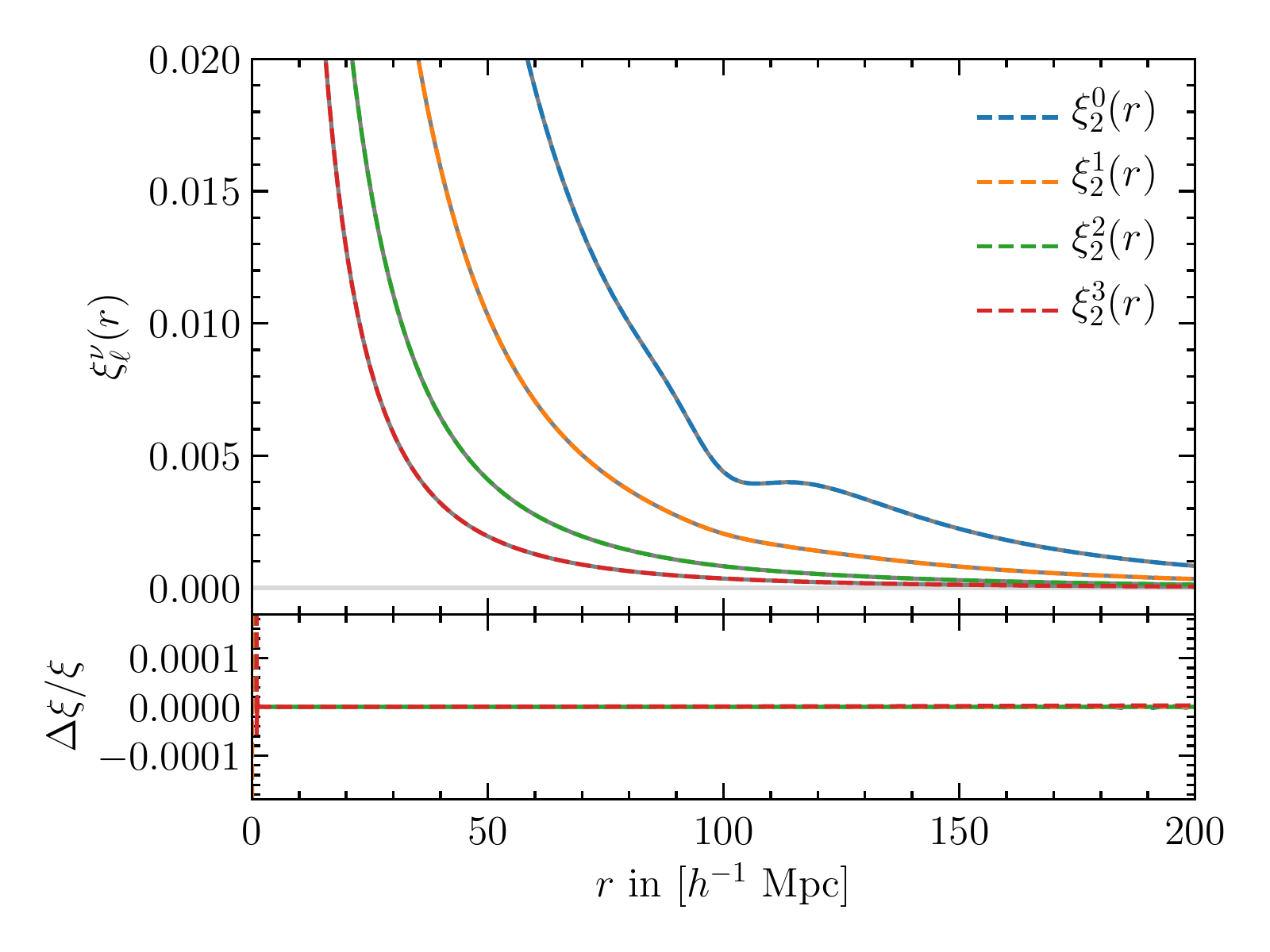}
    \includegraphics[width=0.49\textwidth]{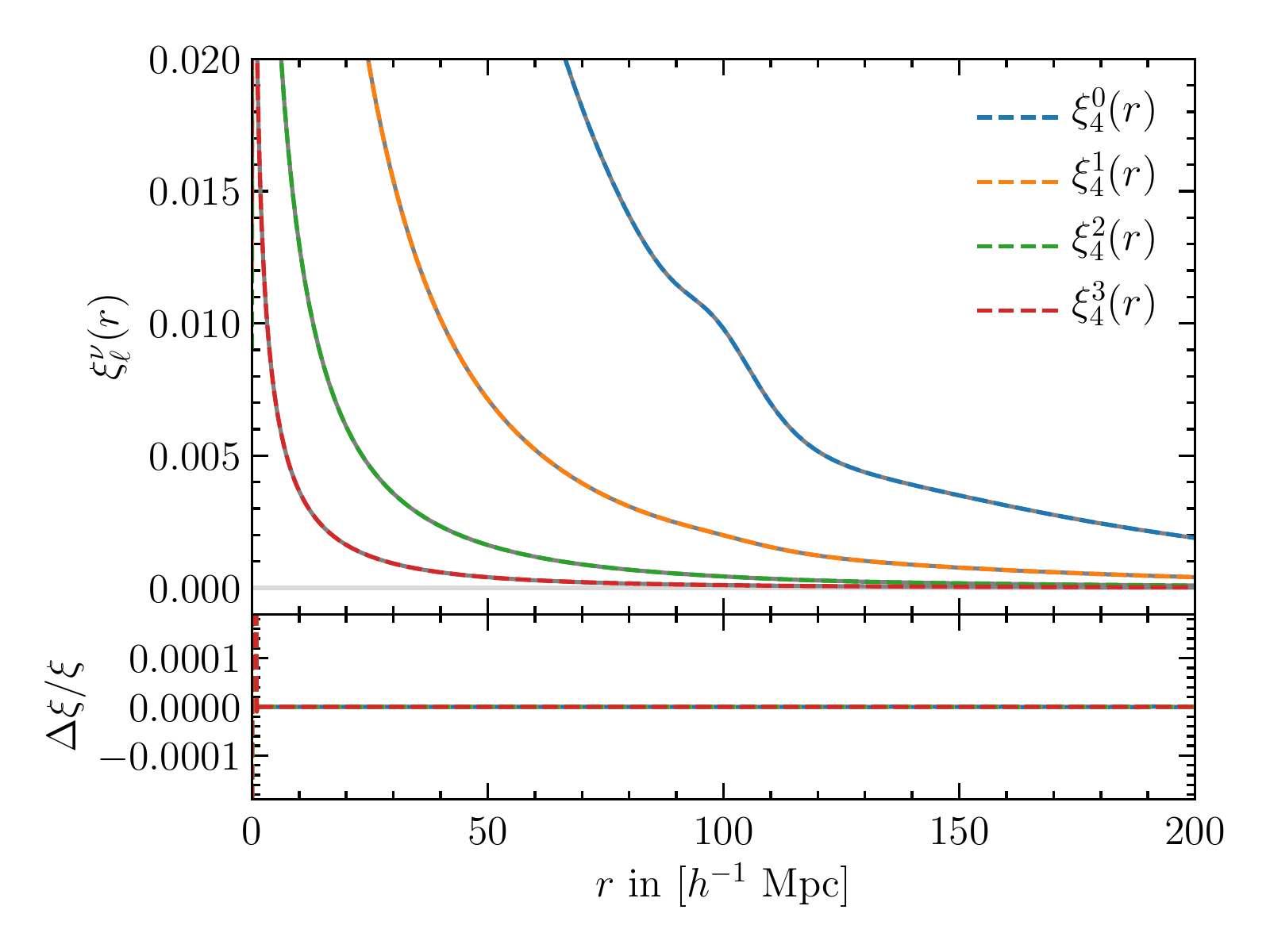}
    \caption{
        The function $\xi_\ell^\nu(r)$ for $\nu=0,1,2,3$ and for
        $\ell=0$ (top left), $\ell=1$ (top right), $\ell=2$ (bottom left), and
        $\ell=4$ (bottom right) as calculated using $q_\nu$ described in
        \refapp{xiq}. We also show the relative residuals to the \texttt{quadosc}
        algorithm.}
    \label{fig:xiellnu}
\end{figure*}

In \reffig{phiqt_NN} we show the dependence of $\phi^q(t)$ on the biasing
parameter $q$ and the number of sampling points $N$ for a given integration
interval $k_\mathrm{min}=\SI{e-5}{\h\per\mega\parsec}$ and
$k_\mathrm{max}=\SI{e3}{\h\per\mega\parsec}$. For all four biasing parameters
($q=0$, $1$, $2$, and $3$ from top left to bottom right), we calculate
$\phi^q(t)$ with four different resolutions ($N=1024$, $N=512$, $N=256$, and
$N=128$) to compare with the benchmark case with $N=10000$ (gray line).
As all the other conditions are the same, any differences from the benchmark
case must be due to aliasing. As expected, the aliasing effects show up
near the Nyquist frequency ($t_\mathrm{Ny}\simeq N\pi/G$, corresponding to the
cutoff) for each case. Among the cases that we study here, the aliasing
effect is largest for the $q=0$ case and smaller for cases with larger $q$
($q=2$ and $q=3$), where the biasing makes the lower-$k$ slope
(wave number smaller than the turnover wave number
$k_\mathrm{to}\sim \SI{0.01}{\h\per\mega\parsec}$) shallower.

As for $N$, we need to choose the grid size $N$ large enough so that the FFTLog
sampling captures the BAO features correctly.
In \reffig{xiNdependence}, we show the dependence of $\xi(r)$ and its second
derivative on $N$. Since the second derivative depends on smaller-scale
structure in the function $P(k)$, it weights larger $t$ in Fourier space
[$\phi^q(t)$] more heavily, and a higher sampling number $N$ is needed to
achieve the same precision as for $\xi(r)$.
We also show the dependence of $w_{\ell\ell}(\chi,\chi)$ on $N$ in
\reffig{wlNdependence} for a power spectrum with BAO (left) and a power spectrum
without BAO (right). As shown from the BAO-less calculation (right panel),
the broad shape is well reproduced even for a very small sampling number
$N\sim32$. The BAO feature, however, is not completely recovered for the
sparse sampling ($N < 512$) cases (see, for example, \reffig{phiqt}).

The top two panels of \reffig{Mlqnut} show the functions $M^{q_\nu}_\ell(t)$
(left) and $M^q_{\ell\ell'}(t,R)$ (right). In each graph the top panel shows
the absolute values, and the bottom panel shows the phase angle of the complex
number. The absolute value rises or falls monotonically depending on the value
of $q$, as shown in the figure. The bottom two panels of \reffig{Mlqnut} show
the full integrands of the integrals in \refeq{xifftlog} (left) and
\refeq{well1d} (right). Note that for the right figure (for two spherical
Bessel functions), we used the biased power spectrum $k^p\,P(k)$ as introduced
in the main text. Note that in all cases the full integrand decreases rapidly
with $t$ for the choices of $q$ shown.

Finally, we checked the prescription that we have adopted in \refsec{xi}:
As a default, we choose $q_\nu=1.9-\nu$ as long as it is within the allowed
range given in \refeq{xi:q}. If the mean results in a $q_\nu$
outside the range, then we use instead
\ba
q_\nu &= \frac13(q_{\nu,\min} + 2 q_{\nu,\max})\,,
\label{eq:qsub}
\ea
where $q_{\nu,\min}$ and $q_{\nu,\max}$ are the boundaries given in
\refeq{xi:q}. In \reffig{xiellnu} we verify that the resulting $\xi_\ell^\nu(r)$
from this choice of $q_\nu$ matches well with the benchmark results from
the \texttt{quadosc} calculation.

\subsection{Aliasing effect in backward FFTLog}
\label{app:discrete_convolution}
\begin{figure}
    \includegraphics[width=0.49\textwidth]{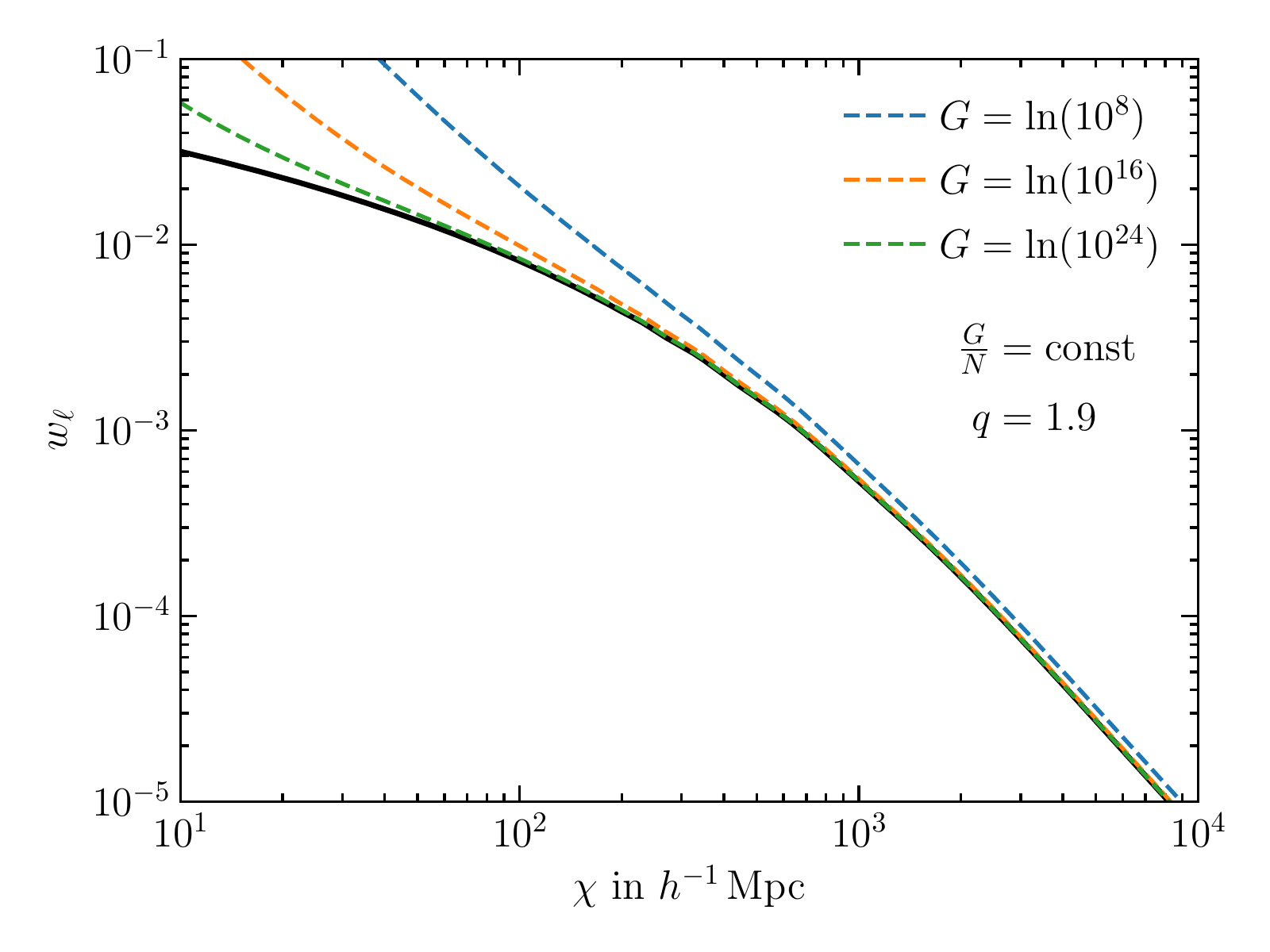}
    \caption{Here we exemplify the dependence of the result of the 2-FAST
    algorithm for $w_{\ell\ell}(\chi,\chi)$ on the integration interval $G$ for
    $\ell=42$ and $q=1.9$. The ratio $G/N$ is kept constant so that the same
    sampling points on the power spectrum are used in all cases. The black line
    is the result from the Lucas algorithm. While the error goes down for
    larger $G$, even extremely large ranges $G$ lead to significant error for
    $\chi<\SI{100}{\per\h\mega\parsec}$. The error here is dominated by the
    $s>0$ terms. To get correct results it is more efficient to adjust $q$. }
    \label{fig:wlerr-G}
\end{figure}
\begin{figure}
    \includegraphics[width=0.49\textwidth]{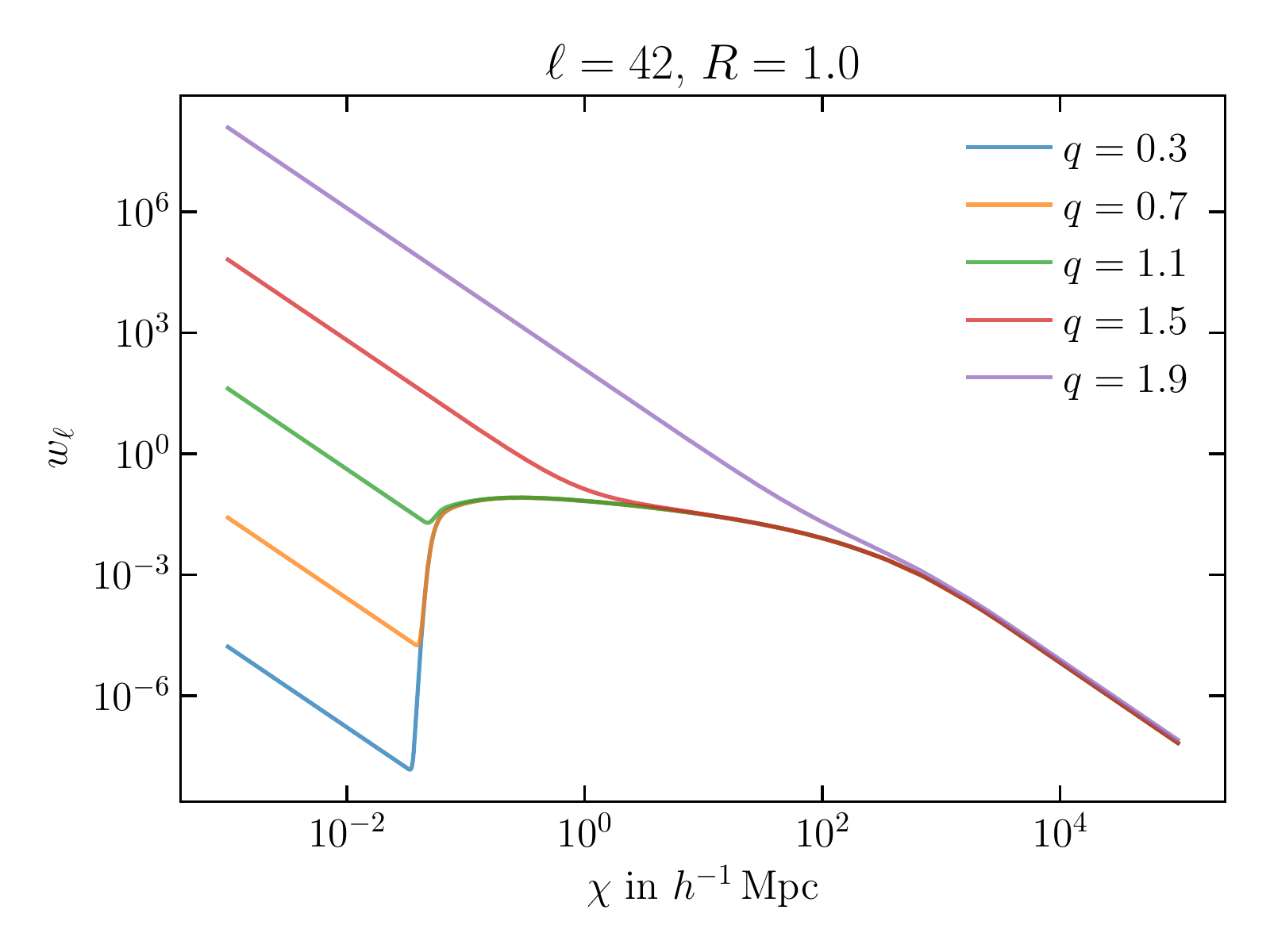}
    \includegraphics[width=0.49\textwidth]{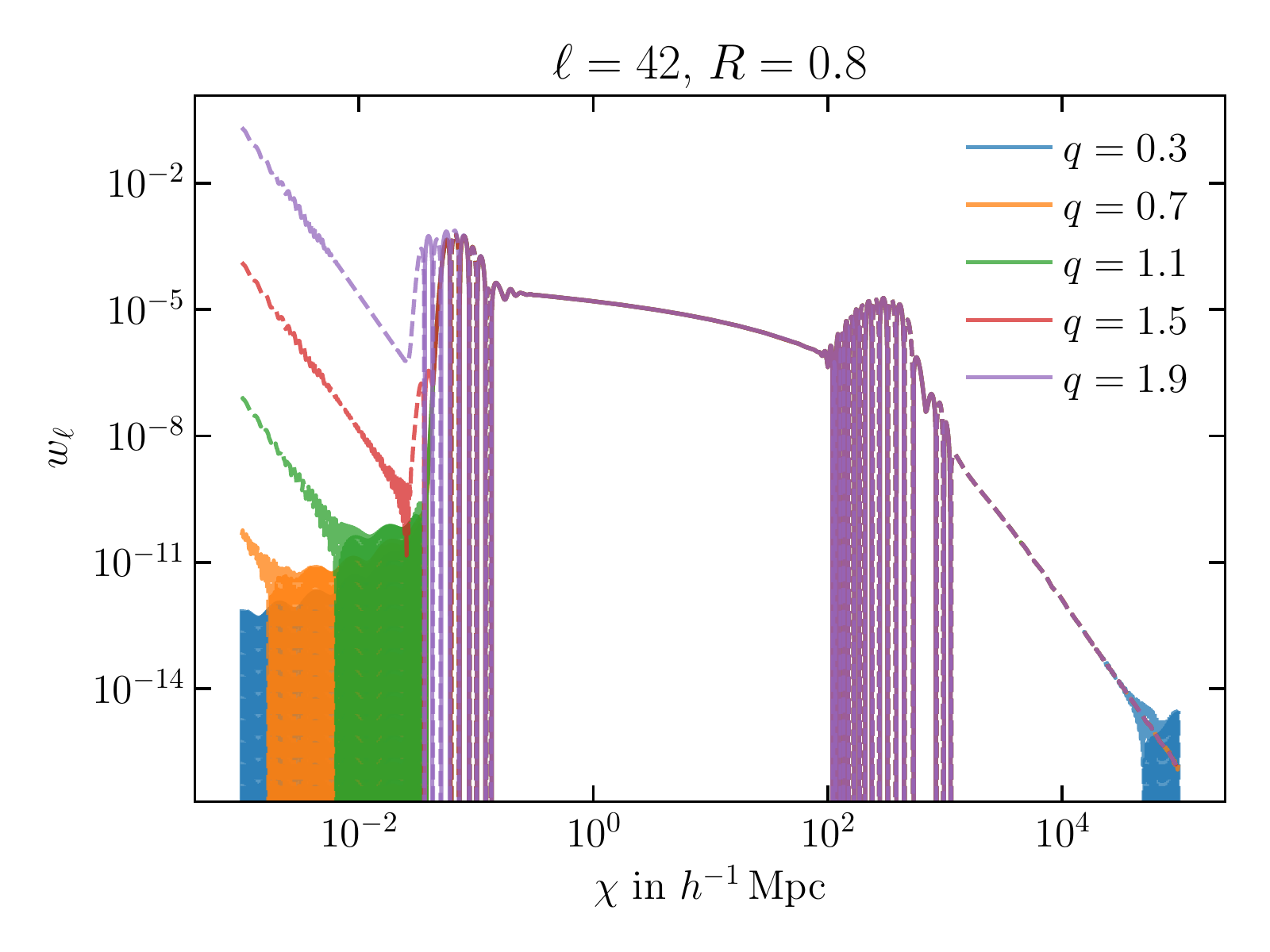}
    \caption{The dependence of the 2-FAST result for $w_{\ell\ell}(\chi,R\chi)$
		 		on the choice of $q$ for $\ell=42$ and $R=1$ (left) and $0.8$ (right).
				For $R=1$, choosing $q\lesssim1.5$ is required to get an accurate
				result for the cosmologically relevant range
				$\chi\gtrsim\SI{100}{\per\h\mega\parsec}$. In contrast, for $R=0.8$,
				the aliasing effect in relevant scales is small for all values of $q$ shown here.
    }
    \label{fig:plotmania-ell42}
\end{figure}
\begin{figure}
    \includegraphics[width=0.49\textwidth]{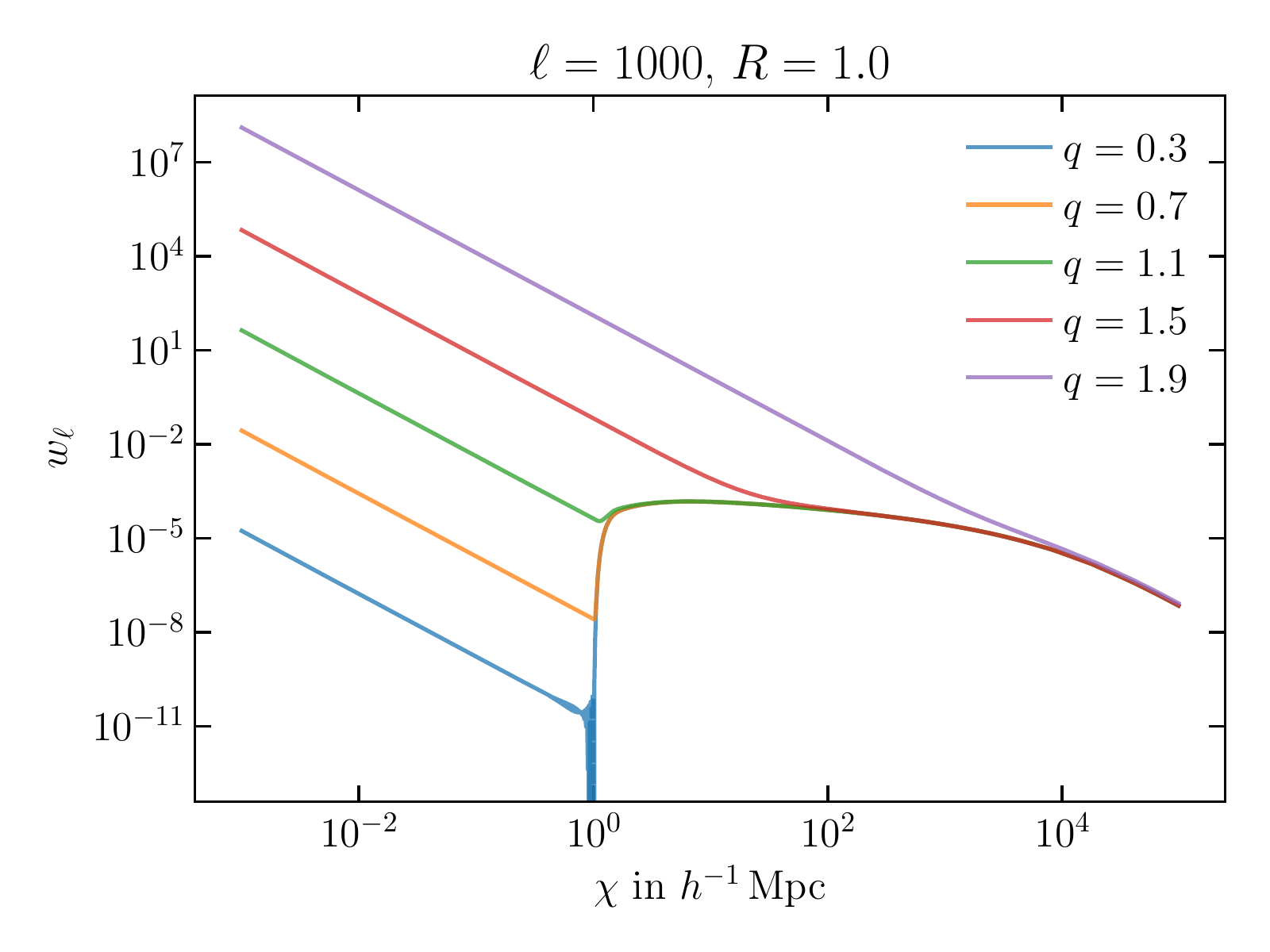}
    \includegraphics[width=0.49\textwidth]{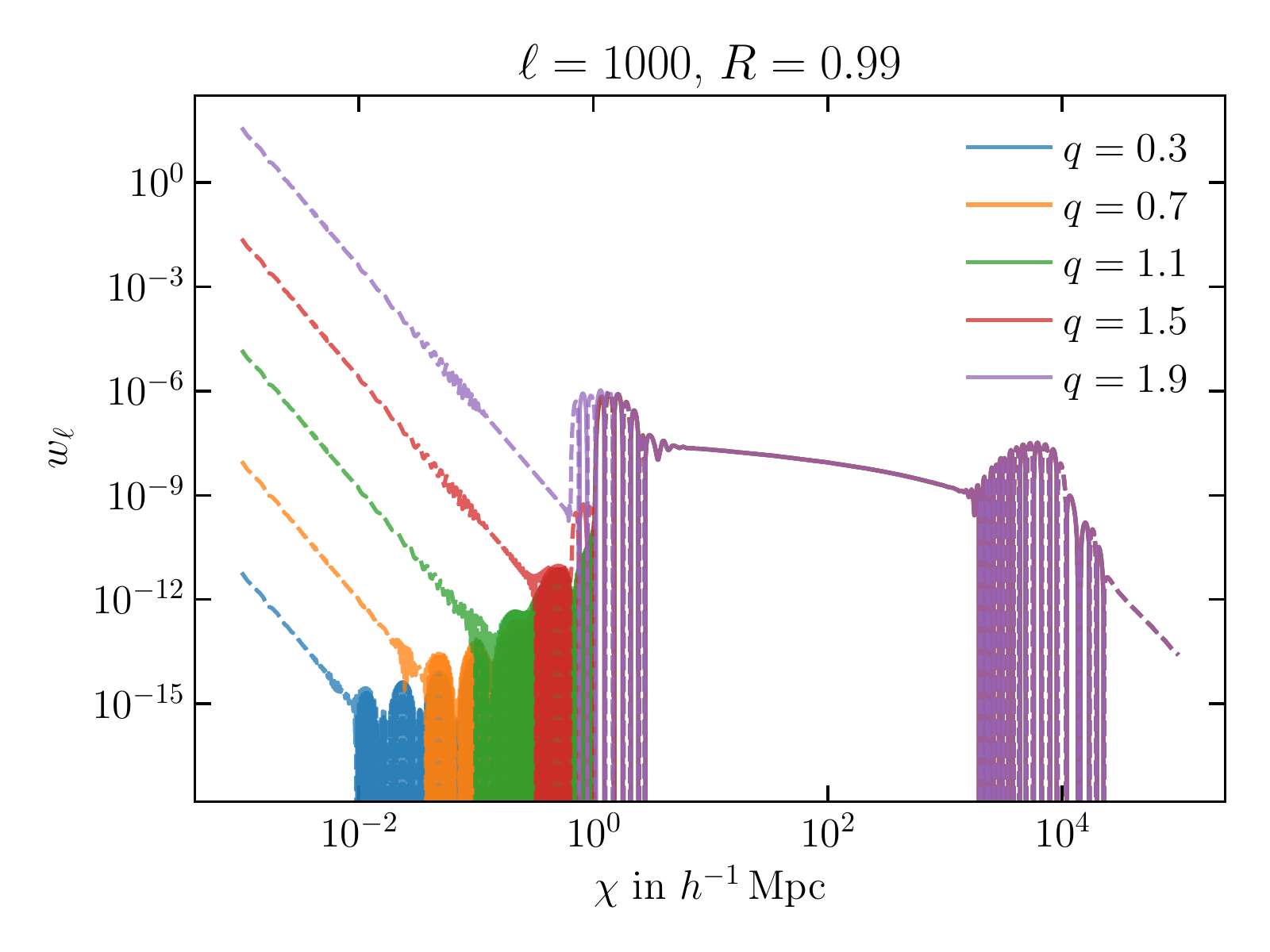}
    \caption{ Same as \reffig{plotmania-ell42},
    except that $\ell=1000$ and $R=1,~0.99$. Once again, for $R=1$, $q=1.1$
    results in a small aliasing effect. For $R=0.99$, any value $q\lesssim1.5$ gives a small aliasing effect over the cosmologically relevant scale.}
    \label{fig:plotmania-ell1000}
\end{figure}
When implementing the 2-FAST algorithm, we calculate the convolution between
the power spectrum $P(k)$ and one or two spherical Bessel functions by the
backward discrete FFTLog transformation of $\phi^q(t) M_{\ell}^{q_\nu}(t)$ or
$\phi^q(t)M_{\ell\ell'}^q(t,R)$, which are, respectively, the analytically
calculated Fourier transformations of one or two spherical Bessel functions. In
this section, we study yet another aliasing effect associated with the discrete
sampling in $t$ space and justify our choice of the biasing parameter $q$.

To avoid clutter, let us consider calculating the following convolution integral:
\be
f(x)
= \int_{-\infty}^{\infty} \dd y\, g(y) \, h(x+y)
= \int_{-\infty}^{\infty} \frac{\dd q}{2\pi}\, \tilde{g}(-q)\,\tilde{h}(q)\,e^{iqx},
\label{eq:continuous_convolution}
\ee
where $\tilde{g}(q)$ and $\tilde{h}(q)$ are the functions in Fourier space.
In order to mock the 2-FAST implementation, we discretize the later integration
as well as the calculation of $\tilde{g}$ by using $N$ sampling points within
the $x$-range of $G$. First, the Fourier transformation of $g(x)$ is
\be
\tilde{g}(q_m)
= \Delta_x \sum_{k=0}^{N-1} g(x_k) \, e^{-ix_kq_m}
= \frac{G}{N} \sum_{k=0}^{N-1} g(x_k) \, e^{-i2\pi mx_k/G},
\ee
where $q_m=m\Delta_q$ and $x_k=k\Delta_x$ with the intervals
$\Delta_x\equiv G/N$ and $\Delta_q\equiv 2\pi/G$ in $x$ space and $q$ space.
On the other hand, the function $\tilde{h}(q)$ is calculated from the Fourier
transformation:
\be
\tilde{h}(q)=\int_{-\infty}^{\infty}\dd x\, h(x) \, e^{-iqx}\,.
\ee
Combining the two, we find that the implementation actually calculates
\ba
f(x_n)
&= \frac{1}{G}\sum_{m=0}^{N-1} \, e^{iq_mx_n} \tilde g(-q_m) \, \tilde h(q_m)
=  \sum_{k=0}^{N-1} g(x_k)  \int_{-\infty}^{\infty}\dd x\, h(x)
\left[\frac{1}{N}\sum_{m=0}^{N-1}
e^{i 2\pi m \;\! \left(x_n+x_k-x\right) / G}
\right]\,.
\label{eq:discrete_convolution_exact}
\ea
Here, the function in square brackets,
\be
W(x) \equiv \frac{1}{N}\sum_{m=0}^{N-1}e^{i2\pi mx/G} =
e^{i\pi \frac{N-1}{G}x}\,\frac{\sin(N\pi x/G)}{N\sin(\pi x/G)},
\ee
can be approximated as
\be
\label{eq:Wx-diracdelta}
W(x) \simeq \frac{G}{N} \sum_{s=-\infty}^{\infty} \delta^D(x+sG)\,,
\ee
which is exact in the $N\to\infty$ limit.
With this approximation, we find that the convolution integral
\refeq{discrete_convolution_exact} becomes
\ba
f(x_n)
&\simeq
\Delta_x
\sum_{k=0}^{N-1}
g(x_k)
\sum_{s=-\infty}^\infty
h(x_n+x_k+sG)\,,
\label{eq:discrete_convolution}
\ea
where $\Delta_x=G/N$, $x_n=n\Delta_x$, and $x_k=k\Delta_x$.
Comparing \refeq{continuous_convolution} with \refeq{discrete_convolution},
we see that the desired convolution corresponds to the $s=0$ case, and
all the other $s$ values cause aliasing. Note that the $s\neq0$ peaks
are separated by
$2\pi/\Delta_q = G$, that is, the total duration of discrete
sampling in $x$ space. In order to suppress this effect, we need to employ a
function $h(x)$ that decays fast so that the aliasing contribution is far
smaller than the desired result at $s=0$.

For the case at hand, $g(\kappa)= e^{(3-q)\kappa}P(k_0e^\kappa)$ is the biased
power spectrum, and $h(\kappa) = e^{q_\nu\kappa}j_\ell(\alpha e^{\kappa})$,
\refeq{xilog}, and $h(\kappa)=e^{q\kappa}j_\ell(\alpha
e^{\kappa})j_{\ell'}(R\alpha e^{\kappa})$, \refeq{welllog2}, for calculating,
respectively, $\xi_\ell^\nu(r)$ and $w_{\ell\ell'}(\chi,\chi')$. The outcomes
of the 2-FAST implementation are then
\ba
\xi_\ell^\nu(r)
&=
\frac{k_0^3 e^{-(q_\nu+\nu)\rho}}{2\pi^2(k_0r_0)^\nu}\,
\Delta_\kappa
\sum_{k} e^{(3-q_\nu-\nu)\kappa}P(k)
\sum_{s=-\infty}^\infty
e^{q_\nu (\kappa + \rho + s G)}\,j_\ell(kr e^{sG})
\\
&=
\frac{\Delta_\kappa}{2\pi^2}
\sum_{k} \frac{k^3P(k)}{(kr)^\nu}
\sum_{s=-\infty}^\infty
e^{q_\nu s G}\,j_\ell(kr e^{sG})\,,
\ea
and
\ba
w_{\ell\ell'}(\chi,\chi')
&=
\frac{2}{\pi}\,
\Delta_\kappa
\sum_{k} k^3P(k)
\sum_{s=-\infty}^\infty
e^{qsG}\,j_\ell(k\chi e^{sG})\,j_{\ell'}(k\chi'e^{sG})\,.
\ea
If the $s=0$ term in the sum dominates over all other terms, then we recover
the desired convolution integral. How big is the aliasing effect?
We estimate the aliasing effect by using a simple approximation of replacing
the spherical Bessel functions by their envelopes.
Using the asymptotic behavior of the spherical Bessel functions,
\ba
\label{eq:jl-asymptote-low}
\lim_{x\rightarrow0} j_\ell(x)
&= \frac{x^\ell}{(2\ell+1)!!}
= \frac{\sqrt{\pi}\,x^\ell}{2^{\ell+1}\,\Gamma(\ell+\frac32)}
\\
\lim_{x\rightarrow\infty} j_\ell(x) &= x^{-1}\sin\big(x-\tfrac{\pi}{2}\ell\big)\,,
\label{eq:jl-asymptote-high}
\ea
we estimate the aliasing effect (denoting that $E$ stands for the error)
as
\ba
\label{eq:err-xi}
E[\xi_\ell^\nu(r)]
&\simeq
\frac{\Delta_\kappa}{2\pi^2}
\sum_{k} \frac{k^3P(k)}{(kr)^\nu}\,
\bigg[
\sum_{s<0}
\frac{(kr)^\ell e^{(q_\nu+\ell)sG}}{(2\ell+1)!!}
+
\sum_{s>0}
\frac{e^{(q_\nu-1)sG}}{kr}
\bigg]\,,
\\
\label{eq:err-harmonic}
E[w_{\ell\ell'}(\chi,\chi')]
&\simeq
\frac{2}{\pi}\,
\Delta_\kappa
\sum_k
k^3P(k)\,
\bigg[
\sum_{s<0}
\frac{(k\chi)^\ell (k\chi')^{\ell'} e^{(q+\ell+\ell')sG}}{(2\ell+1)!!(2\ell'+1)!!}
+
\sum_{s>0}
\frac{e^{(q-2)sG}}{k^2\chi\chi'}
\bigg]\,.
\ea
That is, we expect that the aliasing effects from $s<0$ peaks affect
larger separations (large $r$ or $\chi$) with $r^\ell$ or
$\chi^\ell{\chi'}^{\ell'}$ dependence, and aliasing effects from $s>0$
peaks affect smaller separations with $1/r$ or $1/(\chi\chi')$
dependence.

Although the $r$ and $\chi,\chi'$ dependences are the same as what we
estimated, it turns out, however, that the actual aliasing effect is far
smaller than the estimation above, which is based on the envelope of the
spherical Bessel functions and the approximation \refeq{Wx-diracdelta}.
By examining our implementation of $\xi_\ell^\nu(r)$ and
$w_{\ell\ell'}(\chi,\chi')$, we find that the spherical Bessel functions
oscillate many times over the width of a peak in $W(x)$ so that the
aliasing for these cases has left only a small residual as an error.
One exception is when calculating $w_{\ell\ell'}(\chi,\chi)$ (or $R=1$),
where such a cancellation does not happen because the aliasing contribution
from spherical Bessel functions is positive definite. We, therefore, choose
the biasing parameter $q$ that diminishes the aliasing effect for the $R=1$ case.

In \reffig{wlerr-G} we show the $w_{\ell=\ell'=42}(\chi,\chi)$ with three
different values of $G$, the size of the integration interval. For the same
biasing parameter $q=1.9$. With width $G=\ln(10^8)$ that we adopted when
calculating $\xi_\ell^\nu(r)$, the aliasing effect is clearly visible on all
$\chi$ values that we show here. The aliasing effect does get milder as we
increase the interval $G$. In order to get a reliable result for
$\chi>\SI{e2}{\per\h\mega\parsec}$, however, we need to choose the interval
over 24 orders of magnitude in $k$ space.

We find that biasing the convolved integrand provides a more efficient way of
reducing the aliasing effect. That is, when adopting a smaller $q$ value,
the integrand decays fast enough to suppress the aliasing effect.
For example, we show the $w_{\ell=\ell'=42}(\chi,\chi)$ for five values of $q$
between $0.3$ and $1.9$ in the left panel of \reffig{plotmania-ell42}.
It turns out that one must take $q\lesssim1.5$ in order to suppress the aliasing
effect on cosmologically relevant scales $\chi\gtrsim\SI{100}{\per\h\mega\parsec}$.
On the other hand, the right panel of \reffig{plotmania-ell42} shows that all
values of $q$ yield an accurate calculation of $w_{\ell=\ell'=42}(\chi,\chi)$ on
all cosmologically relevant scales $\chi\gtrsim\SI{100}{\per\h\mega\parsec}$.
Finally, \reffig{plotmania-ell1000} shows the same for $\ell=1000$, and
$R=1,0.99$. Here, too, $q\lesssim1.1$ results in a small error at
$\chi>\SI{100}{\per\h\mega\parsec}$.

\section{Transformation matrix from 2-FAST and trapezoidal method}
\label{app:transmat}
\begin{figure*}
	\includegraphics[width=0.49\textwidth]{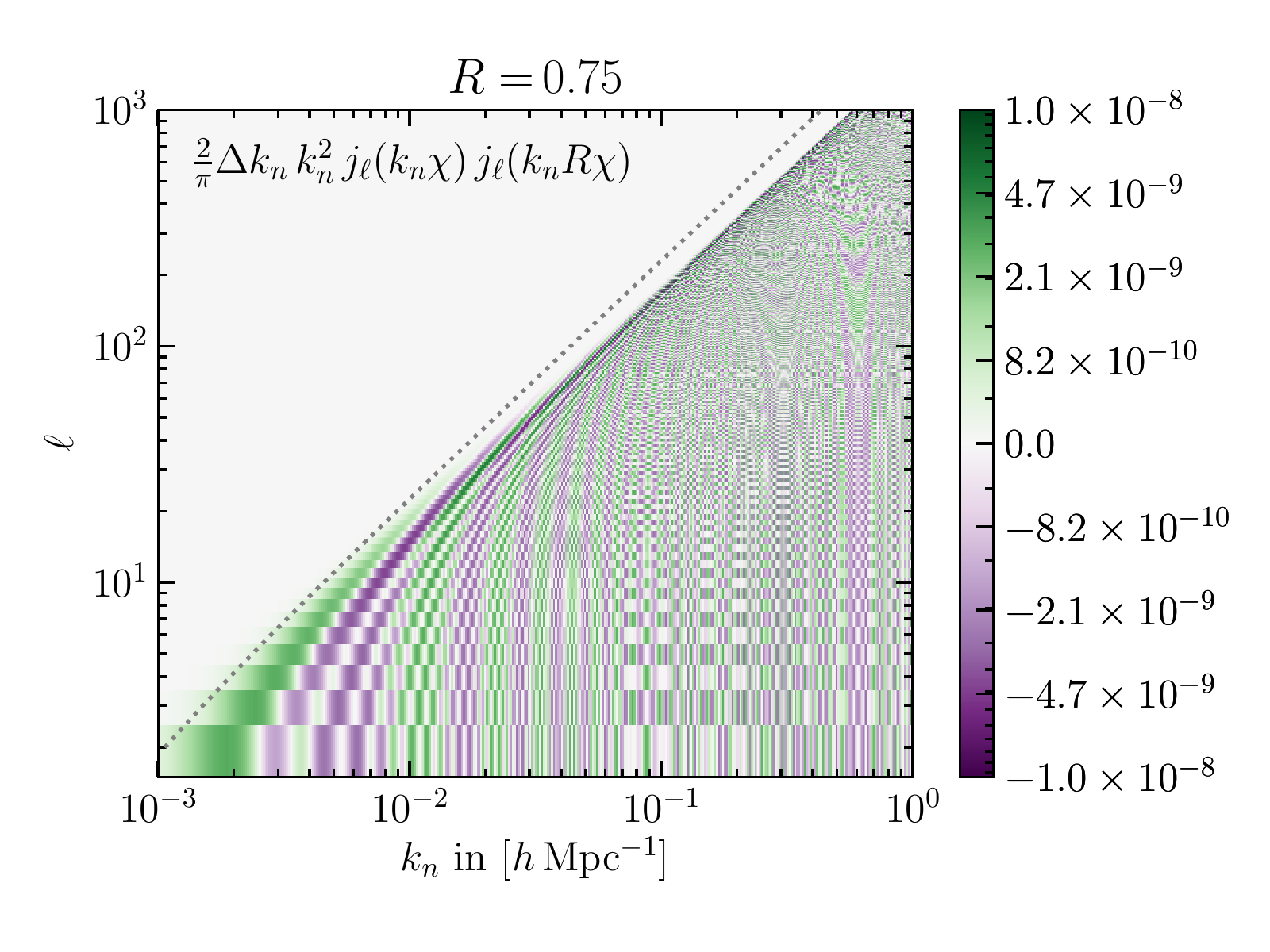}
	\includegraphics[width=0.49\textwidth]{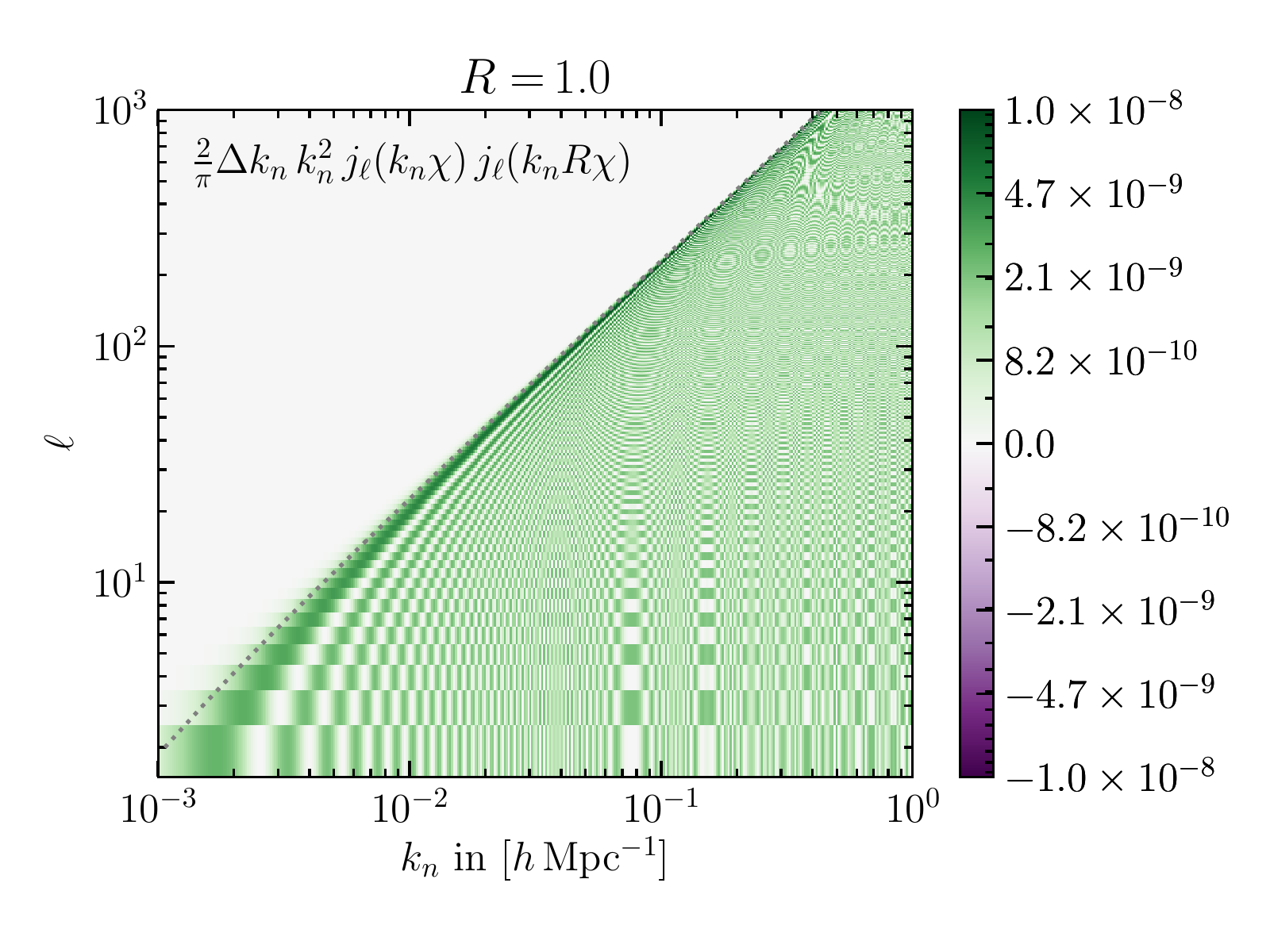}
	\includegraphics[width=0.49\textwidth]{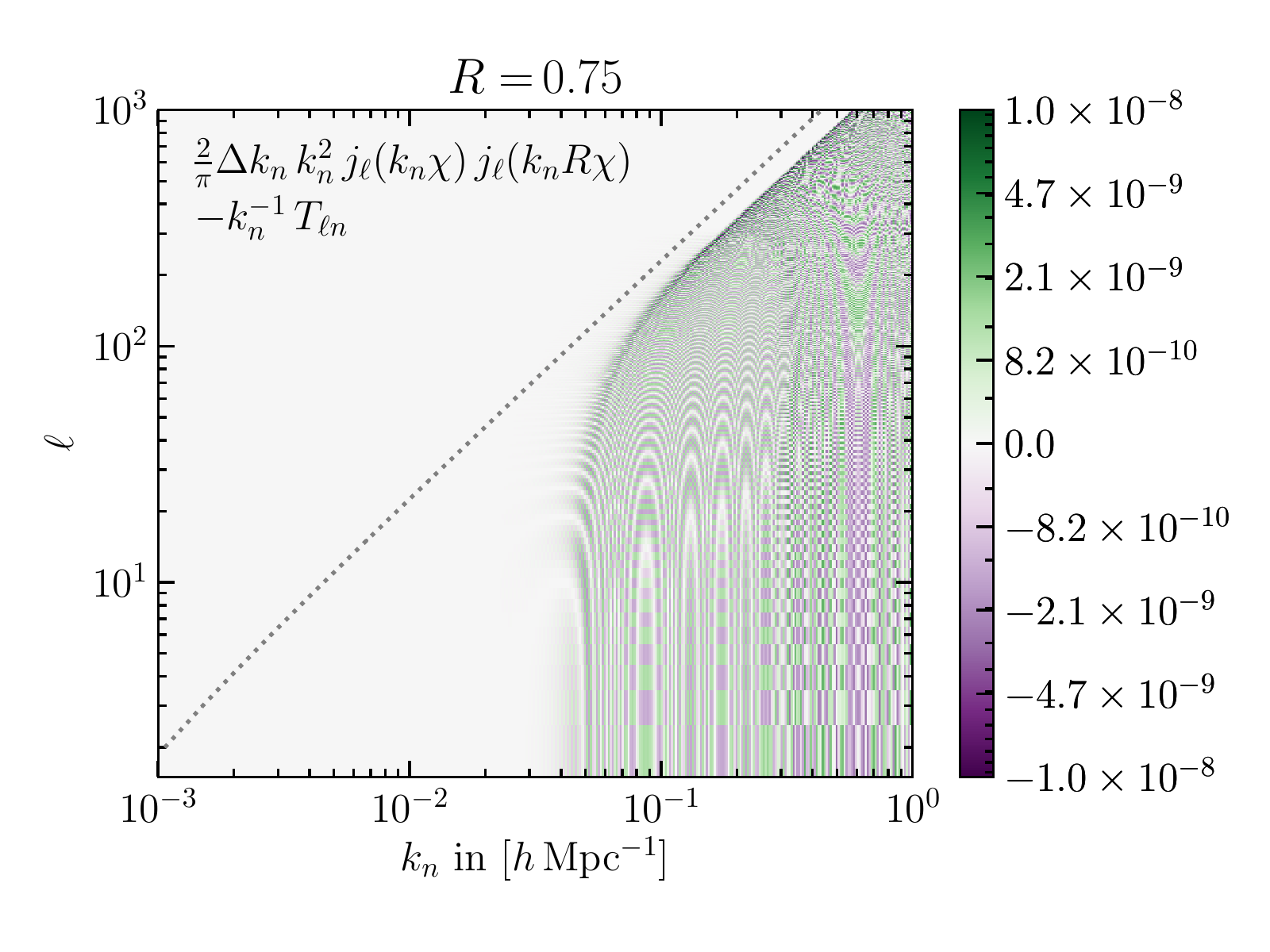}
	\includegraphics[width=0.49\textwidth]{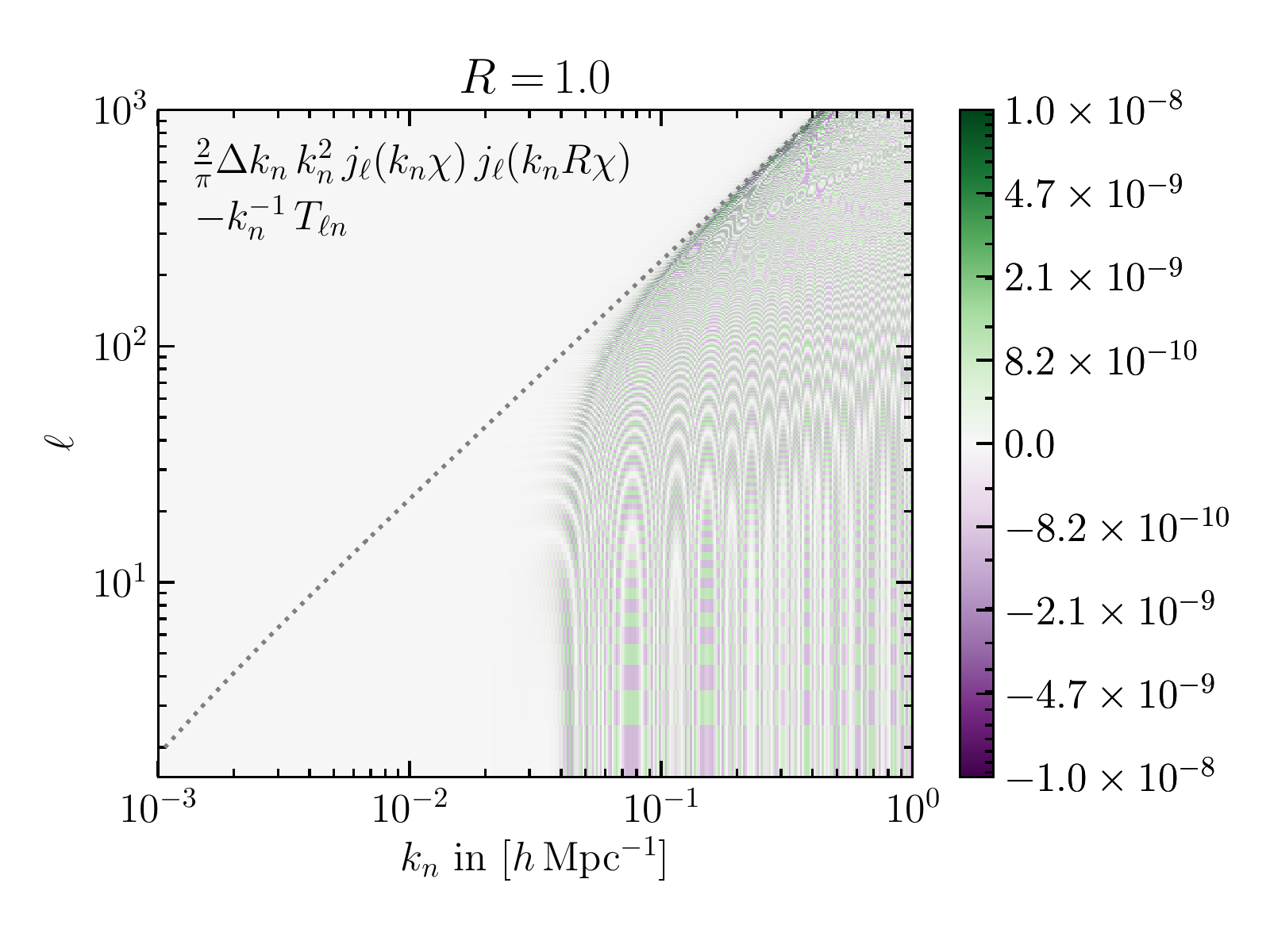}
        \caption{Top panels: Same as \reffig{transmat}, except using the
        traditional method \refeq{traditional}.
        The graph is much more noisy
        than the 2-FAST method, due to undersampling of the oscillations of the
        Bessel functions, which is avoided by the 2-FAST method.
        Bottom panels: To facilitate comparison, we here show the
	difference between the 2-FAST algorithm \refeq{transmat} and the traditional method. At small $k_n$, the
        differences are small, whereas at large $k_n$ the differences are
        the biggest.
        }
	\label{fig:transmat:traditional}
\end{figure*}
\reffig{transmat:traditional} shows the transformation matrix for the case of
two spherical Bessel functions using the traditional method \refeq{traditional}
(left) as well as the difference to the 2-FAST algorithm (right). The two
matrices agree on large scales [$k\simeq \sqrt{\ell(\ell+1)}$], but show quite
different behaviors on small scales (larger $k$). This is because the
traditional trapezoidal method directly samples the multiplication of two
spherical Bessel functions which are very oscillatory, while the 2-FAST
transformation matrix effectively averages over these high-$k$ oscillations. An
accurate integration, therefore, demands much denser sampling for the
trapezoidal method than the 2-FAST method.

\section{Contiguous relations for the Gauss hypergeometric function}
\label{app:hyp2f1-contiguity}
The Gauss hypergeometric function ${}_2F_1(a,b,c,z)$ is defined by the
Gauss series:
\ba
\label{eq:2f1-taylor}
{}_2F_1(a,b,c,z)
&= \sum_{s=0}^\infty \frac{(a)_s(b)_s}{(c)_s}\,\frac{z^s}{s!}
= 1 + \frac{ab}{c}\,z + \frac{a(a+1)\,b(b+1)}{c(c+1)}\,\frac{z^2}{2} + \cdots
\ea
where the Pochhammer symbol is defined as $(a)_s=a(a+1)\cdots(a+s-1)$.
We find that the key to calculating the Gauss hypergeometric function is to
exploit its contiguous relations:
\begin{subequations}
\ba
\label{eq:2F1rA} (c-a)F(a-1,b,c,z) + (2a-c+(b-a)z) F(a,b,c,z) + a(z-1)F(a+1,b,c,z) &= 0 \\
\label{eq:2F1rB} (b-a)F(a,b,c,z) + aF(a+1,b,c,z) - bF(a,b+1,c,z) &= 0 \\
\label{eq:2F1rC} (c-a-b)F(a,b,c,z)+ a(1-z)F(a+1,b,c,z) - (c-b)F(a,b-1,c,z) &= 0 \\
\label{eq:2F1rD} c(a+(b-c)z)F(a,b,c,z) - ac(1-z)F(a+1,b,c,z) + (c-a)(c-b)zF(a,b,c+1,z) &= 0 \\
\label{eq:2F1rE} (c-a-1)F(a,b,c,z) + aF(a+1,b,c,z) - (c-1)F(a,b,c-1,z) &= 0 \\
\label{eq:2F1rF} c(1-z)F(a,b,c,z) - cF(a-1,b,c,z) + (c-b)zF(a,b,c+1,z) &= 0 \\
\label{eq:2F1rG} (a-1+(b+1-c)z)F(a,b,c,z) + (c-a)F(a-1,b,c,z) - (c-1)(1-z)F(a,b,c-1,z) &= 0 \\
\label{eq:2F1rH} c(c-1)(z-1)F(a,b,c-1,z) + c(c-1-(2c-a-b-1)z)F(a,b,c,z)+(c-a)(c-b)zF(a,b,c+1,z) &= 0\,,
\ea
\end{subequations}
that one can find, for example, in \cite{dlmf}.
Note that we can generate more relations by using the symmetry between $a$ and
$b$
\ba
{}_2F_1(a,b,c,z)={}_2F_1(b,a,c,z)
\ea
which follows from the series definition in \refeq{2f1-taylor}.

\section{Computing the Gauss hypergeometric function}
\label{app:hyp2f1}
Computing the Gauss hypergeometric function ${}_2F_1(a,b,c,z)$ with the
parameters that we need in \refeq{Uellellp} is a challenge. In particular,
the first parameter $a=\frac12(-\ell+\ell'+n)$ typically contains a large
imaginary component, for which case we cannot apply the general methods of
calculating ${}_2F_1$ in the literature \citep{michel+2008,johansson2016}.
The method we present here fills this gap for the special case that we face
in computing \refeq{Uellellp}. We accomplish this by using analytical
solutions when $\ell=0$ and recurrence relations that we construct
from the hypergeometric function's contiguous relations \refeqs{2F1rA}{2F1rH}.

In this section, we focus on the recurrence relation along
$\ell\rightarrow\ell+1$ (thick black arrows in \reffig{ellladder}). The recurrence relations along
$\Delta\ell\rightarrow\Delta\ell\pm2$ are needed for a few iterations only, and
so they are not as critical. Hence, we only consider $\Delta\ell=\mathrm{const}$ here.
We shall present the details of the full calculation of
$M_{\ell\ell'}^q$ for general cases with $\Delta\ell=\ell'-\ell\neq \mathrm{const}$ in
\refapp{Mellell-recurrence}. In this section we only consider
$\Delta\ell=4$ when a specific $\Delta\ell$ is needed.

Furthermore, for simplicity, we only treat the
hypergeometric function ${}_2F_1$, without the $\ell$-dependent prefactors in
\refeqs{Mellell'R}{Uellellp}. Nevertheless, the method presented in this
section is the core of evaluating $M_{\ell\ell'}^q$ because the prefactor
merely scales the recurrence relations, without changing the stability
properties.

Comparing with \refeq{Uellellp}, we identify $a$, $b$, $c$, and $z$ as
\ba
\label{eq:a}
a =\,& \frac12n + \frac12\Delta\ell = \frac12 \left(q-1-i t + \Delta\ell\right) \\
b =\,& \ell + \frac12 + \frac12n + \frac12\Delta\ell = c + a - 1 - \Delta\ell \\
\label{eq:c}
c =\,& \ell + \frac32 + \Delta\ell \\
\label{eq:z}
z =\,& R^2\,,
\ea
and we introduce a shorthand notation
\ba
F_\ell[i,j,k] &= {}_2F_1(a+i,b+j,c+k,z)
\ea
to avoid clutter.

Finally, in order to test the accuracy of our results, we have checked out
several software implementations of the Gauss hypergeometric function. One of
the best we have found is the \texttt{Arb} library
\citep{johansson2013arb}.\footnote{\url{http://fredrikj.net/arb/}}
\texttt{Arb} is an arbitrary precision floating point library with automatic
rigorous error bounds. Bindings for the \texttt{Julia} language exist in the
package \texttt{Nemo}.\footnote{\url{https://github.com/Nemocas/Nemo.jl}}

\subsection{Recurrence relation for the \texorpdfstring{$\ell$}{l}-ladder}
\label{app:2f1_recursion}
We have calculated ${}_2F_1$ based on the recurrence relation that relates
$F_{\ell}[0,0,0] =  {}_2F_1(a,b,c;z)$
to
$F_{\ell+1}[0,0,0] =  {}_2F_1(a,b+1,c+1;z) = F_\ell[0,1,1]$.
The key is to exploit the contiguous relations of the Gauss hypergeometric
functions given in \refapp{hyp2f1-contiguity}. Among many possible
relations, we choose a recurrence relation using $F_\ell[0,0,0]$ and
$F_\ell[0,1,0]$. That is, we use
\begin{center}
	\begin{tabular}{l@{ with $(a,b,c)\rightarrow$ }c}
		\refeq{2F1rF} & $(b+1,a,c)$ \\
		\refeq{2F1rE} & $(b+1,a,c+1)$
	\end{tabular}
\end{center}
to find
\ba
\label{eq:b+1a}
(c-a)zF_\ell[0,1,1] &= cF_\ell[0,0,0] - c(1-z)F_\ell[0,1,0] \\
(b+1)F_\ell[0,2,1] &= -(c-b-1)F_\ell[0,1,1] + cF_\ell[0,1,0]\,.
\label{eq:b+1b}
\ea
This way we can compute $F_{\ell+1}[0,0,0]=F_\ell[0,1,1]$ and
$F_{\ell+1}[0,1,0]=F_\ell[0,2,1]$ from $F_\ell[0,0,0]$ and $F_\ell[0,1,0]$.

We then combine the recurrence relations \refeqs{b+1a}{b+1b} with
analytical solutions at $\ell=0$.
We define the function
\ba
g^{\pm}(n,R) &= (1+R)^n \pm (1-R)^n
\ea
Then, for $\Delta\ell=4$ this gives the following analytical solution:
\ba
\label{eq:F000-ell0-dl4}
F_{0,4}[0,0,0]
&= \frac{945}{2m\big(576 - 820 m^2 + 273 m^4 - 30 m^6 + m^8\big) R^9}
\vs&\quad\times
\bigg[
	-5mR\big[21 + (-11+2m^2)R^2\big]\,g^+(m,R)
	\vs&\qquad\quad
	+ \big[105 + 45(-2+m^2)R^2 + (9-10m^2+m^4)R^4\big]\,g^-(m,R)
\bigg]
\\
F_{0,4}[0,1,0]
&= \frac{945}{
	2(-5+n)(-3+n)(-2+n)(-1+n)n(1+n)(2+n)(3+n)(5+n)R^9
}
\vs&\quad\times\bigg[
	\big[105 + 15(-5+3n^2)R^2 + n^2(-4+n^2)R^4\big]\,g^-(-n,-R)
	\vs&\qquad\quad
	- nR\,\big[105 + (-4+n^2)R^2(10+R^2)\big]\,g^+(-n,-R)
\bigg]
\label{eq:F010-ell0-dl4}
\ea
where $m=1-n$. These expressions are numerically unstable for $R\sim0$. For
simplicity one may use 256-bit arbitrary precision floating point arithmetic to
evaluate these. Since these expressions are cosmology independent and only need
to be evaluated once, the performance is not very critical here.

It turns out that, however, unless $z=R^2\simeq1$, the forward-directional
recurrence relation \refeqs{b+1a}{b+1b} are unstable under an injection of small
noise (which happens, for example, due to the numerical round-off error at each
recursion step).

Instead, we find that the reverse, backward-directional recurrence relation,
\ba
\label{eq:b+1a-back}
F_\ell[0,1,0] &= c^{-1}(c-b-1)F_\ell[0,1,1] + c^{-1}(b+1)F_\ell[0,2,1] \\
F_\ell[0,0,0] &= c^{-1}(c-a)zF_\ell[0,1,1] + (1-z)F_\ell[0,1,0],
\label{eq:b+1b-back}
\ea
is quite stable so that the noise decays while the recursion proceeds along
the backward direction of $\ell\to\ell-1$.
This is the basis of Miller's algorithm \cite{bickley+1952} that we have
implemented here. \reffig{miller} shows the performance and error propagation
for the backward recursion.

On the other hand, for the backward recursion, we do not have the luxury of
an analytical expression for the \emph{initial} condition at large
$\ell$ values. The challenge now is, therefore, to find suitable values to
start the recursion with.
Again, the key fact is that the backward recursion is so stable that \emph{any}
deviation from the true value (noise) decays quickly.
The general strategy we adopt, therefore, is to start the recursion from some
large $\ell_\mathrm{seed}$, which is sufficiently larger than the maximum
multipole $\ell_\mathrm{max}$ that we want to calculate the $F_\ell[0,0,0]$ for.
Specifically, we increase $\ell_\text{seed}$ until convergence is reached for
the resulting $F_{\ell_\mathrm{max}}[0,0,0]$ and $F_{\ell_\mathrm{max}}[0,1,0]$
within $10^{-10}$ accuracy. That is, we require
\ba
\frac{||\vec F_{\ell_\max}^{(\mathrm{new}~\ell_\mathrm{seed})} - \vec F_{\ell_\max}^{(\mathrm{old}~\ell_\mathrm{seed})}||}
{||\vec F_{\ell_\max}^{(\mathrm{new}~\ell_\mathrm{seed})}||}
&< \num{e-10}
\label{eq:conv_criteria}
\ea
where we define $\vec F_\ell = (F_\ell[0,0,0], F_\ell[0,1,0])$, and we define
the norm as the Euclidean distance: $||\vec x|| \equiv
\sqrt{\Re(x_1)^2+\Im(x_1)^2+\Re(x_2)^2+\Im(x_2)^2}$, with $x_1=F[0,0,0]$ and
$x_2=F[0,1,0]$, and $\Im(z)$ is the imaginary part of $z$.

In principle, any initial guess for $F_{\ell_\mathrm{seed}}[0,0,0]$ should work.
When choosing a value close to the true value, however, the recursion chain
converges to the true $F_\ell[0,0,0]$ value much faster. In the remainder of
this section, we shall present our implementation of setting up the initial
conditions.

\begin{figure*}
    \includegraphics[width=0.49\textwidth]{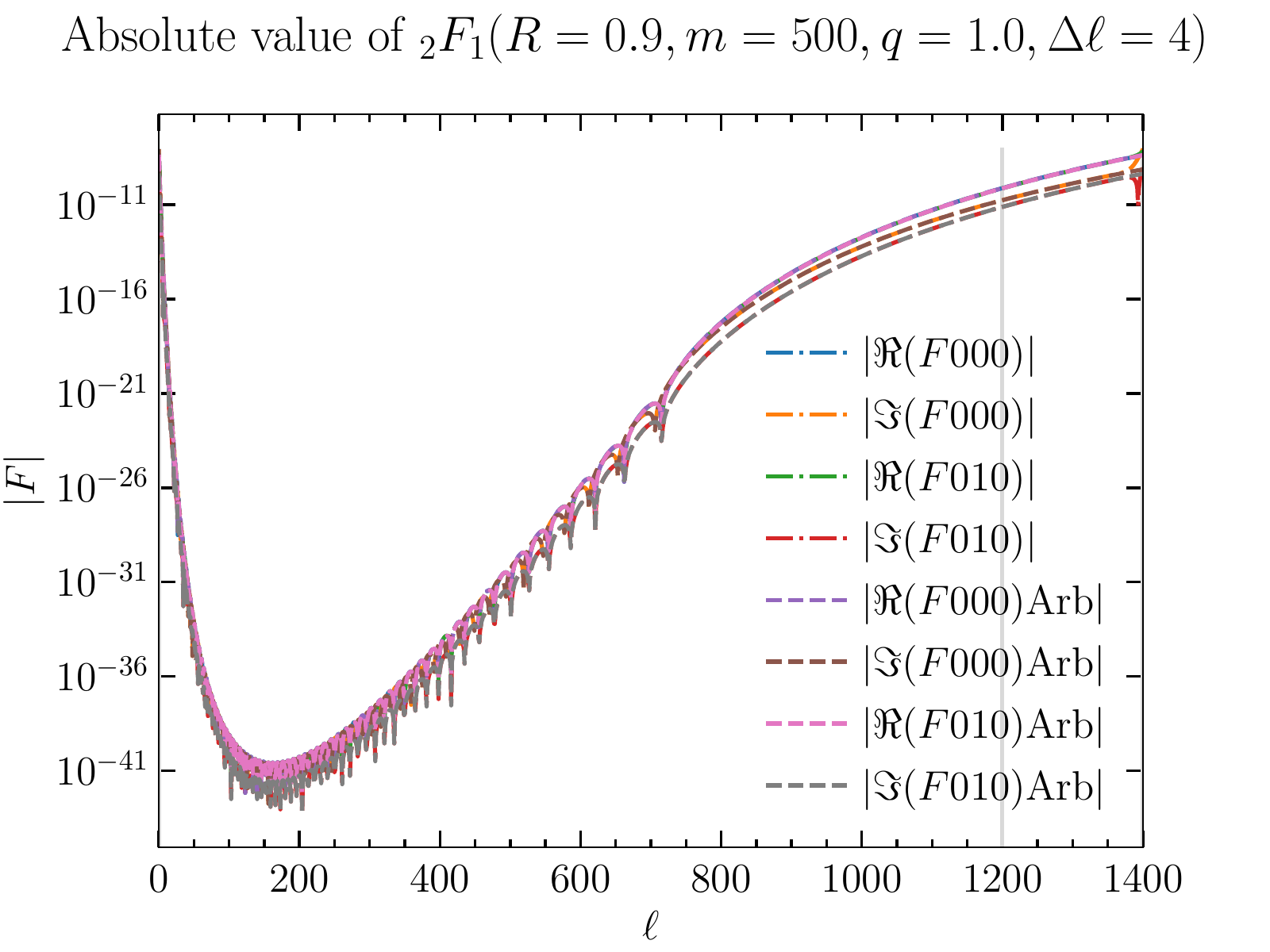}
    \includegraphics[width=0.49\textwidth]{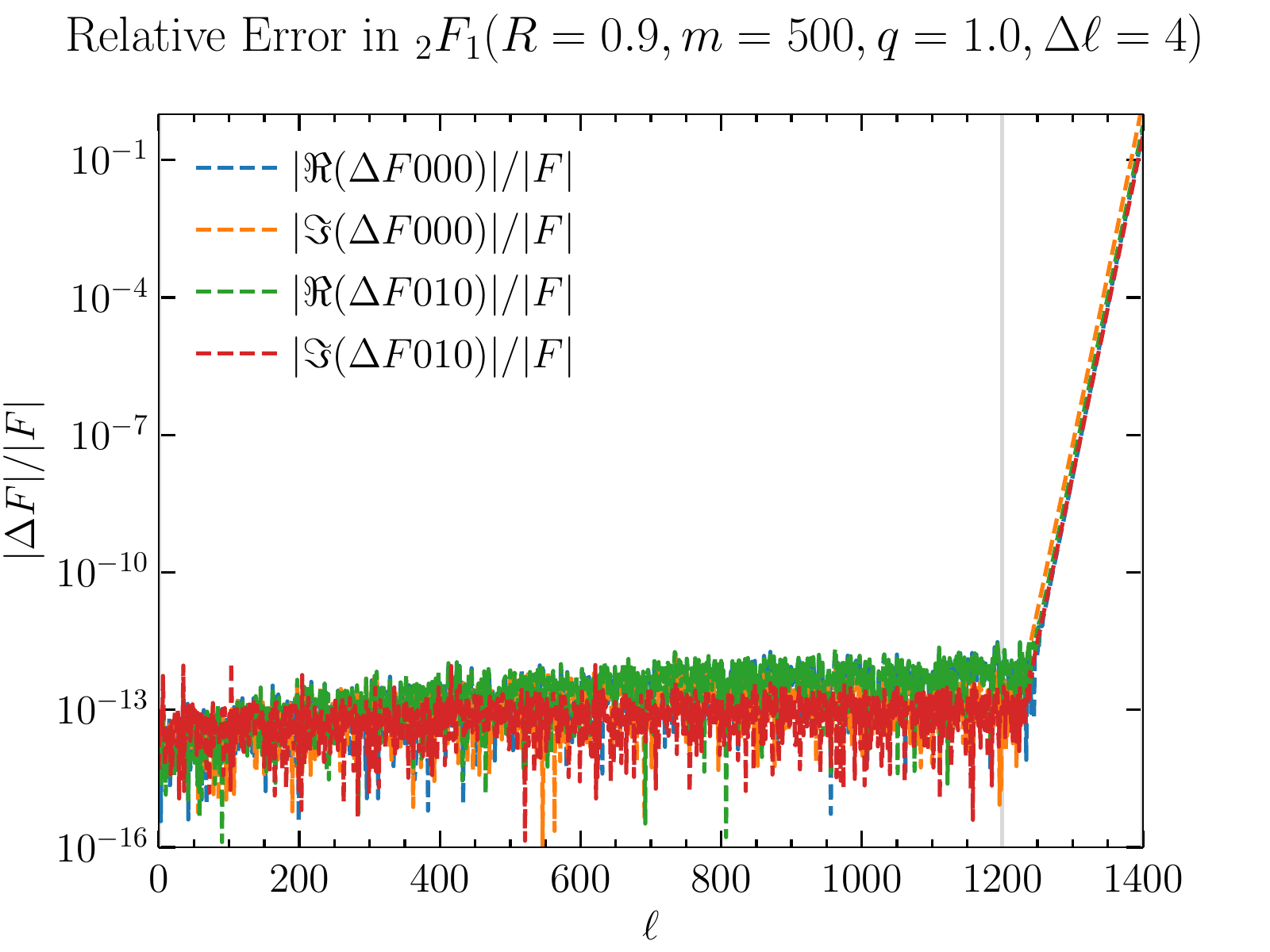}
    \includegraphics[width=0.49\textwidth]{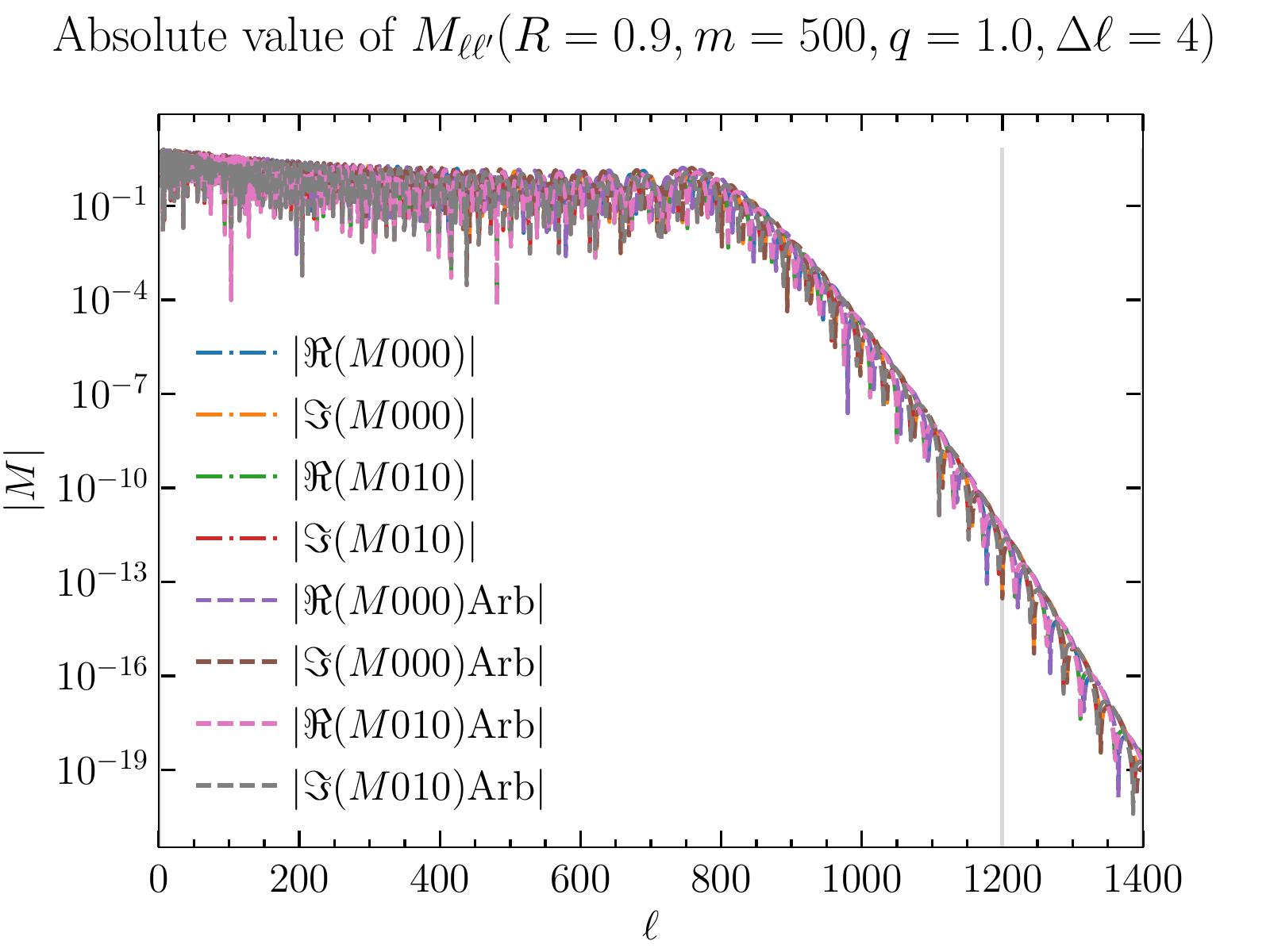}
    \includegraphics[width=0.49\textwidth]{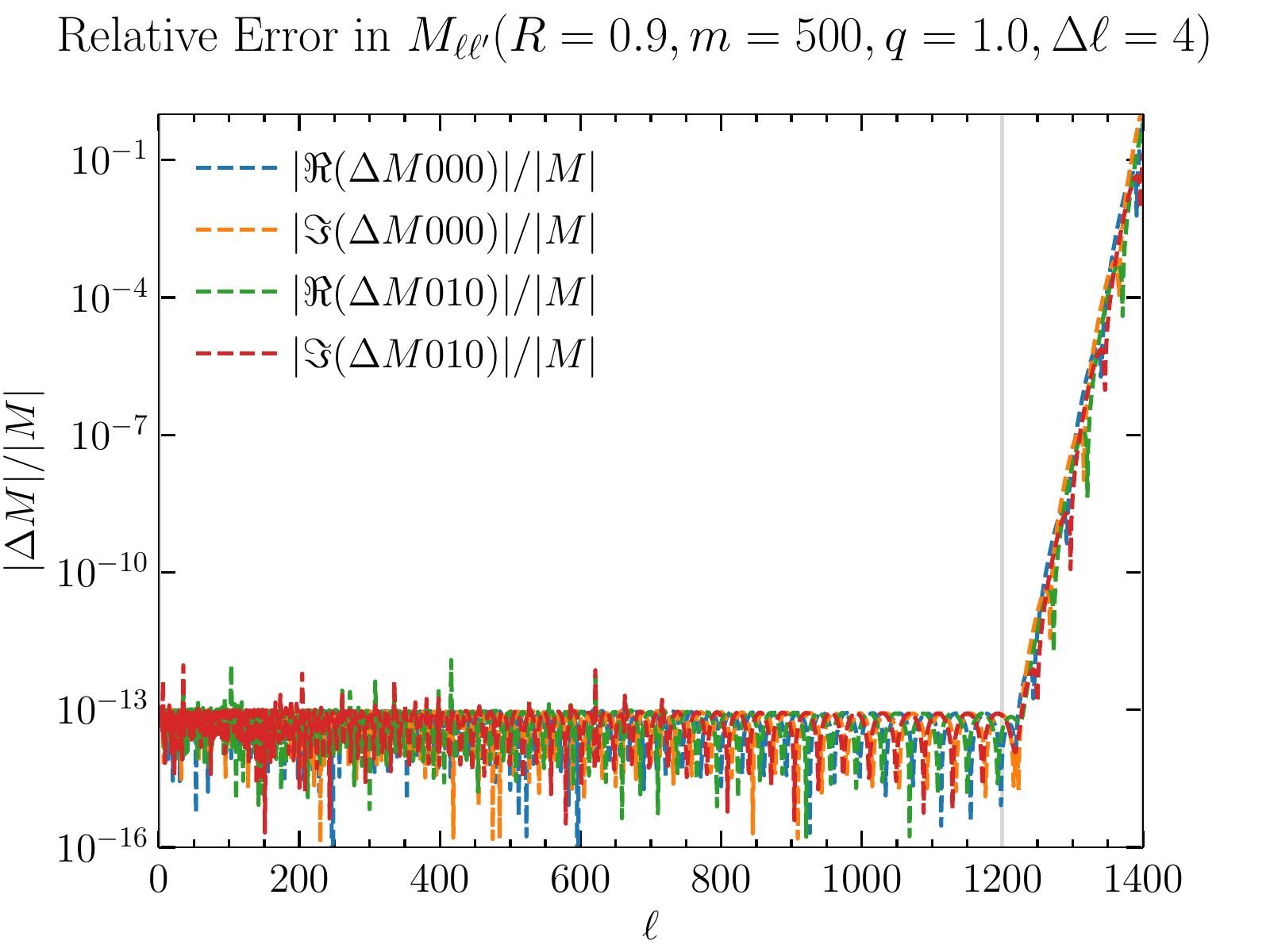}
    \caption{Demonstration of Miller's algorithm. Top Left: The absolute
    value of the real and imaginary parts of the hypergeometric function
    calculated via recurrence relations is compared with the arbitrary
    precision library \texttt{Arb}\citep{johansson2013arb}. Top Right: The
    relative difference between the recurrence relation method and \texttt{Arb} is
    shown. As shown, large errors die out before our target $\ell_\max=1200$
    indicated by the gray vertical line is reached. Bottom Panels: Same
    as the top two panels, but for $M^q_{\ell\ell'}$ instead of ${}_2F_1$.
    }
    \label{fig:miller}
\end{figure*}

\subsection{Setting up the initial condition for the backward recursion for
\texorpdfstring{$z\ll1$}{z<<1}}
\label{app:zlt1IC}
Here we show the method by which we set up the initial condition at some large
$\ell_\mathrm{seed}$ for the $z=R^2\ll1$ case, exploiting that the error in the
initial values will die out when stepping down the $\ell$ ladder.
First, we define
\ba
\vec F_\ell &= \left(\begin{matrix} F_\ell[0,0,0] \\ F_\ell[0,1,0] \end{matrix}\right)\,,
\ea
so that we can represent the recursion relation in matrix form
$\vec F_{\ell+1} = A_\ell\cdot \vec F_\ell$ with
\ba
\label{eq:fell-recursion}
A_\ell &= \left(\begin{matrix}
    \frac{c}{c-a}\,\frac{1}{z}       &  -\frac{c}{c-a}\,\frac{1-z}{z} \\
    \frac{c}{c-a}\,\frac{a-\Delta\ell}{c+a-\Delta\ell}\,\frac{1}{z}
    &   \frac{c}{c+a-\Delta\ell}\big[1+\frac{\Delta\ell}{c-a}\,\frac{1-z}{z}\big]
\end{matrix}\right)
\ea
which follows from \refeqs{b+1a}{b+1b}, and we used $b=c+a-1-\Delta\ell$.
The inverse transformation is then given as
\ba
\label{eq:Anellinv}
A_\ell^{-1}
&= \left(\begin{matrix}
    z - \frac{a}{c} + \frac{\Delta\ell}{c}\,(1-z)
    &  \left[1+\frac{a}{c}-\frac{\Delta\ell}{c}\right](1-z)
    \\
    -\frac{a}{c}+\frac{\Delta\ell}{c}
    &  1+\frac{a}{c}-\frac{\Delta\ell}{c}
\end{matrix}\right)\,.
\ea
Applying the recursion multiple times, we can relate $\vec F_{\ell}$
to $\vec F_{\ell+m}$ as
\ba
\vec F_{\ell+m} &= A_\ell^{[m]} \vec F_\ell \\
\vec F_\ell &= A_\ell^{[-m]} \vec F_{\ell+m}\,,
\ea
where we introduce the notation for the matrix
\ba
\label{eq:Anell-notation-for}
A_\ell^{[m]}  &= A_{\ell+m-1} \cdots A_\ell \\
A_\ell^{[-m]} &= A_\ell^{-1} \cdots A_{\ell+m-1}^{-1}\,.
\label{eq:Anell-notation-back}
\ea
Note that $A_\ell^{[-m]}\,A_\ell^{[m]}=1$. The matrix $A_\ell^{[m]}$ raises
the $\ell$ to $\ell+m$, whereas the matrix $A_\ell^{[-m]}$ lowers $\ell+m$
down to $\ell$.

Just to show the point directly, let us analyze the stability properties in
the $\ell\rightarrow\infty$ limit, where the lowering matrix becomes
\ba
A_\infty^{-1}
&= \left(\begin{matrix}
	z  &  1-z  \\
	0  &  1
\end{matrix}\right),
\ea
whose eigenvalues are $\lambda_1=1$ and $\lambda_2=z$, with eigenvectors
$\vec{x}_1=1/\sqrt{2}(1,1)$ and $\vec{x}_2=(1,0)$. That is, the
$\vec{x}_2$ component of the solution will be suppressed by a factor $z^n$ when
lowering $\ell$ by $n$ steps, whereas the $x_1$ component is the dominant
solution in the backwards direction. This motivates us to define the initial
value of the recursion at $\ell_\mathrm{seed}$ as
\ba
\vec F_{\ell_\text{seed}}
= \frac{\lambda}{\sqrt{2}} \left(\begin{matrix} 1 \\ 1 \end{matrix}\right)
\label{eq:Fseed}
\ea
with a complex constant $\lambda$ that we fix so that
$\vec F_0^\text{seeded}= A_0^{[-\ell_\text{seed}]} \vec F_{\ell_\text{seed}}$
matches the analytical solution \refeqs{F000-ell0-dl4}{F010-ell0-dl4}.

\subsection{Extension to general \texorpdfstring{$z$}{z}} 
\label{app:generalIC}

The method we have described in \refapp{zlt1IC} works best for small $z$
because the error in the initial condition decays as $z^n$. For the
$z=R^2\sim1$ cases, therefore, the error in the initial condition decays slowly
so that one needs to set $\ell_\mathrm{seed}$ much larger than $\ell_\mathrm{max}$.
In the extreme case of $z=1$, the initial error does not die at all.
In this section, we summarize our method of generating the initial condition
at $\ell_\mathrm{seed}$ for these cases.

For the $z\sim1$ cases, we use the fact that the forward recursion is more stable than the
$z<1$ cases. For that, we first attempt to invert the matrix
$A_0^{[-\ell_\text{seed}]}$ to find the matrix for the forward recursion.
Because the matrices $A_{\ell}^{-1}$ are invertible, in principle, the final
matrix $A_0^{[-\ell_\text{seed}]}$ must be invertible as well. If the forward
recursion is stable enough that $A_0^{[-\ell_\text{seed}]}$ is numerically
invertible, we find $\vec{F}_{\ell_\mathrm{seed}}$ from $\vec{F}_0$ and the matrix
$A_0^{[\ell_\text{seed}]}$ for the forward recursion; we then run the backward
recursion in order to clean any possible error caused by numerical round-off.
Often there are cases, however, where, as a result of accumulated
numerical round-off error at each step of the matrix multiplication, the
resulting $A_0^{[-\ell_\text{seed}]}$ ends up singular (noninvertible) or,
for the same reason, numerical infinities appear in the inverted matrix. In
that case, we use the seeding value in \refeq{Fseed}, and then we choose
$\lambda\in\mathbb{C}$ such that we match the analytical solution
\refeqs{F000-ell0-dl4}{F010-ell0-dl4} at $\ell=0$:
\ba
\label{eq:fseed}
\vec F_0
&=
A_0^{[-\ell_\text{seed}]} \,
\frac{\lambda}{\sqrt{2}} \left(\begin{matrix} 1 \\ 1 \end{matrix}\right)\,.
\ea
Whether the inversion of the matrix is successful ($R\sim1$ cases) or not
($R\ll1$ cases), any error introduced by numerical round-off will be corrected
both by the choice of $\lambda$ and by running the backward recursion
from $\ell_\text{seed}$ down to $\ell_\max$. Hence, this approach works for all
values of $z$.

\subsubsection{Special cases}
\label{app:special}
The special cases $m=0$ (the direct current (DC) mode), $z=R^2=1$, and $z=R^2=0$ need to be
handled separately.
\begin{itemize}
\item{\textit{The DC mode}:
With the above approach, we may run into division-by-zero problems when the mode
$m=0$. Then for some choice of $q$ and $\Delta\ell$ it may happen that
$n+\Delta\ell=0$, which implies $a=0$. In this case, however, we have the
trivial solution
\ba
F_\ell[0,0,0] &= 1 \\
F_\ell[0,1,0] &= 1
\ea
for any $\ell$. This follows from \refeq{2f1-taylor}.
}
\item{$z=1$:
For the case $R^2=1$, we can speed up the computation by using the analytical
expression for the initial condition at any $\ell$
\ba
\label{eq:F000-z1}
F_\ell[0,0,0] &= \frac{\Gamma(c)\,\Gamma(c-a-b)}{\Gamma(c-a)\,\Gamma(c-b)}\,.
\ea
For $F_\ell[0,1,0]$ \refeq{F000-z1} does not give a finite answer when $q=1$
and $t=0$. Hence, it is best to require $q\neq1$.
}
\item{$z=0$:
For the case $R=0$, the $w_{\ell\ell'}(\chi,\chi')=0$ for all $\ell'>0$
due to the vanishing of the spherical Bessel function at the origin. Thus no
calculation is required.
}
\end{itemize}

\subsection{Underflow protection}
\label{app:underflow}
The hypergeometric function values (and $M^q_{\ell\ell'}$ as well) may under
some circumstances (especially when the number of sample points $N$ of the
power spectrum is large, e.g. $R=0.99$, $q=0.5$, $m=4100$, $G=23.6$,
$\ell_\max=1200$) suffer from underflow error when represented as a double
precision number. That is, starting at $\ell=0$, which can generally be
represented in doubles, $|{}_2F_1|$ gets smaller and smaller going towards
higher $\ell$, and it eventually hits the underflow value for double precision
numbers $|{}_2F_1|<\num{e-308}$.

We must, therefore, take care of the case where the seeding value becomes
$\vec F_{\ell_\text{seed}}=0$ as a result of the underflow.
We detect this when one or more of the elements in $A_0^{[-\ell_\text{seed}]}$
becomes $\pm{}\infty$ as represented by double precision.
The solution we adopt here is to store the largest
$\ell_\text{finite}<\ell_\max$ where no underflow occurs, and approximate
${}_2F_1\approx0$ for $\ell>\ell_\text{finite}$.

\section{Recursion for the full kernel}
\label{app:Mellell-recurrence}
In this appendix we derive the recurrence relations that are valid not just for
the hypergeometric function, but for the full kernel $M^q_{\ell\ell'}(t,R)$ in
\refeq{Mellell'R}. We also detail the recursions to move towards any even
$\Delta\ell\equiv\ell'-\ell$ (thick gray arrows in \reffig{ellladder}), for the
two cases $R\leq1$ and $R>1$. These two cases need to
be treated separately due to \refeq{Mellellsymmetry}. Note, however, that due
to the symmetry $w_{\ell\ell'}(\chi,\chi')=w_{\ell'\ell}(\chi',\chi)$, only one
of the two cases is actually needed, although for some computations (e.g.
lensing-galaxy cross-correlation, see
\refsec{lensing-galaxy-cross-correlation}) it is convenient to be able to
compute both cases explicitly.

As in \refeqs{a}{z} $a$, $b$, $c$, and $z$ will be the parameters to the
hypergeometric function ${}_2F_1(a,b;c;z)$ in \refeq{Uellellp}.

To get the full kernel $M^q_{\ell\ell'}(t,R)$, the hypergeometric function needs
to be multiplied by the prefactors in \refeqs{Mellell'R}{Uellellp}. The
prefactor is
\ba
A_{\ell,\Delta\ell}
&= (k_0\chi_0)^{it-q} \, 2^{n-2} \, \pi \, R^{\ell+\Delta\ell} \,
\frac{\Gamma\big[\ell + \frac12 + \frac12n + \frac12\Delta\ell\big]}
{\Gamma\big[1 - \frac12 n - \frac12\Delta\ell\big]\,
\Gamma\big(\ell + \frac32 + \Delta\ell\big)}
\label{eq:Aelldell}
\ea
where $\Delta\ell=\ell'-\ell$. We get
\ba
\label{eq:Aellp1}
A_{\ell+1,\Delta\ell}
&= A_{\ell,\Delta\ell}\,R\,\frac{\ell+\frac12+\frac12n+\frac12\Delta\ell}{\ell+\frac32+\Delta\ell}
= A_{\ell,\Delta\ell}\,\frac{bR}{c}
\ea
which adds another factor to our recursion relations
\refeqs{b+1a-back}{b+1b-back}. Since this multiplies all elements in the
recursion matrix by the same number, the ratio of the eigenvalues is the same
as without the prefactor, and hence this does not change the stability
properties of the relation discussed in \refapp{hyp2f1}.

Our recursion relations along $\ell\rightarrow\ell+1$ are stable in the
backward direction for any $\Delta\ell=0,\pm2,\pm4$ (which are all we have
tested). The recursion relations detailed below to change $\Delta\ell$ are
stable in the $\Delta\ell\rightarrow\Delta\ell-2$ direction for $R<1$, and they
are stable in the $\Delta\ell\rightarrow\Delta\ell+2$ direction for $R>1$.
Thus, we perform the $\Delta\ell=\mathrm{const}$ recursions with $\Delta\ell=4$ for
$R<1$, and with $\Delta\ell=-4$ for $R>1$. Note, however, that due to the
symmetry \refeq{Mellellsymmetry}, the initialization with $\Delta\ell=4$ as
described in \refapp{hyp2f1} is also valid for the $R>1$ case.

The relations to move towards $\Delta\ell\neq0$ are determined by
\refeqs{a}{z}. Our recursion relations along $\ell\rightarrow\ell-1$ give us
$F_\ell[0,1,0]$ along with $F_\ell[0,0,0]$ (or their $M^q_{\ell\ell}$
equivalent). However, in the sections below we derive the recursions using the
values $F[0,0,0]$ and $F[0,-1,0]$. To go from $F[0,1,0]$ we use \refeq{2F1rA}
with $a\leftrightarrow b$ to get
\ba
(c-b)F[0,-1,0] &= b(1-z)F[0,1,0] - \big(2b-c+(a-b)z\big)F[0,0,0]
\label{eq:F0m10=F010+F000}
\ea
In this section we will generally drop the subscript $\ell$, and sometimes use
$\Delta\ell=\ell-\ell'$ as a subscript instead. $a$, $b$, and $c$, which are
functions of $\ell$ and $\Delta\ell$, are understood to take on their values at
$F[0,0,0]$ in any given equation. In the remainder of this appendix, we derive
the $\Delta\ell\neq0$ relations.

\subsection{Towards \texorpdfstring{$\Delta\ell=-2$}{dl=-2} when \texorpdfstring{$R<1$}{R<1}}
To calculate $\Delta\ell=-2$ we need $F[-1,-1,-2]$. We start with
\refeq{2F1rE} with $(a,b,c)\rightarrow(a-1,b-1,c-1)$ to get
\ba
F[-1,-1,-2] &= \frac{c-1-a}{c-2}\,F[-1,-1,-1]
+ \frac{a-1}{c-2}\,F[0,-1,-1]
\ea
\refeq{2F1rF} with $(a,b,c)\rightarrow(a,b-1,c-1)$ gives
\ba
F[-1,-1,-1] &= (1-z)\,F[0,-1,-1]
+ \frac{c-b}{c-1}\,z\,F[0,-1,0]
\ea
and \refeq{2F1rE} with $a\leftrightarrow b$ and$(a,b,c)\rightarrow(a,b-1,c)$
gives
\ba
F[0,-1,-1] &= \frac{c-b}{c-1}\,F[0,-1,0] + \frac{b-1}{c-1}\,F[0,0,0]
\ea
Putting the last three together, one at a time, we get
\ba
F[-1,-1,-2] &=
\frac{(c-1-a)(1-z) + a-1}{c-2}\,F[0,-1,-1]
+ \frac{c-1-a}{c-2}\,\frac{c-b}{c-1}\,z\,F[0,-1,0]
\\
&=
\frac{c-b}{c-1}\,F[0,-1,0]
+ \frac{c-2 - (c-1-a)z}{(c-2)(c-1)}\,(b-1)\,F[0,0,0]
\label{eq:recursion-m1m1m2}
\ea

From \refeq{Aelldell} we see that
\ba
A_{\ell,\Delta\ell-2}
&= A_{\ell,\Delta\ell} \, R^{-2} \,
\frac{\big(\ell-\frac12+\Delta\ell\big)\big(\ell+\frac12+\Delta\ell\big)}
{\big(\ell-\frac12+\frac12\Delta\ell+\frac12n\big)\big(1-\frac12\Delta\ell-\frac12n\big)}
\\
&= A_{\ell,\Delta\ell}\,R^{-2}\,\frac{(c-2)(c-1)}{(b-1)(1-a)}
\label{eq:Aelldell-dl=-2}
\ea
where $a$, $b$, and $c$ are evaluated at $\Delta\ell$.
Hence, writing $\tilde F_{\Delta\ell}=A_{\ell,\Delta\ell}\,F$
we get
\ba
\tilde F_{-2}[-1,-1,-2]
&=
\frac{(c-b)(c-2)}{(b-1)(1-a)R^2}\,\tilde F_0[0,-1,0]
+ \frac{c-2 - (c-1-a)z}{(b-1)(1-a)R^2}\,(b-1)\,\tilde F_0[0,0,0]
\\
\tilde F_{-2}[0,0,0] &= \frac{(c-2)(c-1)}{(b-1)(1-a)R^2}\,\tilde F_0[0,0,0]
\ea
which keeps the factor $A_{\ell,\Delta\ell}$ the same for the two
hypergeometric function values, and allows us to use these as the starting
values for a recursion towards $\Delta\ell=-4$.

\subsection{Towards \texorpdfstring{$\Delta\ell=-4$}{dl=-4} when \texorpdfstring{$R<1$}{R<1}}
\label{sec:recursion-dlm4}
We also need to go towards $\Delta\ell=-4$, or $F[-2,-2,-4]$. We do this by
building a recursion from $F[0,0,0]$ and $F[-1,-1,-2]$. We use the following seven
relations,
\begin{center}
	\begin{tabular}{l@{ with $(a,b,c)\rightarrow$ }c}
		\refeq{2F1rE} & $(b-1,a,c)$ \\
		\refeq{2F1rH} & $(a,b-1,c-1)$ \\
		\refeq{2F1rF} & $(a,b-1,c-2)$ \\
		\refeq{2F1rC} & $(a-1,b-1,c-2)$ \\
		\refeq{2F1rE} & $(b-2,a-1,c-2)$ \\
		\refeq{2F1rH} & $(a-1,b-2,c-3)$ \\
		\refeq{2F1rF} & $(a-1,b-2,c-4)$
	\end{tabular}
\end{center}
which result in the following equations:
\ba
\label{eq:F[-2,-2,-4]-first}
(c-1)F[0,-1,-1] &= (c-b)F[0,-1,0] + (b-1)F[0,0,0]
\\
(c-1)(c-2)(1-z)F[0,-1,-2] &= (c-1)(c-2-(2c-2-b-a)z)F[0,-1,-1]
\vs&\quad+ (c-b)(c-1-a)zF[0,-1,0]
\\
(c-2)(1-z)F[0,-1,-2] &= (c-2)F[-1,-1,-2] - (c-1-b)zF[0,-1,-1]
\\
(c-1-b)F[-1,-2,-2] &= (c-a-b)F[-1,-1,-2] + (a-1)(1-z)F[0,-1,-2]
\\
(c-3)F[-1,-2,-3] &= (c-b-1)F[-1,-2,-2] + (b-2)F[-1,-1,-2]
\\
(c-3)(c-4)(1-z)F[-1,-2,-4] &= (c-3)(c-4-(2c-4-a-b)z)F[-1,-2,-3]
\vs&\quad+ (c-2-a)(c-1-b)zF[-1,-2,-2]
\\
(c-4)F[-2,-2,-4] &= (c-4)(1-z)F[-1,-2,-4] + (c-2-b)zF[-1,-2,-3]
\label{eq:F[-2,-2,-4]-last}
\ea
The first three can be solved for $(1-z)F[0,-1,-2]$. Inserting the first into
the second and third we get
\ba
(c-1)(c-2)(1-z)F[0,-1,-2]
&= (c-2-(c-1-b)z)(c-b)F[0,-1,0]
\vs&\quad+ (c-2-(2c-2-b-a)z)(b-1)F[0,0,0]
\\
(c-1)(c-2)(1-z)F[0,-1,-2] &= (c-1)(c-2)F[-1,-1,-2]
- (c-b)(c-1-b)zF[0,-1,0]
\vs&\quad
- (c-1-b)z(b-1)F[0,0,0]
\ea
and then solving the second for $F[0,-1,0]$ and inserting into the third, we get
\ba
(1-z)F[0,-1,-2]
&=
\frac{c-2-(c-1-b)z}{c-2}\,F[-1,-1,-2]
- (c-1-a)z(c-1-b)z\frac{b-1}{(c-1)(c-2)^2}\,F[0,0,0]
\label{eq:F[0,-1,-2]}
\ea
To get $F[-2,-2,-4]$ we start with the last of
\refeqs{F[-2,-2,-4]-first}{F[-2,-2,-4]-last} and then continue using each
upwards in succession. We get
\ba
F[-2,-2,-4]
&=
\frac{c-4-(c-2-a)z}{c-4}\,F[-1,-2,-3]
+ \frac{(c-2-a)(c-1-b)z}{(c-3)(c-4)}\,F[-1,-2,-2]
\\
&=
\frac{c-b-1}{c-3}F[-1,-2,-2]
+ \frac{(c-4-(c-2-a)z)(b-2)}{(c-3)(c-4)}\,F[-1,-1,-2]
\\
&=
\frac{(c-a-2)(c-4 - (b-2)z)}{(c-3)(c-4)}\,F[-1,-1,-2]
+ \frac{(a-1)(1-z)}{c-3}\,F[0,-1,-2]
\\
&=
\frac{(c-a-2)(c-4 - (b-2)z)(c-2) + (c-2-(c-1-b)z)(a-1)(c-4)}{(c-2)(c-3)(c-4)}\,F[-1,-1,-2]
\vs&\quad
- (c-1-a)z(c-1-b)z\frac{(a-1)(b-1)}{(c-1)(c-2)^2(c-3)}\,F[0,0,0]
\\
&=
\frac{(c-2)(c-4) - \big[a(c-b) + b(c-a) - 3c + 4\big]z}{(c-2)(c-4)}\,F[-1,-1,-2]
\vs&\quad
- (c-1-a)z(c-1-b)z\frac{(a-1)(b-1)}{(c-1)(c-2)^2(c-3)}\,F[0,0,0]
\label{eq:Fm2m2m4-recursion}
\ea
The factor $A_{\ell,\Delta\ell}$ can be adjusted according to
\refeq{Aelldell-dl=-2}, where now $a$, $b$, and $c$ are evaluated at
$\Delta\ell=-2$.

\subsection{Towards \texorpdfstring{$\Delta\ell=2$}{dl=2} when \texorpdfstring{$R<1$}{R<1}}
We need $F[1,1,2]$ in terms of $F[0,0,0]$ and $F[0,-1,0]$. To get $F[1,1,2]$ we
use \refeq{2F1rB} with $(a,b,c)\rightarrow(a,b+1,c+2)$ to reduce $a$, that is,
\ba
F[1,1,2] &= \frac{b+1}{a}\,F[0,2,2] - \frac{b-a+1}{a}\,F[0,1,2]
\ea
We get $F[0,2,2]$ from \refeq{2F1rE} with $a\leftrightarrow b$ and
$(a,b,c)\rightarrow(a,b+1,c+2)$:
\ba
F[0,2,2] &= \frac{c+1}{b+1}\,F[0,1,1] - \frac{c-b}{b+1}\,F[0,1,2]
\ea
$F[0,1,1]$ and $F[0,1,2]$ we can get from
\begin{center}
	\begin{tabular}{l@{ with $(a,b,c)\rightarrow$ }c}
		\refeq{2F1rE} & $(b,a,c+1)$ \\
		\refeq{2F1rF} & $(b,a,c)$ \\
	\end{tabular}
\end{center}
which are
\ba
\label{eq:F011}
F[0,1,1] &= -\frac{1}{b}\Big[(c-b)F[0,0,1] - c F[0,0,0]\Big] \\
F[0,0,1] &= -\frac{c}{(c-a)z}\Big[(1-z)F[0,0,0] - F[0,-1,0]\Big]
\label{eq:F001}
\ea
By applying \refeqs{F011}{F001} multiple times we get
\ba
F[1,1,2]
&=
\frac{c+1}{a}\,F[0,1,1]
- \frac{c+1-a}{a}\,F[0,1,2]
\\
&=
\frac{c+1}{a}\,\frac{1}{z}\bigg[F[0,1,1] - F[0,0,1]\bigg]
\\
&=
\frac{c+1}{a}\,\frac{c}{b}\,\frac{1}{z}\bigg[
F[0,0,0] - F[0,0,1]
\bigg]
\\
&=
\frac{c+1}{a(c-a)}\,\frac{c}{b}\,\frac{1}{z^2}\bigg[
(c-az)F[0,0,0] - cF[0,-1,0]
\bigg]
\label{eq:F112-recursion}
\ea
where $a$, $b$, and $c$ are evaluated with the $\Delta\ell$ that corresponds
to $F[0,0,0]$. For the full recursion we need to see how \refeq{Aelldell}
changes. It is
\ba
A_{\ell,\Delta\ell+2}
&= A_{\ell,\Delta\ell}\,R^2\,
\frac{\big(\ell+\frac12+\frac12\Delta\ell + \frac12n\big)\big(-\frac12\Delta\ell-\frac12n\big)}
{\big(\ell+\frac32+\Delta\ell+1\big)\big(\ell+\frac32+\Delta\ell\big)}
\\
&= -A_{\ell,\Delta\ell}\,R^2\,\frac{ab}{c(c+1)}
\ea
Combining with \refeq{F112-recursion} we get
\ba
\tilde F[1,1,2]
&=
\frac{1}{(c-a)z}\bigg[
c\tilde F[0,-1,0]
-(c-az)\tilde F[0,0,0]
\bigg]
\ea

\subsection{Towards \texorpdfstring{$\Delta\ell=4$}{dl=4} when \texorpdfstring{$R<1$}{R<1}}
All we need here is to reverse the relations in \refsec{recursion-dlm4}. That
is, we are given $F[-2,-2,-4]$ and $F[-1,-1,-2]$, and we want to calculate
$F[0,0,0]$. We can then shift the result to get the relation for $F[2,2,4]$
instead. Solving \refeq{Fm2m2m4-recursion} for $F[0,0,0]$ we get
\ba
F[0,0,0]
&=
\frac{(c-1)(c-2)(c-3)}{(a-1)(b-1)(c-4)(c-1-a)z(c-1-b)z}\bigg[
\vs&\qquad
\Big((c-2)(c-4) - \big[a(c-b) + b(c-a) - 3c + 4\big]z\Big)\,F[-1,-1,-2]
\vs&\qquad
- (c-2)(c-4)F[-2,-2,-4]
\bigg]
\ea
Applying the transformation $(a,b,c)\rightarrow(a+2,b+2,c+4)$ we get
\ba
F[2,2,4]
&=
\frac{(c+3)(c+2)(c+1)}{(a+1)(b+1)c(c+1-a)z(c+1-b)z}\bigg[
\vs&\qquad
\Big((c+2)c - \big[(a+2)(c+2-b) + (b+2)(c+2-a) - 3c - 8\big]z\Big)\,F[1,1,2]
\vs&\qquad
- (c+2)cF[0,0,0]
\bigg]
\ea

\subsection{Towards \texorpdfstring{$\Delta\ell=-2$}{dl=-2} when \texorpdfstring{$R>1$}{R>1}}
We use \refeq{Mellellsymmetry} to avoid arguments of the hypergeometric
function $R^2>1$. That means that $\ell$ and $\ell'$ get swapped, and we need
to derive new recurrence relations. The swapping may be done by first writing
$\Delta\ell$ in terms of $\ell$ and $\ell'$, then swapping
$\ell\leftrightarrow\ell'$, and replacing $\ell'=\ell+\Delta\ell$. Then we get
\ba
a &= \tfrac12n - \tfrac12\Delta\ell \\
b &= \ell+\tfrac12 + \tfrac12n + \tfrac12\Delta\ell = c + a - 1 + \Delta\ell \\
c &= \ell + \tfrac32
\ea
In other words,
\ba
(a,b,c) &\rightarrow (a+1,b-1,c) \qquad\text{for } \Delta\ell=-2 \\
(a,b,c) &\rightarrow (a-1,b+1,c) \qquad\text{for } \Delta\ell=2\,.
\ea
Hence, we need $F[1,-1,0]$ and $F[-1,1,0]$ in terms of $F[0,0,0]$ and
$F[0,-1,0]$.

For $\Delta\ell=-2$ we use \refeq{2F1rB} with $b\rightarrow b-1$ to get
\ba
F[1,-1,0] &= \frac{b-1}{a}\,F[0,0,0] - \frac{b-1-a}{a}\,F[0,-1,0]
\ea
To use \refeq{Aelldell} for the full kernel recursion in the case $R>1$, we
need to exchange $\ell$ and $\ell'$. Furthermore, \refeq{Mellellsymmetry} says
that we need to let
\ba
\alpha &\rightarrow R\alpha \\
R &\rightarrow R^{-1}
\ea
Then,
\ba
A^{R>1}_{\ell,\Delta\ell}
&= \alpha^{it-q} \, 2^{n-2} \, \pi \, R^\ell \,
\frac{\Gamma\big[\ell + \frac12 + \frac12\Delta\ell + \frac12 n\big]}
{\Gamma\big[1 + \frac12\Delta\ell - \frac12 n\big]\,\Gamma\big(\ell + \frac32\big)}
\label{eq:AelldellRg1}
\ea
Thus,
\ba
A^{R>1}_{\ell,\Delta\ell-2}
&= A^{R>1}_{\ell,\Delta\ell}
\,\frac{\frac12\Delta\ell-\frac12n}
{\ell - \frac12 + \frac12\Delta\ell + \frac12 n}
\\
&= - \frac{a}{b-1} \, A^{R>1}_{\ell,\Delta\ell}
\ea

\subsection{Towards \texorpdfstring{$\Delta\ell=-4$}{dl=-4} when \texorpdfstring{$R>1$}{R>1}}
For this recursion we need $F[2,-2,0]$ in terms of $F[0,0,0]$ and $F[1,-1,0]$.
We use
\begin{center}
	\begin{tabular}{l@{ with $(a,b,c)\rightarrow$ }c}
		\refeq{2F1rB} & $(a,b-1,c)$ \\
		\refeq{2F1rC} & $(a,b-1,c)$ \\
		\refeq{2F1rC} & $(b-2,a+1,c)$ \\
		\refeq{2F1rB} & $(a+1,b-2,c)$
	\end{tabular}
\end{center}
which results in
\ba
(b-1-a)F[0,-1,0] &= (b-1)F[0,0,0] - aF[1,-1,0] \\
(c-b+1)F[0,-2,0] &= (c-a-b+1)F[0,-1,0] + a(1-z)F[1,-1,0] \\
(c-b+1-a)F[1,-2,0] &= (c-a-1)F[0,-2,0] - (b-2)(1-z)F[1,-1,0] \\
(a+1)F[2,-2,0] &= (b-2)F[1,-1,0] - (b-a-3)F[1,-2,0]
\ea

\subsection{Towards \texorpdfstring{$\Delta\ell=2$}{dl=2} when \texorpdfstring{$R>1$}{R>1}}
We want $F[-1,1,0]$ from $F[0,0,0]$ and $F[0,-1,0]$.
We use
\begin{center}
	\begin{tabular}{l@{ with $(a,b,c)\rightarrow$ }c}
		\refeq{2F1rA} & $(b,a,c)$ \\
		\refeq{2F1rC} & $(b,a,c)$ \\
		\refeq{2F1rC} & $(a-1,b+1,c)$
	\end{tabular}
\end{center}
which results in
\ba
b(1-z)F[0,1,0] &= (c-b)F[0,-1,0] + (2b-c+(a-b)z)F[0,0,0] \\
(c-a)F[-1,0,0] &= (c-b-a)F[0,0,0] + b(1-z)F[0,1,0] \\
(c-a-b)F[-1,1,0] &= (c-b-1)F[-1,0,0] - (a-1)(1-z)F[0,1,0]
\ea
From \refeq{AelldellRg1} we get
\ba
A^{R>1}_{\ell,\Delta\ell+2} &= A^{R>1}_{\ell,\Delta\ell}
\,\frac{\ell+\frac12+\frac12\Delta\ell+\frac12n}
{1 + \frac12\Delta\ell - \frac12n}
\\
&= A^{R>1}_{\ell,\Delta\ell}
\,\frac{b}{1-a}
\ea

\subsection{Towards \texorpdfstring{$\Delta\ell=4$}{dl=4} when \texorpdfstring{$R>1$}{R>1}}
For this recursion we need $F[-2,2,0]$ in terms of $F[0,0,0]$ and $F[-1,1,0]$.
We use
\begin{center}
	\begin{tabular}{l@{ with $(a,b,c)\rightarrow$ }c}
		\refeq{2F1rB} & $(a-1,b,c)$ \\
		\refeq{2F1rC} & $(b,a-1,c)$ \\
		\refeq{2F1rC} & $(a-2,b+1,c)$ \\
		\refeq{2F1rB} & $(a-2,b+1,c)$
	\end{tabular}
\end{center}
which results in
\ba
(b-a+1)F[-1,0,0] &= bF[-1,1,0] - (a-1)F[0,0,0] \\
(c-a+1)F[-2,0,0] &= (c-a-b+1)F[-1,0,0] + b(1-z)F[-1,1,0] \\
(c-b-a+1)F[-2,1,0] &= (c-b-1)F[-2,0,0] - (a-2)(1-z)F[-1,1,0] \\
(b+1)F[-2,2,0] &= (b-a+3)F[-2,1,0] + (a-2)F[-1,1,0]
\ea

\section{Angular power spectrum with redshift-space distortion}
\label{app:cl}
To calculate redshift-space distortion (RSD) we use the well-known equation
\ba
C_\ell
&= \int \dd{}z\, W(z)\,D(z) \int \dd{}z'\,W'(z')\,D(z')
\big[
bb'w_{\ell,00}
- bf'w_{\ell,02}
- fb'w_{\ell,20}
+ ff'w_{\ell,22}
\big]
\label{eq:Cell-wljjrr}
\ea
where $W(z)$ and $W'(z')$ are window functions, $D(z)$ and $D(z')$ are
growth factors, $b$ and $b'$ are linear biases, $f$ and $f'$
dimensionless linear growth rates, and $w_{\ell,jj'}(\chi,\chi')$ were defined
in \refeq{wljj}.
Using the recurrence relation for spherical Bessel-$j$ functions
\ba
j'_\ell(x) &= \frac{\ell}{2\ell+1} j_{\ell-1}(x) - \frac{\ell+1}{2\ell+1}
j_{\ell+1}(x)
\ea
which results in
\ba
j''_\ell(x)
&= f_{-2} \, j_{\ell-2}(x)
+ f_0 \, j_\ell(x)
+ f_2 \, j_{\ell+2}(x)
\ea
where
\ba
f_{-2} &= \frac{\ell (\ell-1)}{(2\ell-1)(2\ell+1)} &
f_0 &= -\frac{2\ell^2 + 2\ell - 1}{(2\ell-1) (2\ell+3)} &
f_2 &= \frac{(\ell+1) (\ell+2)}{(2\ell+1)(2\ell+3)}\,.
\ea
We can express the terms in \refeq{Cell-wljjrr} in terms of
$w_{\ell\ell'}(\chi,\chi')$ in the following way:
\ba
w_{\ell,00}(\chi,\chi') &= w_{\ell,\ell}(\chi,\chi')
\\
w_{\ell,02}(\chi,\chi') &= \left(\begin{matrix}f_{-2} & f_0 & f_2\end{matrix}\right)
\left(\begin{matrix}
    w_{\ell,\ell-2}(\chi,\chi') \\
    w_{\ell,\ell}(\chi,\chi')  \\
    w_{\ell,\ell+2}(\chi,\chi')
\end{matrix}\right)
\\
w_{\ell,20}(\chi,\chi') &= \left(\begin{matrix}f_{-2} & f_0 & f_2\end{matrix}\right)
\left(\begin{matrix}
    w_{\ell-2,\ell}(\chi,\chi') \\
    w_{\ell,\ell}(\chi,\chi')  \\
    w_{\ell+2,\ell}(\chi,\chi')
\end{matrix}\right)
\\
w_{\ell,22}(\chi,\chi')
&= \left(\begin{matrix}f_{-2} & f_0 & f_2\end{matrix}\right)
\left(\begin{matrix}
    w_{\ell-2,\ell-2}  &  w_{\ell-2,\ell}  &  w_{\ell-2,\ell+2} \\
    w_{\ell,\ell-2}    &  w_{\ell,\ell}    &  w_{\ell,\ell+2} \\
    w_{\ell+2,\ell-2}  &  w_{\ell+2,\ell}  &  w_{\ell+2,\ell+2}
\end{matrix}\right)
\left(\begin{matrix}f_{-2} \\ f_0 \\ f_2\end{matrix}\right)
\ea
where the $w_{\ell\ell'}(\chi,\chi')$ are given by \refeq{well}.

Finally, note that
\ba
w_{\ell=i+j,\ell'=i+k}
&= w_{\ell=i+j,\Delta\ell=k-j}
\ea
for any integers $i,j,k$. This means that we can calculate all
$w_{\ell\pm2,\ell\pm2}(\chi,\chi')$ from $w_{\ell,\ell\pm(0,2,4)}(\chi,\chi')$ with
$\ell\pm2$, as indicated in \reffig{ellladder} by the gray squares.

\section{The Lucas~1995 algorithm}
\label{app:lucas1995}
As a benchmark calculation of the integrals over two Bessel functions, we use
the algorithm proposed by \citet{lucas1995}. We use it because it
takes into account the entire integration range from $k=0$ to $k=\infty$ to
high accuracy. Here, we summarize the algorithm applied to two spherical
Bessel functions.

The idea of \citet{lucas1995} is to add and subtract a product of
Bessel-$Y$ functions such that the product of Bessel-$J$ functions splits into
two summands, each of which is asymptotically proportional to a sine function.
That is,
\ba
j_\ell(k\chi) \, j_{\ell'}(k\chi')
&= h_1(k; \ell, \ell', \chi, \chi')
+ h_2(k; \ell, \ell', \chi, \chi')
\ea
where
\ba
h_1(k; \ell, \ell', \chi, \chi')
&= \tfrac12\big[
    j_\ell(k\chi) \, j_{\ell'}(k\chi')
    - y_\ell(k\chi) \, y_{\ell'}(k\chi')
\big]
\\
h_2(k; \ell, \ell', \chi, \chi')
&= \tfrac12\big[
    j_\ell(k\chi) \, j_{\ell'}(k\chi')
    + y_\ell(k\chi) \, y_{\ell'}(k\chi')
\big]\,.
\ea
The functions $h_1$ and $h_2$ behave asymptotically like sine functions. That
is, for $k\gg1$
\ba
h_1(k; \ell, \ell', \chi, \chi')
&\sim \frac{1}{2 \chi\chi'k^2}\,\cos\big[(\chi+\chi')k - \tfrac\pi2(\ell+\ell'+1)\big]
\\
h_2(k; \ell, \ell', \chi, \chi')
&\sim \frac{1}{2 \chi\chi'k^2}\,\cos\big[(\chi-\chi')k + \tfrac\pi2(\ell-\ell')\big]\,.
\ea
Then for large $k$, the $h_1$ and $h_2$ terms are integrated between
successive zeros, and the resulting alternating series is summed via a
series acceleration. The series acceleration effectively integrates to
$k=\infty$. We use the Levin u-transform as described in
\cite{NR3rd}, which is also our algorithm of choice in
\refsec{xifftlog-accuracy}. We use the same \texttt{quadosc} algorithm as
summarized there.

The case $\chi=\chi'$ makes the function $h_2$ nonoscillatory for large $k$.
This case is thus treated specially, by integrating the $h_2$ term to infinity
via Gauss-Kronrod integration, and applying the \texttt{quadosc} algorithm to the
$h_1$ term only.

For the evaluation of the spherical Bessel-$j$ and spherical Bessel-$y$ functions we use the
Bessel function implementations included in the \texttt{Julia} programming
language version 0.5.

For small $k$, the spherical Bessel-$y$ functions tend towards infinity, which
can lead to catastrophic cancellation in the summation of the series. Hence,
for the first few zeros, the integral is calculated directly via adaptive
Gauss-Kronrod integration without the splitting into $h_1$ and $h_2$. We found
that doing this for the first $\ell^2$ approximate zeros works fairly well,
although for large $\ell$ that puts the burden of the calculation on the
Gauss-Kronrod integration.

This procedure works well when $\ell$ is small. For large $\ell\gtrsim200$ the
adaptive Gauss-Kronrod integration needs to integrate over many oscillations,
making the integration slow and possibly fail. Further investigation may
reveal the exact nature of the problem. However, we find that reducing the
relative error
tolerance to \num{e-10} seems to work very well, and it is our choice in this
paper.

\section{Generalized Limber approximation}
\label{app:general_limber}
In order to be applicable to the lensing-convergence-galaxy cross-correlation
\refeq{cl2-kappa-g-chi}, we must extend the Limber approximation as written in
\refsec{psipsi-limber} to the cases $\ell\neq\ell'$.

From the result of Ref.~\citep{loverde/afshordi:2008},
\ba
\label{eq:limber-core}
\int_0^\infty dr\,f(r)\,J_\nu(kr)
&=
k^{-1}\,f\!\left(\tfrac{\nu}{k}\right) + \orderof\big(f''\!\left(\tfrac{\nu}{k}\right)\big)\,,
\ea
where $\nu=\ell+\tfrac12$ and $J_\nu(kr)$ are Bessel functions.
Defining $c'_1=b'_g$ and $c'_2=f'$, and
\ba
A &= \tfrac32\Omega_m H_0^2\,\ell(\ell+1)
\\
\varphi_i(\chi,\chi') &=
\frac{1}{\chi}\,
\frac{\chi_\star-\chi}{\chi_\star}\,
(1+z)\,
D(z)D(z')\,
c_i'\,
\frac{1}{\sqrt{\chi\chi'}}
\\
j_\ell(k\chi)
&=
\sqrt{\frac{\pi}{2k\chi}}\,
J_{\ell+\frac12}(k\chi)
\\
\nu &= \ell + \tfrac12
\\
\nu' &= \ell' + \tfrac12\,,
\ea
and using the approach in \refapp{cl}, we see that \refeq{cl2-kappa-g-chi}
contains terms of the form
\ba
C^i_{\ell\ell'}(\chi_\star,\chi')
&=
A
\int_0^\infty dk\,k^{-1}\,P(k)\,
J_{\nu'}(k\chi')
\int_0^{\chi_\star} d\chi\,
\varphi_i(\chi,\chi')
J_\nu(k\chi)\,.
\ea
Applying \refeq{limber-core} twice, we get
\ba
C^i_{\ell\ell'}(\chi_\star,\chi')
&=
A
\int_0^\infty dk\,k^{-2}\,
\varphi_i\left(\frac{\nu}{k}\chi',\chi'\right)\,
P(k)\,
J_{\nu'}(k\chi')
\nonumber\\
&=
\frac{A\,\chi'}{\nu'^2}\,
\varphi_i\left(\frac{\nu}{\nu'}\,\chi',\chi'\right)\,
P\left(\frac{\nu'}{\chi'}\right)\,.
\ea
Then to get the full power spectrum, we use the same approach as in
\refapp{cl}.

\end{widetext}

\end{document}

%% file: flow-ell-ladder2.tex
\begin{tikzpicture}[
        scale=0.66, transform shape,
        wlc/.style={
            circle,
            inner sep=0,
            minimum size=7mm,
	    text height=1.5ex,
	    text depth=0.1ex,
            draw=black,
            fill=white,
        },
        wl/.style={
            circle,
            inner sep=0,
            minimum size=7mm,
	    text height=1.5ex,
	    text depth=0.15ex,
            draw=black,
            fill=black!20
        },
    ]
    \newcommand{\lmax}{6}
    \pgfmathtruncatemacro\lmaxpp{\lmax+2}
    \pgfmathtruncatemacro\lmaxppp{\lmaxpp+1}
    \pgfmathtruncatemacro\lmaxppdl{\lmaxpp+4}
    \pgfmathtruncatemacro\lmaxppmdl{\lmaxpp-4}
    \draw[->] (0,0) -- (10,0) node[right]{\Large $\ell$};
    \draw[->] (0,0) -- (0,10) node[above]{\Large $\ell'$};
    \draw (\lmaxpp,0.1) -- +(0,-0.2) node[below]{\Large $\ell_{\rm max}$};
    \foreach \x in {2,...,\lmax} {
        \node at (\x,\x) [draw=black,line width=1,draw opacity=0.2,rectangle,minimum size=4cm,fill=black,fill opacity=0.2] {};
    }
    \foreach \x in {0,...,\lmaxpp} {
        \ifthenelse {\x < 2}{
            \node[wl] (w\x\x) at (\x,\x) {\x,\x};
        }{
            \ifthenelse{\x > \lmax}{
                \node[wl] (w\x\x) at (\x,\x) {\x,\x};
            }{
                \node[wlc] (w\x\x) at (\x,\x) {\x,\x};
            }
        }
        \pgfmathtruncatemacro\y{\x + 2}
        \node[wl] (w\x\y) at (\x,\y) {\x,\y};
        \draw[line width=2,<-,black!70] (w\x\x) -> (w\x\y);
        \pgfmathtruncatemacro\z{\x + 4}
        \node[wl] (w\x\z) at (\x,\z) {\x,\z};
        \draw[line width=2,<-,black!70] (w\x\y) -> (w\x\z);
        \pgfmathtruncatemacro\y{\x - 2}
        \node[wl] (w\x\y) at (\x,\y) {\x,\y};
        \draw[line width=2,->,black!70] (w\x\x) -- (w\x\y);
        \pgfmathtruncatemacro\z{\x - 4}
        \node[wl] (w\x\z) at (\x,\z) {\x,\z};
        \draw[line width=2,->,black!70] (w\x\y) -- (w\x\z);
    }
    \draw[line width=3,->] (\lmaxppp,\lmaxpp+4) node[right]{\large $R\leq1$} -- (w\lmaxpp\lmaxppdl);
    \draw[help lines,line width=3,dashed,->] (\lmaxppp,\lmaxpp-4) node[right]{\large $R>1$} -- (w\lmaxpp\lmaxppmdl);
    \foreach \x [remember=\x as \y (initially 0)] in {1,...,\lmaxpp} {
        \pgfmathtruncatemacro\xdl{\x+4}
        \pgfmathtruncatemacro\ydl{\y+4}
        \draw[line width=3,<-] (w\y\ydl) -- (w\x\xdl);
    }
\end{tikzpicture}

%% file: flowchart_2f1-v2.tex
\newcommand{\mybox}[2]{%
    \begin{minipage}{#1}
        \centering
        #2%
    \end{minipage}
}

\begin{tikzpicture}[
        node distance=5mm,
        terminal/.style={
            rectangle,minimum size=6mm,rounded corners=3mm,
            thick,draw=black!50,
            top color=white,bottom color=black!20,
        },
        nonterminal/.style={
            rectangle,
            minimum size=6mm,
            thick,
            draw=red!50!black!50,
            top color=white,
            bottom color=red!50!black!20, 
        },
        >=stealth,thick,black!50,text=black,
        h skip loop/.style = {to path={-- ++(0,#1) -| (\tikztotarget)}},
        v skip loop/.style = {to path={-- ++(#1,0) |- (\tikztotarget)}},
        hv path/.style = {to path={-| (\tikztotarget)}},
        vh path/.style = {to path={|- (\tikztotarget)}},
        hvh path/.style = {to path={-| ($(\tikztotarget) + (-1.8,0)$) |- (\tikztotarget)}},
        vhv path/.style = {to path={|- ($(\tikztotarget) + (0,2.1)$) -| (\tikztotarget)}},
        mpath/.style = {line width=2},
    ]
    \newcommand{\Mphiw}{2.2cm}
    \newcommand{\ffttext}{\mybox{1.2cm}{$\mathrm{FFT}_{t\rightarrow\chi}$}}
    %
    \node (start) [terminal] {\mybox{1.5cm}{Start}};
    %
    \node (Aellcalc) [nonterminal,below=of start.south west,anchor=north east] {\mybox{2.5cm}{
        Compute \\ $A_{\ell=0,\Delta\ell=\pm4}$ \\ \refeq{Aelldell}}};
    \node (fmaxcalc) [nonterminal,below=of start.south east,anchor=north west] {\mybox{2.5cm}{
        Compute \\
        ${}_2F_{1,{\ell=0,\Delta\ell=\pm4}}$ \\
        \refeqs{F000-ell0-dl4}{F010-ell0-dl4}}};
    \coordinate (Mphimultxcoord) at ($0.5*(Aellcalc) + 0.5*(fmaxcalc)$);
    \node (Mphimult) [nonterminal,below=of Mphimultxcoord |- Aellcalc.south]
        {\mybox{0.5cm}{$\times$}};
    \node (Mll0) [terminal,below=of Mphimult] {\mybox{2.5cm}{
        $M^{q}_{\ell=0,\Delta\ell=\pm4}(t,R)$}};
    \node (Millerplus) [nonterminal,below=of Mll0] {\mybox{2.6cm}{
        Compute $M^q_{\ell_\max,\Delta\ell=\pm4}(t,R)$ via \refapp{hyp2f1}}};
    \node (Mlmax) [terminal,below=of Millerplus] {\mybox{2.6cm}{
        $M^{q}_{\ell_\max,\Delta\ell=\pm4}(t,R)$}};
    %
    \node (Mll) [terminal,below=of Mlmax,yshift=-6.3mm] {\mybox{\Mphiw}{
        $M^{q}_{\ell,\Delta\ell=\pm4}(t,R)$}};
    \node (elltoellm1) [nonterminal,left=of Mll,xshift=2.5mm] {\mybox{2.5cm}{
        $\ell\rightarrow\ell-1$ \\ via \refeq{Aellp1} and \\ \refeqs{b+1a-back}{b+1b-back}}};
    %
    \node (dlcalc) [nonterminal,right=of Mll,anchor=north west,yshift=3mm]
        {\mybox{1.4cm}{\vspace{-4mm}
            \ba
            \Delta\ell&\rightarrow+4             \nonumber \\[1.1ex]
            \Delta\ell&\rightarrow+2             \nonumber \\[1.1ex]
            \Delta\ell&\rightarrow\phantom{\pm}0 \nonumber \\[1.1ex]
            \Delta\ell&\rightarrow-2             \nonumber \\[1.1ex]
            \Delta\ell&\rightarrow-4             \nonumber
            \ea
        }};
    \node (dlp4) [terminal,right=of dlcalc.north east,yshift=-3mm]
        {\mybox{\Mphiw}{$M^q_{\ell,\Delta\ell=4}(t,R)$}};
    \node (dlp2) [terminal,below=of dlp4,yshift=5mm]
        {\mybox{\Mphiw}{$M^q_{\ell,\Delta\ell=2}(t,R)$}};
    \node (dle0) [terminal,below=of dlp2,yshift=5mm]
        {\mybox{\Mphiw}{$M^q_{\ell,\Delta\ell=0}(t,R)$}};
    \node (dlm2) [terminal,below=of dle0,yshift=5mm]
        {\mybox{\Mphiw}{$M^q_{\ell,\Delta\ell=-2}(t,R)$}};
    \node (dlm4) [terminal,below=of dlm2,yshift=5mm]
        {\mybox{\Mphiw}{$M^q_{\ell,\Delta\ell=-4}(t,R)$}};
    \node (pkm4) [nonterminal,right=of dlm4] {\mybox{0.6cm}{$\times$}};
    \node (pkm2) at (pkm4 |- dlm2) [nonterminal] {\mybox{0.6cm}{$\times$}};
    \node (pke0) at (pkm4 |- dle0) [nonterminal] {\mybox{0.6cm}{$\times$}};
    \node (pkp2) at (pkm4 |- dlp2) [nonterminal] {\mybox{0.6cm}{$\times$}};
    \node (pkp4) at (pkm4 |- dlp4) [nonterminal] {\mybox{0.6cm}{$\times$}};
    %
    \node (phi) [terminal,above=of pkp4] {\mybox{2cm}{$\phi^q(t)$}};
    \node (pkfft) [nonterminal,above=of phi] {\mybox{2cm}{
        $\mathrm{FFTlog}_{k\rightarrow t}$}};
    \node (pkq) [terminal,above=of pkfft]{\mybox{2cm}{$k^{3-q}P(k)$}};
    \node (fftm4) [nonterminal,right=of pkm4]{\ffttext};
    \node (fftm2) [nonterminal,right=of pkm2]{\ffttext};
    \node (ffte0) [nonterminal,right=of pke0]{\ffttext};
    \node (fftp2) [nonterminal,right=of pkp2]{\ffttext};
    \node (fftp4) [nonterminal,right=of pkp4]{\ffttext};
    \node (prem4) [nonterminal,right=of fftm4] {\mybox{0.6cm}{$\times$}};
    \node (prem2) [nonterminal,right=of fftm2] {\mybox{0.6cm}{$\times$}};
    \node (pree0) [nonterminal,right=of ffte0] {\mybox{0.6cm}{$\times$}};
    \node (prep2) [nonterminal,right=of fftp2] {\mybox{0.6cm}{$\times$}};
    \node (prep4) [nonterminal,right=of fftp4] {\mybox{0.6cm}{$\times$}};
    \node (prefactor) [terminal,above=of prep4]
        {$4k_0^3G^{-1}\left(\dfrac{\chi}{\chi_0}\right)^{-q}$};
    \node (wm4) [terminal,right=of prem4]{\mybox{1.9cm}{$w_{\ell,\ell-4}(\chi,R)$}};
    \node (wm2) [terminal,right=of prem2]{\mybox{1.9cm}{$w_{\ell,\ell-2}(\chi,R)$}};
    \node (we0) [terminal,right=of pree0]{\mybox{1.9cm}{$w_{\ell,\ell}(\chi,R)$}};
    \node (wp2) [terminal,right=of prep2]{\mybox{1.9cm}{$w_{\ell,\ell+2}(\chi,R)$}};
    \node (wp4) [terminal,right=of prep4]{\mybox{1.9cm}{$w_{\ell,\ell+4}(\chi,R)$}};
    \graph[use existing nodes] {
        start ->[hv path] {Aellcalc, fmaxcalc} ->[vh path] Mphimult -> Mll0
        -> Millerplus -> Mlmax -> Mll;
        \foreach \x in {m4,m2,e0,p2,p4} {
            dl\x ->[mpath] pk\x ->[mpath] fft\x ->[mpath] pre\x ->[mpath] w\x;
        };
        pkq -> pkfft -> phi ->[mpath] pkp4;
        prefactor ->[mpath] prep4;
    };
    \draw[<-,mpath] ($(Mll.north west) + (0.3,0)$) |- ($(elltoellm1.north) + (0,0.4)$)
        -- (elltoellm1.north);
    \draw[->,mpath] ($(Mll.south west) + (0.3,0)$) -- +(0,-0.8) -| (elltoellm1.south);
    \draw[->,mpath] (Mll.east) -- (Mll.east -| dlcalc.west);
    \draw[->,mpath] ($(Mll.south east) - (0.3,0)$) |- (dlcalc.west |- dlp2);
    \draw[->,mpath] (dlcalc.east |- dlp4) -- (dlp4);
    \draw[->,mpath] (dlcalc.east |- dlp2) -- (dlp2);
    \draw[->,mpath] (dlcalc.east |- dle0) -- (dle0);
    \draw[->,mpath] (dlcalc.east |- dlm2) -- (dlm2);
    \draw[->,mpath] (dlcalc.east |- dlm4) -- (dlm4);
\end{tikzpicture}